\documentclass[review]{elsarticle}

\usepackage{hyperref}
\usepackage{color}
\usepackage{amsmath}
\usepackage{stmaryrd}
\usepackage{graphicx}
\usepackage{subcaption}
\usepackage[export]{adjustbox}

\newcommand{\vect}[1]{\boldsymbol{{#1}}}

\newcommand{\dd}[1]{\ \text{d}{#1}}

\newcommand{\jump}[1]{\llbracket #1 \rrbracket}
\newcommand{\avg}[1]{\left\{ #1 \right\}}
\newcommand{\len}{\ell}

\newcommand{\comment}[1]{\ignorespaces}

\newtheorem{remark}{Remark}

\hypersetup{
    colorlinks, linkcolor=black,
    citecolor=black
}

\begin{document}

\begin{frontmatter}

  \title{A fully-coupled computational framework for large-scale \\ simulation of fluid-driven
    fracture propagation\\ on parallel computers}

\author[aa]{Bianca Giovanardi}

\author[bb]{Santiago Serebrinsky}
\author[aa]{Ra\'ul Radovitzky\corref{mycorrespondingauthor}}
\cortext[mycorrespondingauthor]{Corresponding author}
\ead{rapa@mit.edu}

\address[aa]{Massachusetts Institute of Technology, \\77 Massachusetts Ave, Cambridge, MA
  02139, U.S.A.}
\address[bb]{Y-TEC, YPF Tecnolog\'ia, S.A.\\Calle Baradero, Ensenada, 1925, Buenos Aires, Argentina}

\begin{abstract}
  The propagation of cracks driven by a pressurized fluid emerges in several areas of
  engineering, including structural, geotechnical, and petroleum engineering.  In this paper,
  we present a robust numerical framework to simulate fluid-driven fracture propagation that
  addresses the challenges emerging in the simulation of this complex coupled nonlinear
  hydro-mechanical response.  We observe that the numerical difficulties stem from the strong
  nonlinearities present in the fluid equations as well as those associated with crack
  propagation, from the quasi-static nature of the problem, and from the \emph{a priori}
  unknown and potentially intricate crack geometries that may arise. An additional challenge is
  the need for large scale simulation owing to the mesh resolution requirements and the
  expected 3D character of the problem in practical applications.  To address these challenges
  we model crack propagation with a high-order hybrid discontinuous Galerkin / cohesive zone
  model framework, which has proven massive scalability properties, and we model the
  lubrication flow inside the propagating cracks using continuous finite elements, furnishing a
  fully-coupled discretization of the solid and fluid equations.  We find that a
  conventional Newton-Raphson solution algorithm is robust even in the presence of crack
  propagation. The parallel approach for solving the linearized coupled problem consists of
  standard iterative solvers based on domain decomposition.  The resulting computational
  approach provides the ability to conduct highly-resolved and quasi-static simulations of
  fluid-driven fracture propagation with unspecified crack path.  We conduct a series of
  numerical tests to verify the computational framework against known analytical solutions in
  the toughness and viscosity dominated regimes and we demonstrate its performance in terms of
  robustness and parallel scalability, enabling simulations of several million degrees of
  freedom on hundreds of processors.

\comment{ 
The numerical framework allows fluid-filled cracks to propagate on the interelement boundary 
of the solid finite element mesh, hence triggering the fluid-solid coupling mechanisms
in a natural way, through suitable integrals on the elements 
boundary.  More precisely, a Discontinuous Galerkin finite element 
discretization of the solid enables a direct description of the crack opening between 
adjacent mesh elements and induces a lower dimensional finite element discretization for 
the fluid problem. Crack propagation is then modeled via hybrid cohesive-lubrication interface 
elements, whose cohesive traction-separation law is activated upon fracture initiation and 
lubrication flow is activated upon complete failure. This approach avoids the need to propagate 
topological changes in the mesh as cracks develop and propagate and, thus, enables massive 
scalability in parallel computations.}

\end{abstract}


\end{frontmatter}

\section{Introduction}

Fluid-driven fracture propagation concerns several areas of engineering, 
including structural, geotechnical, and petroleum engineering. 
In recent times there has been a flourishing of research geared at 
delivering computational tools for simulating fluid-driven fracture propagation,
mostly driven by the oil industry, 
where numerical simulations can be used to support the design of field operations 
by providing a physics-based framework to complement legacy approaches based on 
accumulated experience and statistical inference, therefore helping reduce economic risk.
The interested reader can consult, for example, \cite{Lecampion:2017} for a recent 
review of numerical methods that have been proposed to tackle this problem.

\comment{
Hydraulic fracturing is a technique that has been used for decades in the oil industry to
enhance production of formerly inaccessible hydrocarbons in unconventional reservoirs.
Numerical
simulations of hydraulic fracture can support the design of the operations by providing a sound
physics-based framework and therefore can help reduce the economical risk in oil and gas
exploration for energy production.  Driven by the interest in unconventional reservoirs, there
has been in the past few decades a flourishing of research geared at delivering powerful tools
for simulating fluid-driven fracture propagation.}

A thorough mathematical description of the complex physical problem associated with the
injection of pressurized fluid in a deformable solid requires the consideration of several
coupling mechanisms among the fluid flow, the deformation of the solid material and the 
possible propagation of cracks as a result of material failure \cite{EconomidesBook:2000}. 
The injection of fluid results in a pressure applied on the crack walls.
Material deformation and crack propagation increase the volume available to the fluid and
therefore affect the pressure distribution as well as the fluid flow. A basic mathematical
model for this problem involves the elastostatics equations which govern the deformation and 
stress field in the solid, the equations for the fluid flow inside the cracks, and a proper model for
describing crack propagation, which mathematically renders the problem of the moving boundary
type.  Depending on the specific application, additional refinements of the model may include the
consideration of leakage of fluid off the fracture walls into the adjacent porous rock, the transport of
proppant and the chemical reactions involved, and the effects of temperature on the rheology of the
injected fluids.

\comment{Under laminar flow conditions, which can be shown to characterize most industrial 
applications \cite{Zia2017}, the lubrication equation for flows in thin films can be employed to 
model the fluid in a one codimensional manifold to the rock environment.}
\comment{See \cite{Zia2017} for an analysis of the propagation regimes under turbulent flow.}

Early analytical progress in the description of fluid-driven fracture propagation 
was made by Khristianovic, Geertsma and de Klerk
\cite{Khristianovich:1955,Geertsma:1969}, who devised a one dimensional analytical model for a
plane-strain straight crack whose propagation is driven by the injection of an
incompressible viscous fluid at a constant rate. This model, referred to as KGD, solves the
coupled equations of elastostatics for the solid, lubrication flow and mass conservation for the fluid.
\comment{The KGD model does not explicitly describe the fluid leaving the crack
and flowing into the surrounding medium, but instead adds a time dependent sink term given by
Carter's equation \cite{Howard:1957}.}  
\comment{A generalization of the lubrication equation accounting explicitly for a fluid mass 
exchange between the crack and the surrounding medium is devised in \cite{Giovanardi:2018}.}  
\comment{Finally, a quasi-static crack propagation condition at the crack tip based on results from 
linear elastic fracture mechanics \cite{rice:1968} is used as a moving boundary condition to close the
model together with the trivial initial conditions.} 
Further systematic studies by Detournay \emph{et al.} have provided a complete picture of the simplified 
and analytically-tractable one-dimensional problem \cite{Adachi:2001,Adachi:2002,Detournay:2004,
Garagash-asme:2005,Bunger:2005,Adachi:2008}.
Specific advances included the identification and classification of asymptotic regimes where analytical 
solutions were found, each presenting unique characteristics in terms of near tip asymptotic behavior,
including the possible occurrence of boundary layers in certain regimes.

It is well-established that computational models can complement and significantly expand the 
extent of the analyses available to analytical methods by providing a full-field, albeit
discretized, description of the problem of interest. 
\comment{3D, general geometries, heterogeneities, non-brittle cases.}
In the case of fluid-driven fracture propagation, the coupled elasto-hydrodynamics problem is 
typically discretized using the finite element method to solve the elastic problem 
\cite{Boone:1990,Carrier:2012,Settgast:2017}, while finite differences \cite{Boone:1990}, finite 
elements \cite{Carrier:2012}, or finite volumes \cite{Settgast:2017} have been employed for modeling 
the lower dimensional fluid flow. 
\comment{finite element method is preferred 
to displacement-discontinuity methods (DDM) \cite{Gordeliy:2011,Lecampion:2007} 
to solve the elastic problem when material properties are heterogeneous, or in the 
presence of solid nonlinearities \cite{Lecampion:2017} } 
\comment{
For example, DDMs have been successfully employed to 
simulate benchmarks under specific propagation regimes and in transitions between 
different regimes \cite{Gordeliy:2011, Lecampion:2007}, to simulate the formation of 
complex fracture networks in layered media \cite{Zhang:2017} and the interaction 
between hydraulic and natural fractures \cite{Weng:2011}.}
It bears emphasis that even in the absence 
of crack propagation the hydro-mechanical
coupling is extremely strong, since the ability of the fluid to flow in the fracture depends
nonlinearly on the fracture opening. 

Different numerical strategies have been proposed to describe fracture propagation 
in the solid and to evolve the fluid computational domain as cracks propagate. 
A recent review paper \cite{Lecampion:2017} has described these approaches, 
identifying their main advantages and limitations. 


Cohesive zone models (CZM) have been widely used for their sound fracture mechanics 
basis and their ease of implementation as interface elements within finite element frameworks.
In the context of fluid-driven fracture propagation, the effectiveness of this approach has been 
verified in the case of an \emph{a priori} known straight crack
\cite{Boone:1990,Carrier:2012,ChenZ:2009,Sarris:2011,Hunsweck:2013}. 

The extended finite element method (XFEM), proposed
in \cite{Belytschko:1999} and \cite{moes:1999} as a technique to allow arbitrary crack
propagation without the need of re-meshing, has successfully been applied in the framework 
of fluid-driven fracture propagation in 2D benchmarks either in combination with CZM
\cite{Khoei:2015,Mohammadnejad:2016} or with classical mixed-mode 
propagation criteria  based on the stress field at the crack tip \cite{Gordeliy:2013}. 
The extremely challenging implementation of XFEM in the case of a pressurized crack 
propagating in 3D geometries was done in references \cite{Gupta:2014,Gupta:2016,Gupta:2018}.
\comment{
The extremely challenging extension of XFEM to deal with pressurized
cracks in 3D geometries was performed in \cite{Gupta:2014}, where the pressure field was 
assumed to be known, in \cite{Gupta:2016}, where the hydro-mechanical coupling was solved in 
the absence of propagation and, more recently, in \cite{Gupta:2018} where the hydro-mechanics 
was coupled with a crack propagation criterion based on linear elastic fracture mechanics.}
XFEM methods have been verified against analytical solutions for plane-strain
\cite{Khoei:2015} and penny-shaped \cite{Gupta:2018} impermeable cracks. 
\comment{
The numerical accuracy of the XFEM in various propagation regimes can be improved by 
the use of tailored tip enrichment functions, see for example \cite{Lecampion:2009} and 
\cite{Gordeliy:2015}.}
The main shortcoming of the XFEM approach lies in cases where there is merging or branching of 
several cracks \cite{Karihaloo:2003}. In addition, the suitability of this method for large-scale simulations, 
and therefore 3D scenarios, remains uncertain due to its unproven scalability 
\cite{seagraves:2010, Lecampion:2017}.



Another approach for describing crack propagation along arbitrary crack paths that has 
recently received significant attention is the phase-field model 
\cite{francfort:1998,bourdin:2000,Bourdin:2007}. 
Instead of modeling the displacement discontinuity explicitly, phase-field models represent
the crack surfaces by a scalar field that is responsible for distinguishing between cracked and 
uncracked states of the material and whose evolution is obtained by energy-minimization
principles. 
Phase-field models are straightforward to implement and have been widely used to simulate 
complex fracture geometries, as the description of crack branching and merging is naturally 
accounted for in the model. 
In the context of fluid-driven fracture propagation, however, the absence of an explicit description of 
the displacement jump constitutes a severe limitation, as the crack opening is a fundamental 
variable in the fluid flow equation that strongly controls the nonlinearity in the hydro-mechanical 
coupling. 
In a recent paper it was shown that in the limited case where the fluid pressure field is given,
the phase field energy functional can be cleverly formulated to account for the work of the 
pressurizing fluid without the need to compute the crack opening \cite{Wheeler:2014}. 
\comment{
This framework has been successfully verified in \cite{Wheeler:2014} against 
Sneddon's benchmark \cite{Sneddon:1946,Sneddon:1969}, that is in the trivial case of a 
given and uniform fluid pressure. }
However, in realistic scenarios the fluid pressure field is not known and the lubrication 
equation must be solved, which requires that the crack opening be reconstructed precisely.
Several heuristic techniques have been proposed to approximate the displacement jump based 
on the phase field, see for example \cite{Verhoosel:2013,Landis:2014,Miehe:2015,Lee:2016}, 
although a sound procedure to do so has not been developed yet \cite{Lecampion:2017}.
\comment{enable phase-field descriptions of fluid-driven fracture propagation}
Recent attempts combine the benefit of the smeared phase field formulation with a 
sharp description of the crack \cite{Giovanardi:2017}, but this approach has not been
applied to fluid-filled fracture yet.

In this paper we introduce a computational framework that addresses some of the most important
challenges in the modeling of fluid-driven fracture propagation: the emergence of arbitrary
crack geometries, including crack branching, coalescence, and interaction with pre-existing
cracks, and the requirement of massive parallel scalability owing to the inherently
three-dimensional nature of the problem in practical applications.  Some of the existing
limitations and remaining challenges were also identified in \cite{Lecampion:2017}.  We follow
a fully-coupled formulation of the mathematical model and a numerical
discretization similar to \cite{Carrier:2012}, which employed a combination of finite elements
(2D for the solid, 1D for the fluid) and a cohesive zone model to describe crack propagation.
The approach was proven very promising as it was verified against the analytical solutions
available in several regimes. However, the computational framework assumed a predefined
straight crack in 2D propagating under plane-strain conditions. Here we propose a number of
modifications to that approach, allowing for unspecified fracture path and enabling high
resolution simulations in 3D.  To this end, we adopt the Discontinuous Galerkin / Cohesive Zone
Model (DG/CZM) framework originally proposed in \cite{seagraves:2010,Seagraves:2015a} for the
massively parallel simulation of dynamic fracture and fragmentation of brittle solids.  In that
approach, the flux and stabilization terms arising at interelement boundaries from the DG
formulation prior to fracture are enforced via interface elements.  Upon the onset of fracture,
the Traction-Separation Law (TSL) governing the fracture process becomes operative without the
need to insert a new cohesive element.  A key feature of the method is that it avoids the need
to propagate topological changes in the mesh as cracks and fragments develop, which enables the
indistinctive treatment of crack propagation across processor boundaries and, thus, the
scalability of the method to thousands of processors and billion degrees of freedom in explicit
dynamics calculations \cite{seagraves:2010}.  For the discretization of the lubrication
equation we adopt, as in \cite{Carrier:2012}, a standard continuous Galerkin finite element
formulation whose support is the solid interface elements that have experienced complete
failure.

The application of the DG/CZM approach to fluid-driven fracture propagation has recently 
gained some traction in the community \cite{Serebrinsky:2016,Hirmand:2019}. In those 
contributions the authors tackle the dynamic problem with either explicit \cite{Serebrinsky:2016} 
or implicit \cite{Hirmand:2019} time 
integration and a weak coupling of the elasticity and fluid flow equation.
Both approaches introduce severe restrictions 
on the stable time step and, therefore, impose limitations on the accessible time scales, 
which are in practical terms reduced to subseconds.
Although such limitations can sometimes be mitigated by the use of different techniques, 
including dynamic relaxation, mass scaling, or introduction of artificial numerical dissipation, 
such type of approach is most useful in specific applications that require the description of the stress 
waves emanating from the crack tips as they propagate, for example in micro seismic analysis.
In the case of field operations of hydraulic fracturing, whose time scales are in the order of 
minutes or hours and where one is not interested in dynamic effects as the process is inherently 
quasi-static (fracture growth is stable because it is volume controlled \cite{Lecampion:2017}), 
a quasi-static simulation method is desirable if not mandatory.
The main focus of this paper is to enable robust quasi-static simulations of fluid-driven fracture 
propagation. 

The system of strongly nonlinearly coupled equations resulting from the discretization is
assembled and solved in parallel with a robust fully-coupled iterative algorithm. A
fully-coupled treatment of the coupled system is paramount in the solution of the quasi-static
problem. It properly exposes the strong nonlinearity of the coupled system owing to the cubic
dependence of the conductivity in the flow equations. It also clearly exposes the parabolic
character of the system, thus allowing for natural inflow boundary conditions in a
mathematically and numerically-sound manner. Previous approaches based on staggered iteration
between the fluid and solid problems \cite{Boone:1990, Khoei:2015,Vahab:2018,Hirmand:2019} are
mired with numerical problems whose root is the inconsistent linearization of the coupled
nonlinear problem into two weakly-coupled linear ones, and the artificial ellipticity inherited
in this case by the fluid problem, which forces the need for an also-artificial essential
boundary condition on the pressure. It
bears emphasis that all these issues disappear in the fully-coupled treatment of the coupled
quasi-static system, as the problem retains its parabolic nature and the system only requires a
single boundary condition which can be indistinctively of a Dirichlet or Neumann nature. We
will further show via numerical examples that the staggered approach fails even in the case of
applied pressure near the crack tip, due to the incorrect numerical treatment of the strong
nonlinear dependence of the inflow boundary condition on the crack opening at the mouth.

\comment{
This basic approach has been explored for application to fluid-filled fracture in 
\cite{Serebrinsky:2016}, where an explicit treatment of the hydro-mechanics in time 
and the addition of artificial dynamics to the inherently quasi-static problem 
limited the accuracy and robustness of the approach. 
In this work we employ quasi-static implementation of the DG/CZM framework, which we    
extend to account for the fluid flow inside the fractures in a fully-coupled fashion.}
\comment{A previous attempt to extend this method to problems in fluid-driven crack 
propagation consisted of an extension of the predominantly quasi-static problem to 
the dynamic regime, an explicit time integration of the governing equations.
which is limited by CFL stability see \cite{Lecampion:2017}.}
\comment{Within that framework we enable the arbitrariness of crack paths via 
hybrid DG/CZM - lubrication interface elements placed on all solid elements interfaces, 
where lubrication flow is activated upon complete interface failure.}

We conduct a series of numerical tests to verify the computational framework against
known analytical solutions for plane-strain and axisymmetric cracks in both the 
viscosity and toughness dominated regimes.
An important finding is that the proposed method results in fluid pressure distributions 
that are in excellent agreement with the theoretical predictions, even in the case of a 
\emph{zig-zag} fracture path.
This settles a concern that has been recently raised \cite{Lecampion:2017}.
We also use the case of the interaction of a propagating fluid-driven crack with a 
pre-existing dry crack 
as a benchmark to demonstrate that the proposed computational framework is able to 
describe fracture branching and merging robustly.
We finally demonstrate the parallel scalability of the framework up to 30 million degrees
of freedom on a 700 processors distributed memory machine.

The present work is structured as follows. In Section \ref{sec:governingEquations} we formulate 
the model with the relevant governing equations. In Section \ref{sec:finiteElements} we present 
the discretizations of the solid and the fluid, with particular focus on the
hybrid formulation of the solid (DG/CZM) and fluid (conventional continuous finite elements) and 
on the fully-coupled iterative algorithm to solve the resulting nonlinear coupled discrete system.
In Section \ref{sec:results} we verify the framework against available analytic solutions and we 
present results demonstrating its robustness and parallel scalability.
Conclusions are drawn in Section \ref{sec:conclusions}.
\section{Governing equations} \label{sec:governingEquations}

For completeness, we review the governing equations that describe the 
hydro-mechanical coupling of a fractured solid when a pressurized fluid is allowed to flow 
within the crack walls, see also \cite{Carrier:2012}.
We consider an impermeable brittle elastic medium and assume a fluid is injected in the solid 
cracks at a constant rate. 
The stress field in the solid and the fluid flow are tightly coupled in the following way: 
the fluid applies a pressure on the crack walls of the solid, 
whereas the crack geometry (opening and length) constitutes the fluid domain and 
therefore affects the fluid flow. The description of this two-way coupling requires the formulation
of the governing equations for the elastic deformation of the solid material, the laminar flow of the 
fluid in the cracks, and the propagation of cracks resulting from the stress field in the solid.

\begin{figure}[h]
\centering
\includegraphics[width=0.4\textwidth]{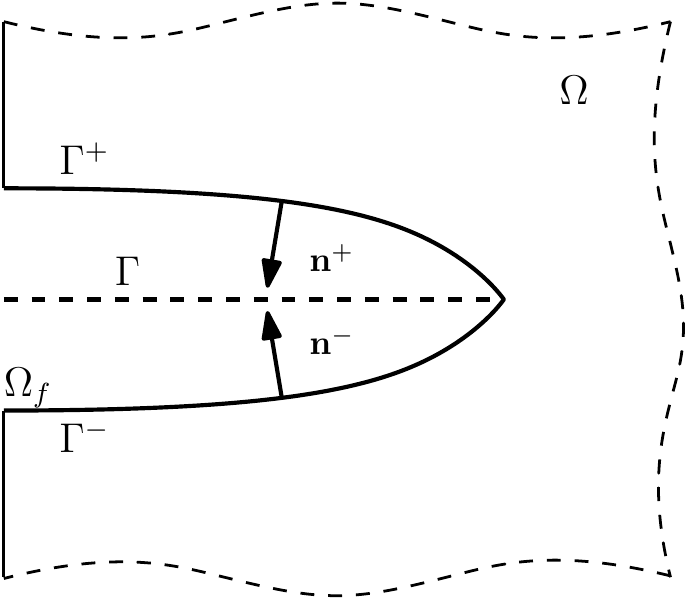}
\caption{\emph{Sketch of the crack geometry in 2D. The crack lips are 
$\Gamma^+$ and $\Gamma^-$ and the medium line of the crack $\Gamma$.}}
\label{fig:crackPlusMinus}
\end{figure}

We denote with $\Omega \subset  {\rm I\!R}^d, d = 2, 3$ the solid domain and with
$\Gamma^+, \Gamma^- \subset \partial \Omega$ the crack lips, as in Figure \ref{fig:crackPlusMinus}.
As the fluid exerts a pressure $p$ on the crack lips, the following equations state the 
equilibrium of the solid:

\begin{equation} \label{eq:elastostatics}
\left\{
\begin{aligned}
- \nabla \cdot \vect{\sigma} &= \vect{0}  \qquad &\mbox{ in }& \Omega \\
\vect{\sigma} \vect{n} &= -p \vect{n} \qquad &\mbox{ on }& \Gamma^+ \cup \Gamma^- \\
\vect{\sigma} \vect{n} &= \bar{\vect{t}} \qquad &\mbox{ on }& \partial\Omega_{N} \\
\vect{u} &= \vect{\bar{u}} \qquad &\mbox{ on }& \partial\Omega_{D}
\end{aligned}
\right. ,
\end{equation}
where $\vect{\sigma}$ is the Cauchy stress tensor and $\vect{u}$ the displacement 
field of the solid. In \eqref{eq:elastostatics}, $\partial\Omega_{N}$ and $\partial\Omega_{D}$ denote 
a partition of $\partial\Omega \setminus \left( \Gamma^+ \cup \Gamma^- \right)$, where far field 
traction $ \bar{\vect{t}}$ and imposed displacement $\vect{\bar{u}}$ are prescribed, respectively.

Small deformations and linear elasticity are commonly employed to model the behavior of the solid 
prior to fracture. More precisely, the stress field $\vect{\sigma}$ is related to the deformation by 
$\vect{\sigma} = \mathcal{C} \vect{\varepsilon}(\vect{u})$, where $\mathcal{C}$ is the 
fourth order elasticity tensor and 
$\vect{\varepsilon}(\vect{u}) = \frac{1}{2} \left( \nabla \vect{u} + \nabla \vect{u}^T \right)$ is 
the infinitesimal strain.

The fluid domain $\Omega_f  \subset  {\rm I\!R}^d$, is 
delimited by $\Gamma^+$ and $\Gamma^-$ as shown in Figure 
\ref{fig:crackPlusMinus}.  
The assumptions of laminar flow and 
and small crack openings $w$ relative to the crack length $\ell$
justify the use of Reynolds lubrication theory to model the fluid flow.
\comment{
Reynolds lubrication equation can be obtained by integrating the momentum balance equations 
of Stokes system across the crack, plugging the fluid's velocity in the continuity equation and 
integrating the continuity equation across the crack, as described in \cite{Hamrock:1994}.}
For the purpose of this paper, we do not account for fluid leaving the crack and flowing into the 
surrounding medium and assume the fractured solid to be impermeable. 
The model can be extended to account for fluid leak-off adding a time dependent sink term given 
by Carter's equation \cite{Howard:1957} as in \cite{Carrier:2012} or employing a generalization 
of the lubrication equation that explicitly models fluid mass exchange between the crack and the 
surrounding medium \cite{Giovanardi:2018}.
The governing equations for the fluid flow read: 

\begin{equation} \label{eq:lubrication}
\left\{
\begin{aligned}
- \nabla_{\Gamma} \cdot \left( \frac{w^3}{12 \mu} \nabla_{\Gamma} p \right) 
	&= - \frac{\partial w}{\partial t} &\mbox{ in } \Gamma \\
\frac{w^3}{12 \mu} \nabla_{\Gamma} p \cdot \vect{n} &= Q_0 &\mbox{ on } \partial\Gamma_{in}
\end{aligned}
\right. ,
\end{equation}
where $\Gamma  \subset  {\rm I\!R}^{d-1}$ is the medium crack surface identifying 
the fluid manifold, $w$ denotes the opening of the crack, and $\nabla_{\Gamma}$ 
the tangential gradient operator on $\Gamma$.  
In \eqref{eq:lubrication}, $\partial\Gamma_{in}\subset  {\rm I\!R}^{d-2},$ is the inflow part of the boundary of $\Gamma$ 
where fluid is injected at a constant volumetric rate $Q_0$, and $\mu$ denotes the fluid 
dynamic viscosity. 

The coupling of the fluid flow and the elastic deformations of the surrounding medium are 
manifested in the mechanical equilibrium requirement at the fluid-solid interface, Equation 
\eqref{eq:elastostatics}, and in the continuity requirement that the crack opening be equal 
to the normal component of the displacement jump:
\begin{equation} \label{eq:displacementJump}
w = \jump{\vect{u}} \cdot \vect{n}_{\Gamma},
\end{equation}
being $\jump{\bullet}$ the jump operator $\jump{\bullet} := \bullet^+ - \bullet^-$ and 
$\vect{n}_{\Gamma} =\vect{n}^- \simeq - \vect{n}^+$ the normal to $\Gamma$.
While in Equation \eqref{eq:elastostatics} the coupling with the fluid flow
appears only in the boundary conditions,
the coupling of Equation \eqref{eq:lubrication} with the solid mechanics is three-fold:
1) through the (cubic) dependence of the conductivity coefficient in the lubrication 
equation on the opening;
2) through the modification of the fluid domain $\Gamma$ as the crack propagates; 
3) through the contribution of the local time rate of change of the opening as a sink 
term in the lubrication equation.

Note that no boundary condition is prescribed
at the crack tip, as Equation \eqref{eq:lubrication} 
becomes degenerate for $w = 0$ and the natural boundary condition 
$w^3 \nabla_{\Gamma} p \cdot \vect{n} = 0$ is identically satisfied 
\cite{Adachi:2008}. 
The absence of Dirichlet boundary conditions has raised doubts about the well posedness of the 
boundary value problem \eqref{eq:lubrication} \cite{Boone:1990, Khoei:2015, Vahab:2018}. 
However, the issue of the nonuniqueness of 
pressure fields satisfying Equation \eqref{eq:lubrication}, which results from the formulation of an 
elliptic boundary value problem without essential boundary conditions, 
 disappears when this equation is
fully coupled with the solid mechanics through \eqref{eq:elastostatics} and 
\eqref{eq:displacementJump}, that is when the parabolic character of the equation is 
exposed.

As more and more fluid is injected in the cracks, the fluid pressure increases. 
A higher fluid pressure in the neighborhood of the crack tip leads to a higher opening stress, 
which may lead to crack propagation. We assume that the fluid pressure is 
at all times balanced 
by the stress field in the solid, so that the resulting crack propagation is quasi-static. 
As in \cite{Carrier:2012},
we employ a Cohesive Zone Model (CZM), originally proposed in \cite{barenblatt:1962} and 
\cite{barenblatt:1959}, to describe the quasi-static fracture propagation. 
We employ a linear traction separation law (TSL) \cite{camacho:1996} to relate the normal 
traction $\vect{t}$
\comment{$\vect{t} = (\vect{\sigma} \vect{n}_{\gamma} \cdot \vect{n}_{\gamma}) \vect{n}_{\gamma}$, 
where $\vect{\sigma}$ is the Cauchy stress} to the 
crack opening $w$ as follows:
\begin{equation}\label{eq:tsl}
\vect{t}(w)  = \sigma_c \left( 1 - \frac{w}{\delta_c} \right) \vect{n}_{\Gamma},
\end{equation}
where $\delta_c$ is the length of the cohesive zone and $\sigma_c$ is the critical stress for 
which the TSL is activated. Note that with this choice the fracture energy $G_c$ is given by
$$
G_c = \frac{1}{2} \sigma_c \delta_c.
$$

\comment{
The propagation of the crack is assumed to be quasi-static. The fracture
is assumed to be filled with the injected fluid from the inlet to the crack tip. 
We point out that we do not consider in the analysis the presence of a fluid lag, which can be 
neglected at sufficient confining stresses \cite{Garagash:1999}. Finally, we assume a laminar flow 
inside the crack and a small crack opening compared to the crack length, hypotheses behind 
lubrication theory.
Even in the simplifying framework where no poroelasticity and no leakage of the fluid outside 
the crack are considered, the model that we describe is able to capture the main coupling
mechanisms between the rock and the fluid.}

\comment{
Note that, for a given crack length, both systems \eqref{eq:elastostatics} and 
\eqref{eq:lubrication} are linear in their own primal variable, whereas \eqref{eq:lubrication}
is nonlinear in its coupling with \eqref{eq:elastostatics} due to the nonlinear dependence 
of the conductivity on the crack opening displacement.}

\section{Computational framework} \label{sec:finiteElements}
The numerical formulation for the coupled problem \eqref{eq:elastostatics}-\eqref{eq:lubrication}-%
\eqref{eq:tsl} consists of several ingredients and includes the spatial discretization of 
the solid mechanics and of the fluid flow. The main characteristic of the solid spatial discretization  
is that we generalize the DG/CZM framework proposed in \cite{seagraves:2010,Seagraves:2015a} 
for the dynamic propagation of dry cracks to the case of quasi-static propagation of fluid-filled cracks. 
The DG/CZM method is based on the combination of a discontinuous Galerkin formulation 
of the continuum problem and a cohesive zone model of fracture. In this framework,
the flux and stabilization terms arising from the DG formulation at
interelement boundaries are enforced via interface elements, much like in the conventional
intrinsic cohesive element approach, albeit in a way that guarantees consistency and
stability prior to fracture. Upon the onset of fracture, the TSL governing the
fracture process becomes operative without the need to insert a new cohesive element. 
The main appeal of this approach lies in the consistent weak enforcement of traction continuity 
at interelement boundaries prior to fracture and the ease of introduction of the cohesive 
model as fracture occurs, which allows to model fracture propagation on arbitrary meshes.

\begin{figure}[h]
\centering
\includegraphics[width=\textwidth]{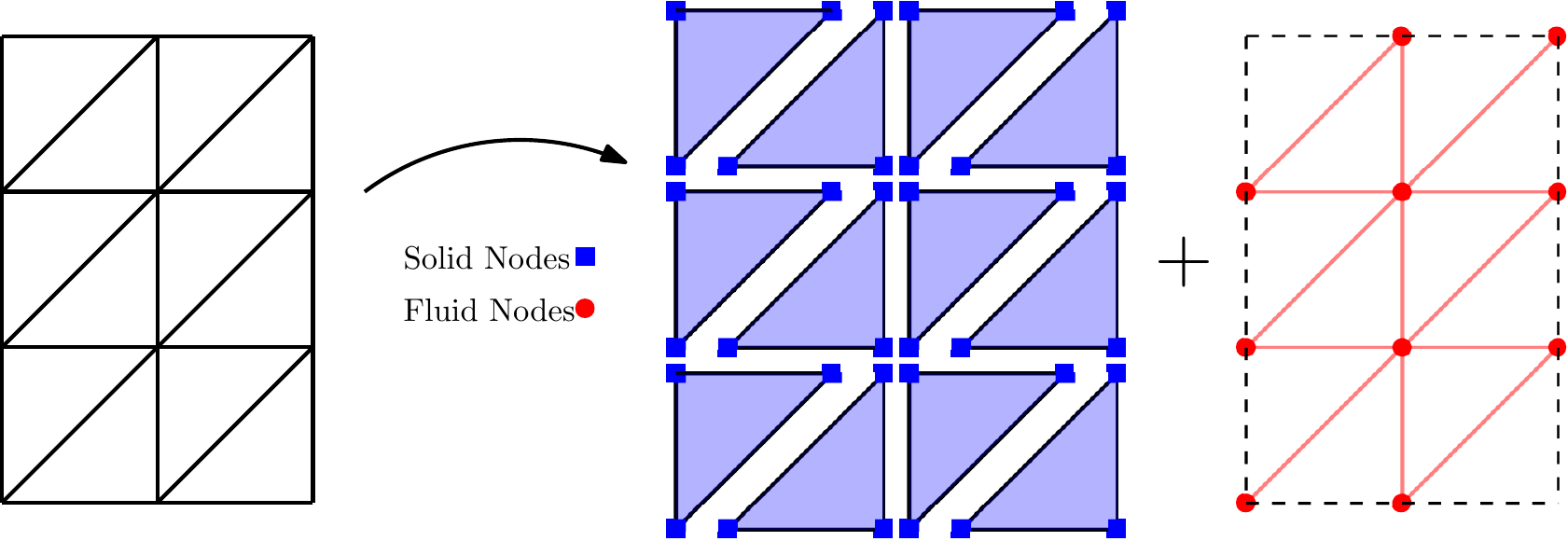}
\caption{\emph{
The solid is described on bulk elements and the fluid on the wireframe of 
the solid discretization. While the bulk finite elements are discontinuous,
the interface finite elements are continuous.}}
\label{fig:hybridCGDG}
\end{figure}

To describe the fluid flow into the newly created channels as the cracks 
propagate and expand we adopt a continuous Galerkin finite element discretization 
whose support is the set of (d-1)-dimensional interface elements. 
Figure \ref{fig:hybridCGDG} shows a schematic 
of the domain: the DG discretization of the bulk material in blue, and all the channels 
in which the fluid could potentially flow as cracks propagate in the red wireframe.

We introduce a partition $\left\{ \Omega_h^e \right\}_{e = 1, ..., N_e}$ of the reference computational domain $\Omega_h$,
and we denote with $\partial_{I}\Omega_h := \cup_{e = 1}^{N_e} \partial \Omega_h^e \setminus \partial \Omega_h$ 
the internal boundary induced by the partition. We denote by $\left\{ \partial\Omega_h^i \right\}_{i = 1, ..., N_i} \subset \partial_{I}\Omega_h$,
the set of elements belonging to the interelement boundary of the partition  $\left\{ \Omega_h^e\right\}$.
The semi-discrete displacement and pressure fields are approximated as 
$\vect{u_h} = \sum_{j=1}^{N_u} u_j \vect{\varphi}_j$ and 
$p_h = \sum_{j=1}^{N_p} p_j \eta_j$, respectively, where $\left\{\vect{\varphi}_i\right\}_{i=1, ..., N_u}$ and 
$\left\{{\eta}_i\right\}_{i=1, ..., N_p}$ are bases of $\vect{\mathcal{U}_h}$ and $\mathcal{P}_h$, 
the conventional interpolation spaces of discontinuous Galerkin and continuous Galerkin finite 
elements, respectively: 
$$
\begin{aligned}
\vect{\mathcal{U}_h} &= 
\left\{ \vect{v_h} \in \vect{L^2(\Omega_h)}: 
\quad \left. \vect{v_h} \right|_{\Omega_h^e} \in \vect{{\rm I\!P}^k (\Omega_h^e)} \ 
\quad \forall e = 1, ... , N_e \right\}, \\
\mathcal{P}_h &= \left\{ q_h \in C^0(\partial_{I}\Omega_h):  
\quad \left. q_h \right|_{\partial\Omega_h^i } \in {\rm I\!P}^k (\partial\Omega_h^i) \ 
\quad \forall i = 1, ... , N_i \right\}.
\end{aligned}
$$

We also introduce, for the sake of notation, the discrete crack opening 
$w_h := \jump{\vect{u_h}} \cdot \vect{n}_{\Gamma}$ on $\partial_{I}\Omega_h$. 
The discrete formulation of \eqref{eq:elastostatics}-\eqref{eq:tsl} is obtained testing 
equations \eqref{eq:elastostatics} and \eqref{eq:lubrication}
against $\vect{v_h}$ and $q_h$, respectively, resulting in:
\begin{equation} \label{eq:weakElastoStatics}
\begin{aligned}
&\sum_{e} \int_{\Omega_h^e} \mathcal{C} \vect{\varepsilon}(\vect{u_h}) : \vect{\varepsilon}(\vect{v_h}) \dd{V} \\
&+ \int_{\partial_{I}\Omega_h} 
(1 -{\alpha_h}) \left( \avg{\mathcal{C} \vect{\varepsilon}(\vect{u_h})} \jump{\vect{v_h}} \cdot \vect{n}_{\Gamma}
+ \frac{b}{h} 
\jump{\vect{v_h}} \otimes \vect{n}_{\Gamma} : \mathcal{C} : \jump{\vect{u_h}} \otimes \vect{n}_{\Gamma} 
\right) \dd{S}\\
&+ \int_{\partial_{I}\Omega_h} 
\left( {\alpha_h}\ \vect{t}(w_h) - {\beta_h}\ p_h \vect{n}_{\Gamma} \right) \cdot \jump{\vect{v_h}}  \dd{S} + \\ 
 &= \sum_{e} \int_{\partial \Omega_h^e \cap \partial \Omega_N} 
 \bar{\vect{t}} \cdot \vect{v_h} \dd{S}
 \qquad \forall \vect{v_h} \in \vect{\mathcal{U}_h},
\end{aligned}
\end{equation}
\begin{equation}\label{eq:weakPressure}
\int_{\partial_{I}\Omega_h} 
{\beta_h} \left( \frac{w_h^3}{12\mu} \nabla_{\Gamma} p_h \cdot \nabla_{\Gamma} q_h 
+ \frac{\partial w_h}{\partial t} q_h  \right) \dd{S}
= \int_{\partial\Gamma_{in}} Q_0 q_h \dd{L} \qquad \forall q_h \in \mathcal{P}_h,
\end{equation}
where $\avg{\bullet} :=\frac{1}{2} \left(\bullet^+ + \bullet^-\right)$ denotes the average operator.
In Equation \eqref{eq:weakElastoStatics}, the first term represents the usual contribution of the bulk 
stresses to the virtual work. The second line of Equation \eqref{eq:weakElastoStatics}
consists of the various internal boundary terms resulting from the DG discretization, including the 
consistency and the stabilization terms \cite{noels:2006, noels:2007b}: the first term ensures the 
consistency of the numerical scheme and comes as a 
boundary contribution from the integration by parts on each element $\Omega_h^e$, whereas the 
second terms weakly enforces displacement continuity at element interfaces with a stabilization 
parameter $b$ relative to the mesh size $h$.
As in \cite{seagraves:2010}, the factor ${\alpha_h}$ is added to control the activation of the TSL: 
${\alpha_h} = 0$ before the onset of fracture and ${\alpha_h} = 1$ after the stress reaches the 
prescribed critical value $\sigma_c$. 
\comment{ 
More precisely, fracture is detected by monitoring the DG tractions exchanged at the interelement 
boundaries at all integration points:
\begin{equation} \label{eq:fractureActivation}
{\alpha_h} (t) = \left\{ 
\begin{aligned}
1 && \text{if} && \max_{\tau < t} \vect{t}(w_h(\tau)) \cdot \vect{n}_{\Gamma} > \sigma_c \\
0 &&  && \text{otherwise} \\
\end{aligned}
 \right. .
\end{equation}}
The third line of Equation \eqref{eq:weakElastoStatics} is therefore active after the onset of fracture and 
accounts for the virtual work of the cohesive tractions as well as the natural boundary condition 
applied by the fluid. 
Similarly to ${\alpha_h}$, ${\beta_h}$ is a binary field that allows to activate the fluid 
domain when actual flow conditions are achieved, \emph{e.g.} outside the fluid-lag region 
behind the crack tip. This allows to incorporate different models of fluid-lag, as described below.
Finally, the last line of Equation \eqref{eq:weakElastoStatics} accounts for the natural boundary 
condition in the remote boundary.

\comment{ 
\begin{equation} \label{eq:fluidActivation}
{\beta_h} (t) = \left\{ 
\begin{aligned}
1 && \text{if} && \max_{\tau < t} w_h(\tau) > \delta_c \\
0 &&  && \text{otherwise} \\
\end{aligned}
 \right. .
\end{equation}}

It bears emphasis that, regardless of the occurrence of fracture propagation, the 
hydro-mechanical coupling of equations \eqref{eq:weakElastoStatics} and \eqref{eq:weakPressure} 
is extremely stiff due to the nonlinear dependence of the lubrication conductivity on the 
crack opening displacement.
In order to minimize the issues of numerical robustness we adopt a 
fully-coupled solution strategy.
We propose an iterative algorithm to advance the 
coupled problem \eqref{eq:weakElastoStatics}-\eqref{eq:weakPressure} in time.
We employ a first order discretization of the time derivative in Equation \eqref{eq:weakPressure},
where the displacement and pressure fields are treated implicitly. More precisely, we discretize
$\frac{\partial w_h}{\partial t}$ at time $t^{n+1}$ as:
$$
\left. \frac{\partial w_h}{\partial t} \right|_{t^{n+1}} \approx \frac{w_h^{n+1} - w_h^n}{\Delta t},
$$
where $\Delta t$ is the time step for the integration.
The resulting fully-discrete system reads:

\begin{equation} \label{eq:discreteElastoStatics}
\begin{aligned}
&\sum_{e} \int_{\Omega_h^e} \mathcal{C} \vect{\varepsilon}(\vect{u_h^{n+1}}) : \vect{\varepsilon}(\vect{v_h}) \dd{V}
+ \int_{\partial_{I}\Omega_h} 
\left( {\alpha_h^{n+1}}\ \vect{t}(w_h^{n+1}) - {\beta_h^n}\ p_h^{n+1} \vect{n}_{\Gamma} \right) \cdot \jump{\vect{v_h}} \dd{S} \\ 
&+ \int_{\partial_{I}\Omega_h} 
(1 -{\alpha_h^{n+1}}) \left( \avg{\mathcal{C} \vect{\varepsilon}(\vect{u_h^{n+1}})} \jump{\vect{v_h}} \cdot \vect{n}_{\Gamma}
+ \frac{b}{h} 
\jump{\vect{v_h}} \otimes \vect{n}_{\Gamma} : \mathcal{C} : \jump{\vect{u_h^{n+1}}} \otimes \vect{n}_{\Gamma} 
\right) \dd{S}\\
 &= \sum_{e} \int_{\partial \Omega_h^e \cap \partial \Omega_N} 
 \bar{\vect{t}} \cdot \vect{v_h} \dd{S} \qquad \forall \vect{v_h} \in \vect{\mathcal{U}_h},\\
\end{aligned}
\end{equation}

\begin{equation}\label{eq:discreteLubrication}
\begin{aligned}
&\int_{\partial_{I}\Omega_h} \beta_h^{n} \left( \frac{(w_h^{n+1})^3}{12\mu} \nabla_{\gamma} p_h^{n+1} \cdot \nabla_{\gamma}q_h + \frac{w_h^{n+1}}{\Delta t} q_h  \right)\dd{S} \\
&= \int_{\partial\Gamma_{in}} Q_0 q_h \dd{L} + \int_{\partial_{I}\Omega_h} \beta_h^{n} \frac{w_h^{n}}{\Delta t} q_h \dd{S}  \qquad \forall q_h \in \mathcal{P}_h,
\end{aligned}
\end{equation}
where $\vect{u_h^{n+1}}$, $p_h^{n+1}$ are the unknown displacement and pressure at time
$t^{n+1}$ and $w_h^{n}$ is the opening at time $t^n$.

System \eqref{eq:discreteElastoStatics}-\eqref{eq:discreteLubrication} constitutes a
fully-coupled nonlinear algebraic system in the nodal unknowns
$\vect{U} = \left\{ u_i \right\}_{i=1}^{N_u}$ and
$\vect{P} = \left\{ p_i \right\}_{i=1}^{N_p}$, whose solution propagates their known values at
time $t^n$ to time $t^{n+1}$. A straightforward linearization of
\eqref{eq:discreteElastoStatics}-\eqref{eq:discreteLubrication} yields a linear system of
equations of the form:

\begin{equation} \label{eq:fully-coupled-fully-discrete}
\left[
\begin{aligned}
A & \quad B \\
C & \quad D
\end{aligned}
\right]
\left[
\begin{aligned}
\vect{U}\\
\vect{P}
\end{aligned}
\right] = 
\left[
\begin{aligned}
\vect{E} \\ 
\vect{F}
\end{aligned}
\right].
\end{equation}

To solve the nonlinear problem we adopt a Newton-Raphson scheme in which the linear system
\eqref{eq:fully-coupled-fully-discrete} is solved to advance from the nonlinear iteration $k$
to $k+1$ until convergence, starting from initial guesses ${w}^{n+1}_0 = {w}^n$ and
${p}^{n+1}_0 = {p}^n$.  Convergence is evaluated in the Euclidean norm of the relative
increments of both unknown nodal arrays. We find that, even in the presence of the fracture
propagation nonlinearity, the Newton-Raphson iterative strategy performed on the fully-coupled
system works robustly as long as we do not update $\beta_h$ until the end of the time step.

We have implemented the proposed computational approach in
a framework for large-scale simulations in computational mechanics.
Parallel scalability is achieved by partitioning the computational domain 
among the participating processors using the ParMETIS library \cite{METIS}. 
Each processor is then responsible for storing and maintaining the solution information for its portion of the mesh 
and for assembling its portion of the linear system \eqref{eq:fully-coupled-fully-discrete} resulting from the discretization.
The efficient and robust solution of the linearized system in each nonlinear iteration was explored using 
a variety of parallel iterative solvers with preconditioners based on domain decomposition, 
using the PETSc library \cite{PETSc:3.7}. We found that a parallel preconditioned 
iterative solver based on domain decomposition combined with a direct solver within
each subdomain provides excellent robustness and parallel scalability,
as we show in Section \ref{sec:results}.

\comment{ 
It should be carefully noted that the fluid activation parameter $\beta_h$ effectively removes the degeneracy of 
\eqref{eq:lubrication} at the crack tip, as it avoids the possibility of activating the fluid domain where the opening 
is vanishing.}
\comment{
From the numerical standpoint, the distinction between $\alpha_h$ and $\beta_h$
allows to distinguish the fluid front from the crack tip.}
\begin{remark}
  It is well-known that a lag region exists between the fluid front and the crack front, where
  the fluid pressure is equal to the vapor pressure $p_v$ for an impermeable crack, or to the
  pore pressure for a crack in a porous medium \cite{Lecampion:2017}.  It is possible to
  compute the lag size either analytically in the simple case of a straight crack, or to embed
  the computation of the lag in the computational approach, as in
  \cite{Lecampion:2007,Vahab:2018}.  \comment{
    For example, in \cite{Lecampion:2007} the lag size was computed from the quasi-static
    propagation condition, and in \cite{Vahab:2018} the lag zone was included in the
    formulation as a flow-free segment on which the vapor pressure is exerted.}  The factors
  controlling the size of the fluid lag are now well-understood: it has been shown that the lag
  size vanishes exponentially with the increase of nondimensional toughness
  \cite{Garagash:1999}, and decreases for increasing confining stresses \cite{Garagash:2006}.
  Furthermore, it has been recently shown that under prevailing conditions in field operations
  of hydraulic fracture the fluid lag is small and fully embedded in the fracture cohesive zone
  \cite{Garagash:2019}. For this reason, in this paper we limit our attention to the case in which 
  the fluid lag is negligible and assume that the fluid permeates the cohesive zone, as was 
  done in \cite{Boone:1990, Khoei:2015, Hirmand:2019}.  
  More precisely, we activate the fluid domain when a given crack opening threshold $\delta_f$ 
  is achieved and choose $\delta_f << \delta_c$, being $\delta_c$ the critical opening of the 
  cohesive law. Assuming that the fluid and
  crack fronts coincide however, exposes the degeneracy of the fluid equations at the crack tip, due to the vanishing
  opening, which results in a nonintegrable fluid pressure singularity. From the numerical
  standpoint, this results in the ill-conditioning of the jacobians, which we handle by
  imposing a cut-off to the fluid pressure to the vapor pressure, as was done multiple times 
  in the past \cite{Boone:1990, Khoei:2015, Hirmand:2019}.
\end{remark}

\section{Results} \label{sec:results}


In order to assess the numerical properties of the computational framework, we conduct a series of 
numerical tests. 

\subsection{Verification of the fully-coupled algorithm} 
\label{sec:results:verification-given-length}

In order to verify the proposed computational approach, we conduct a series of numerical tests
that attempt to replicate the analytical solutions available for a wide range of physical
regimes for fluid-driven straight cracks.  Specifically we consider a straight crack immersed
in a solid under plane strain conditions subject to fluid pressure resulting from inflow
(Neumann) boundary conditions. This problem was originally formulated in
\cite{Khristianovich:1955,Geertsma:1969} and historically referred to as the \emph{KGD}
model. Depending on the relative values of the model parameters, \emph{i.e.} the plane-strain
Young's modulus $E'$, the critical fracture energy release rate $G_c$, the fluid injection rate
$Q_0$, and the fluid viscosity $\mu$, two different propagation regimes can emerge for an
impermeable crack where analytical solutions have been found \cite{Detournay-ijg:2004}.  As
part of the derivation of the solution of the KGD problem it has been found that two
nondimensional parameters dictate the specific solution regime: the nondimensional toughness
$\mathcal{K}$ and the nondimensional viscosity $\mathcal{M}$:
$$
\mathcal{K} = 
\left(12 \mu Q_0 \left( E' \right)^{3}  \right)^{-\frac{1}{4}} 
\frac{8}{\sqrt{2 \pi}}K_{IC},
$$
and
$$
\mathcal{M} = 
12 \mu  Q_0	
\left( E' \right)^3 
\left( \frac{8}{\sqrt{2 \pi}}K_{IC} \right) ^{-4},
$$
where $K_{IC} = \sqrt{G_c E'}$ is the critical stress intensity factor under plane strain 
and mode I propagation conditions \cite{irwin:1957}.
Physically $\mathcal{K}$ and $\mathcal{M}$ represent the importance of fracture and 
viscous flow, respectively, as the main dissipation mechanism. 
It should be noted that $\mathcal{K} = \mathcal{M}^{-1/4}$, which highlights the fact 
that these two mechanisms are actually in competition \cite{Detournay-ijg:2004}.

\begin{figure} [h]
\centering
\includegraphics[width=0.6\textwidth]{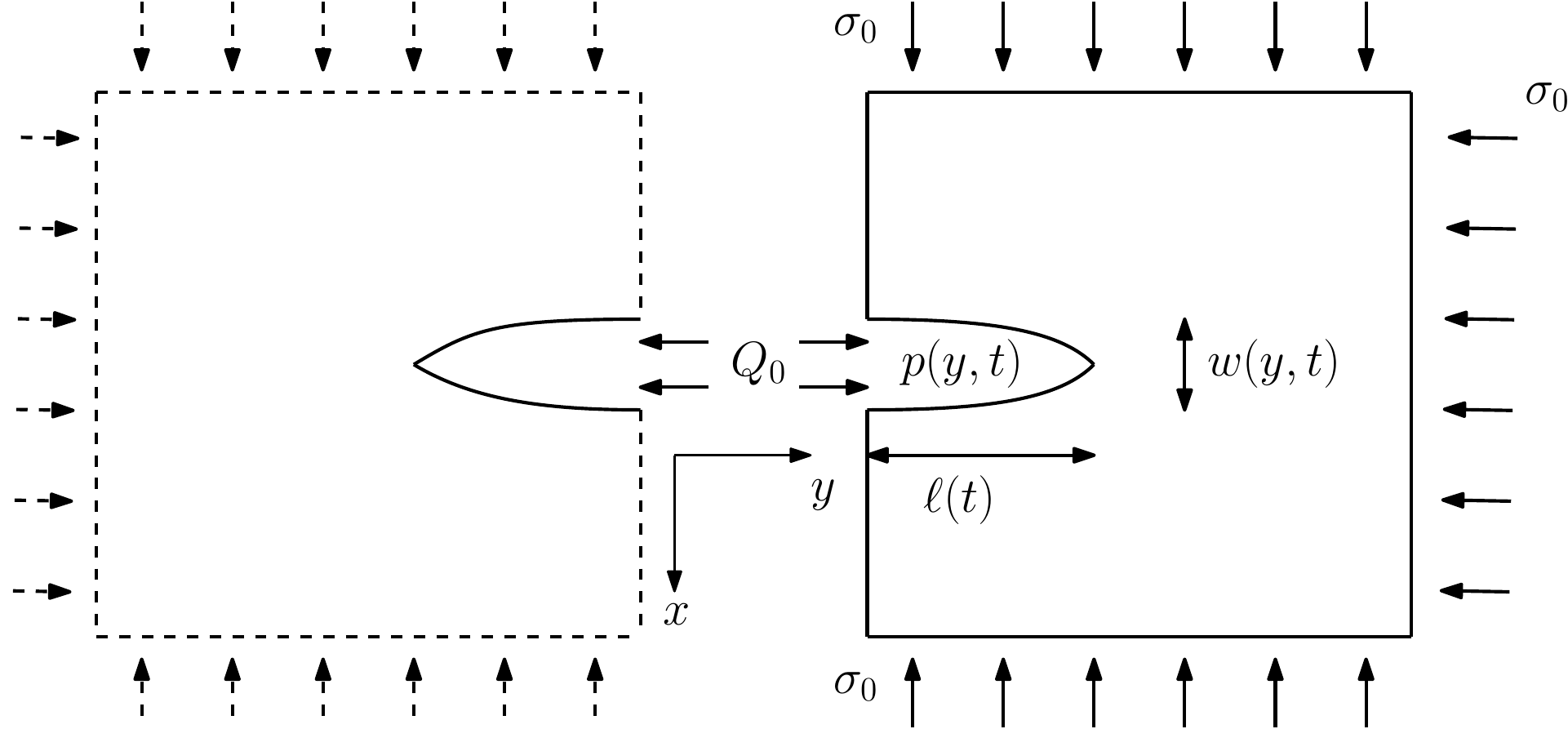}
\caption{\emph{The plain-strain configuration of the fluid-driven fracture. 
Exploiting symmetry, we focus on half of the domain.}}
\label{fig:hydraulicFractureConfiguration}
\end{figure}

In the case that viscous effects dominate (\emph{viscosity-dominated} regime, large
$\mathcal{M}$, low $\mathcal{K}$) the analytical solution presents a near-tip boundary layer
\cite{Garagash-asme:2005}.  This is due to incompatibility of the near-tip behavior of the
zero-toughness solution with the tip asymptote prescribed by linear elastic fracture mechanics
for any small, yet nonzero, toughness. The zero-toughness solution is known to have a weakly
singular near-tip asymptote characterized by the $2/3$-power dependence of the fracture opening
on the distance from the tip, while the pressure field is singular at the tip with a $-1/3$
power.  Conversely, when the main dissipation mechanism is fracture propagation
(\emph{toughness-dominated} regime, low $\mathcal{M}$, large $\mathcal{K}$), the zero-viscosity
pressure solution is uniform, whereas a tip-singularity is present in the pressure only in
higher order terms \cite{Garagash:2000}.  Threshold values delimiting the two regimes have been
obtained by Garagash, Detournay \emph{et al.}  with the result that
$\mathcal{K} < \mathcal{K}_0 = 0.70$ corresponds to the viscosity-dominated regime
\cite{Garagash-asme:2005}, and $\mathcal{K} > \mathcal{K_{\infty}} = 4.13$ corresponds to the
toughness-dominated regime \cite{Garagash:2000}.  For intermediate values of $\mathcal{K}$,
transition between regimes occurs, although it is important to note that a plane-strain
fracture cannot transition from one regime to one another during its time evolution, since
$\mathcal{K}$ and $\mathcal{M}$ do not depend on time \cite{Detournay-ijg:2004}.

\begin{table}[h]
\centering
\begin{tabular}{|c|c|c|} 
\hline
 & viscosity regime & toughness regime \\
\hline
$E$ & $17 \ G Pa$ & $17 \ G Pa$ \\
\hline
$\nu$ & $0.2$ & $0.2$ \\
\hline
$\mu$ & $0.1 \ Pa \ s$ & $10 \ \mu Pa \ s$ \\
\hline
$Q_0$ & $0.001 \ m^2 \ s^{-1}$  & $0.001 \ m^2 \ s^{-1}$  \\
\hline
$G_c$ & $120 \ Pa \ m$ & $120 \ Pa \ m$ \\
\hline
\hline
$\mathcal{K}$ & $0.51$ & $5.1$ \\
\hline
$\mathcal{M}$ & $14$ & $0.0014$ \\
\hline 
\end{tabular}
\caption{\emph{The parameters used in the simulations with the 
corresponding nondimensional toughness $\mathcal{K}$ and viscosity $\mathcal{M}$.
The left-hand set of parameters corresponds to the viscosity-dominated regime, whereas 
the right-hand set of parameters corresponds to the toughness-dominated regime.}}
\label{tab:parametersHM}
\end{table}

In order to test the computational framework in both regimes, we select two representative
cases corresponding to the parameters shown in Table \ref{tab:parametersHM}. Once the
parameters are chosen, the analytical solutions provide the crack length, pressure
distribution, and crack opening distribution as a function of time, respectively
$\len(t), p(y,t), w(y,t)$. We extract the functional forms of these solutions from
\cite{Garagash-asme:2005} in the case of the viscosity-dominated regime, and from
\cite{Garagash:2000} in the case of the toughness-dominated regime.

As an initial step, we verify that the computed and theoretical pressure and opening
distributions are consistent for a crack of given length $\bar{\len}$ embedded in the computational
mesh, by performing the calculations of one time step, which advance the solutions from time $t^n$
to time $t^{n+1}$. This enables the assessment of the properties of the solution of the coupled system
\eqref{eq:discreteElastoStatics}-\eqref{eq:discreteLubrication} in isolation from the crack propagation 
algorithm, which will be tested in Section \ref{sec:results:verification-propagation}. 
It bears emphasis that in order to match the analytical solution
at time $t^{n+1}$, Equation \eqref{eq:discreteLubrication} requires the analytical opening $w^n$ at 
time $t^n$, which is
maintained fixed as a source term during the nonlinear iteration process. By contrast, we have
freedom in the choice of the initial guesses of the pressure and opening fields 
${p}^{n+1}_0$ and ${w}^{n+1}_0$: although they could be adopted from the previous time step 
(\emph{i.e.} ${w}^{n+1}_0 = {w}^n$ and ${p}^{n+1}_0 = {p}^n$), 
we have found that the convergence is very fast and robust even if the initial guess is very
far from the analytic solution. In order to demonstrate robustness, our calculation 
below starts from uniform distributions ${w}^{n+1}_0 = w^0$ and ${p}^{n+1}_0 = p^0$.

\comment{and the analytical opening corresponding to the previous
load step as required by the right hand side of Equation \eqref{eq:discreteLubrication}:
$w^n$ at a time $t^n = \bar{t} - \Delta t$, where $\bar{t}$ is obtained from the analytical
solution as the time that corresponds to a crack of length $\bar{\len}$.
}

For each regime, we test the algorithm for four initial crack lengths $\bar{\len} = 5 m, 6 m, 7m, 8m$. 
To make sure that the computational domain is large enough to support the infinite 
hypothesis, we construct the computational domain $[0, 45m] \times [0, 60m]$. We exploit
the symmetry of the problem and apply half the injection rate
$\frac{Q_0}{2}$. The mesh is refined in the proximity of the crack tip, where the mesh size
is $h = 0.1 m$, the polynomial order of the finite elements is $p=2$. 
We have found the choice of time step $\Delta t = 1 s$ for the viscosity-dominated regime and 
$\Delta t = 0.1 s$ for the toughness dominated regime and for the chosen values for the physical 
parameters the result converge to the analytical solution showing that the right hand side of 
equation \eqref{eq:lubrication} is properly resolved.
For simplicity, we adopt $\bar{\vect{t}} = \vect{0}$ (no far-field stress).	
Finally, in both regimes, the uniform initial guesses for pressure and opening are $p^0 = 1 MPa$ 
and $w_0 = 2 mm$.

\begin{figure}[h]
\includegraphics[width=0.5\textwidth]
{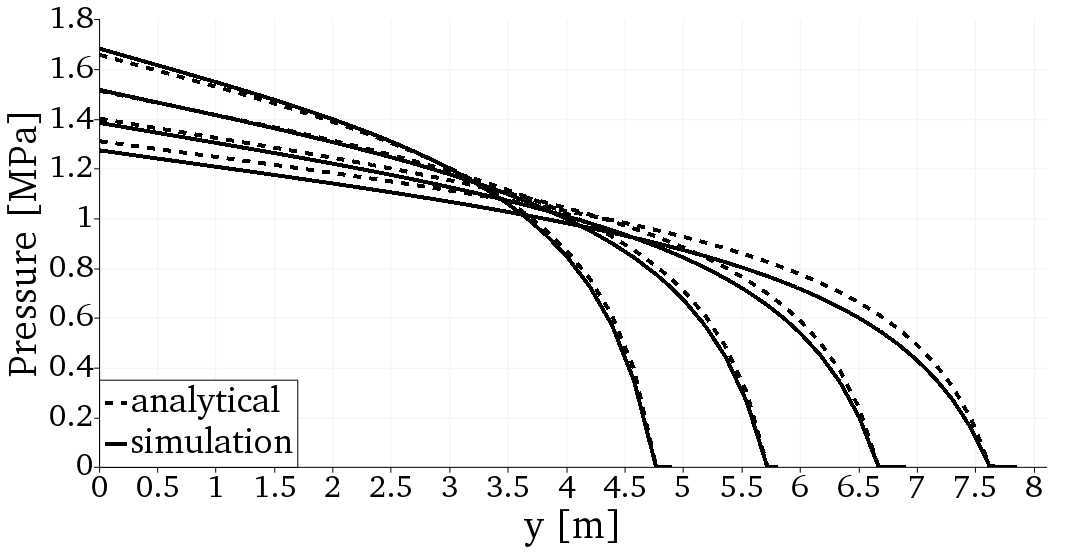} 
\includegraphics[width=0.5\textwidth]
{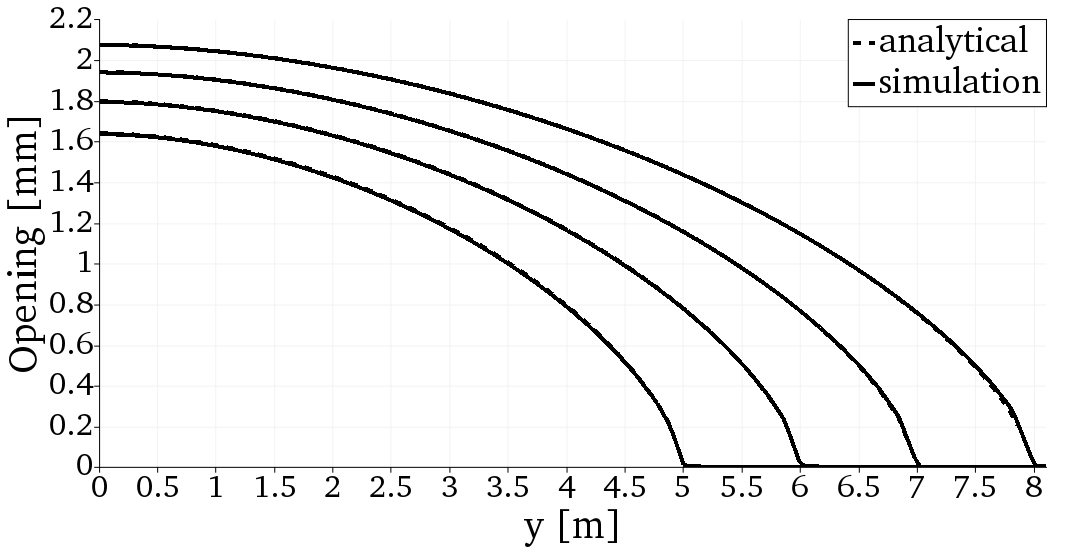} 
\caption{\emph{
Comparison of the computational predictions against the analytical 
solutions in the viscosity-dominated regime from \cite{Garagash-asme:2005}
for several crack lengths $5m$, $6m$, $7m$, $8m$.
}}
\label{fig:analytic-given-length-viscosity}
\end{figure}

\begin{figure}[h]
\includegraphics[width=0.5\textwidth]
{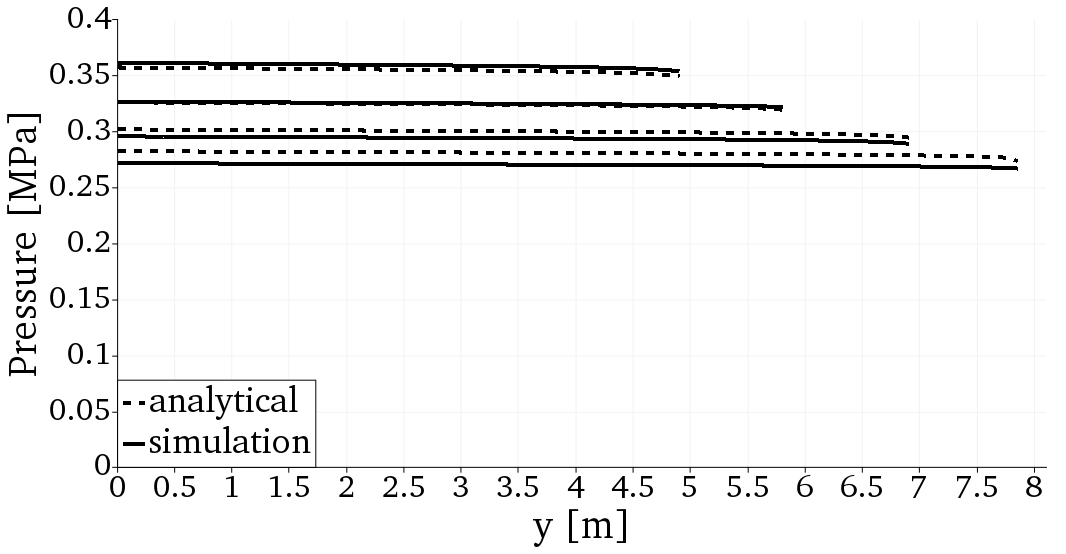} 
\includegraphics[width=0.5\textwidth]
{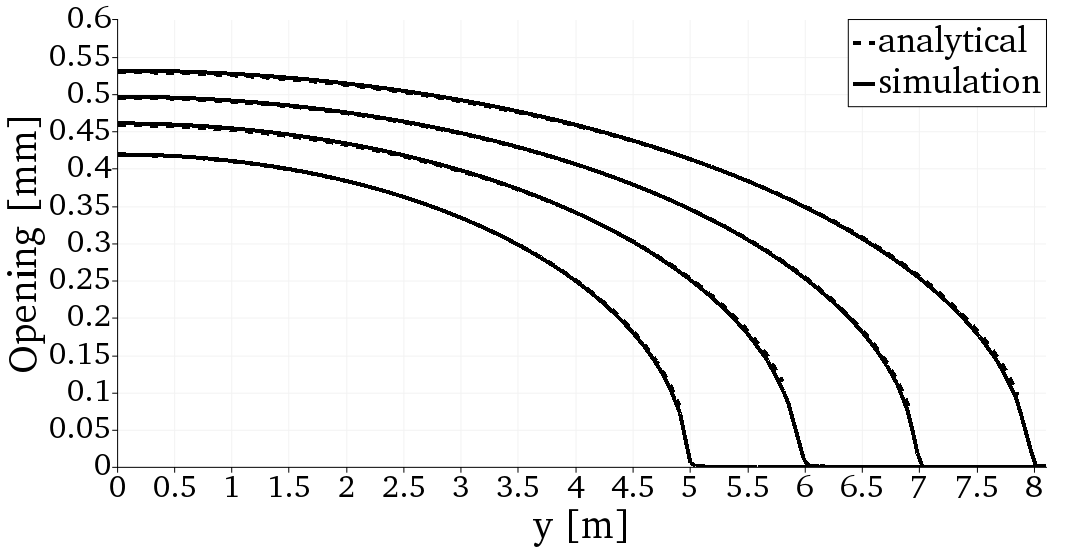} 
\caption{\emph{
Comparison of the computational predictions against the analytical 
solutions in the toughness-dominated regime from \cite{Garagash:2000}
for several crack lengths $5m$, $6m$, $7m$, $8m$.
}}
\label{fig:analytic-given-length-toughness}
\end{figure}

Figures \ref{fig:analytic-given-length-viscosity} and \ref{fig:analytic-given-length-toughness}
show comparisons of the numerical solutions with the analytical ones in the viscosity 
and toughness-dominated regimes, respectively. We observe that the
numerical solution is in excellent agreement with the analytical predictions. 
We also observe that as the initial crack increases the assumption of infinite domain is not 
represented by the computational domain.

\comment{
This choice of discretization parameters $h$ and $\Delta t$ was sufficient to obtain spatially
converged fluid pressure field and opening field under the assumption of given crack length.}

\subsection{Verification of the crack propagation algorithm}
\label{sec:results:verification-propagation}

In our next test we evaluate the performance, robustness, and accuracy of the crack 
propagation algorithm. 
Describing numerically the nonlinearities coming from the softening response as the crack 
propagates, on top of the nonlinearity of the fluid-solid coupling, is extremely challenging,
especially under quasi-static conditions.
In addition, we explore the possible implications of the restrictions in the crack paths 
imposed by the DG/CZM approach. To be more precise, in the proposed framework cracks 
are allowed to propagate on the solid element interfaces.
Hence, a given mesh provides
a finite number of potential crack paths, which may or may not include the exact crack path. 
The concern about whether a potential \emph{zig-zag} path of the crack
can produce a reasonable fluid pressure distribution has been raised in \cite{Lecampion:2017}.
We therefore consider two different meshes: one mesh 
contains the straight analytical crack path, the other mesh is unbiased.

We consider the same case in the viscosity-dominated regime from section \ref{sec:results:verification-given-length}, 
except that in this case we consider a far-field stress of the form $\bar{\vect{t}}  = - \sigma_0 \vect{n}$
to replicate more realistic field conditions with $\sigma_0 = 4.0 \ MPa$. 
Following \cite{Carrier:2012}, we set the fracture stress in Equation \eqref{eq:tsl} to $\sigma_c = 1.0 MPa$.
The initial crack length is $\bar{\len} = 0.5 m$, which corresponds to $t^0 = 0.4 s$.
Both computational meshes are uniformly refined along the expected crack path direction, with a 
characteristic mesh size $h = 0.1m$, and coarsened elsewhere.
The polynomial order of the finite elements is $p=3$. 

\comment{ 
Specifically, we show that our framework does not
require \emph{a priori} knowledge of the crack path.}

\comment{
To investigate the ability of our computational framework to produce results in agreement 
with theoretical predictions even when the given mesh does not include the exact crack path 
(\emph{i.e.} on an arbitrary mesh), we test our framework on two different meshes: one mesh 
contains the straight analytical crack path, the other mesh is unbiased. 
}

The solution procedure is done according to the nonlinear solution described in Section \ref{sec:finiteElements}
with a time step $\Delta t = 0.025 s$ until the final time of $60 \ s$, that is a final crack length of 
approximately $15 \ m$.
We observed that the Newton-Raphson procedure converges robustly in typically 3 to 5 iterations in time steps 
where the crack does not propagate, and in 5-10 iterations when the crack propagates. 
\comment{We have not investigated time step ...}

\begin{figure}[h]
\begin{subfigure}{\textwidth}
\centering
\adjincludegraphics[width=0.35\textwidth, trim={{.51\width} 0 0 {.88\height}},clip]{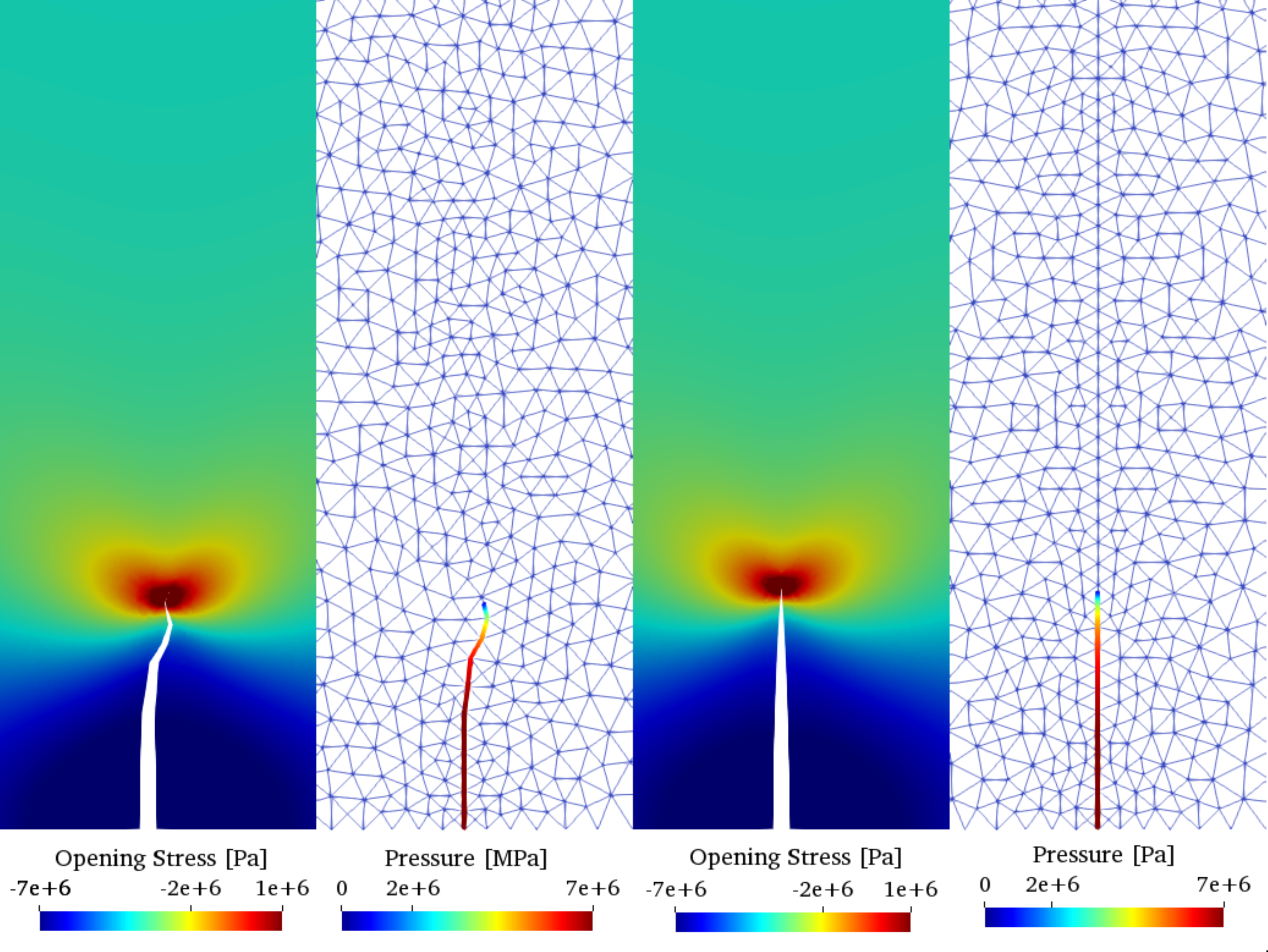}
\hspace*{0.5cm}
\adjincludegraphics[width=0.35\textwidth, trim={0 0 {.51\width} {.88\height}},clip]{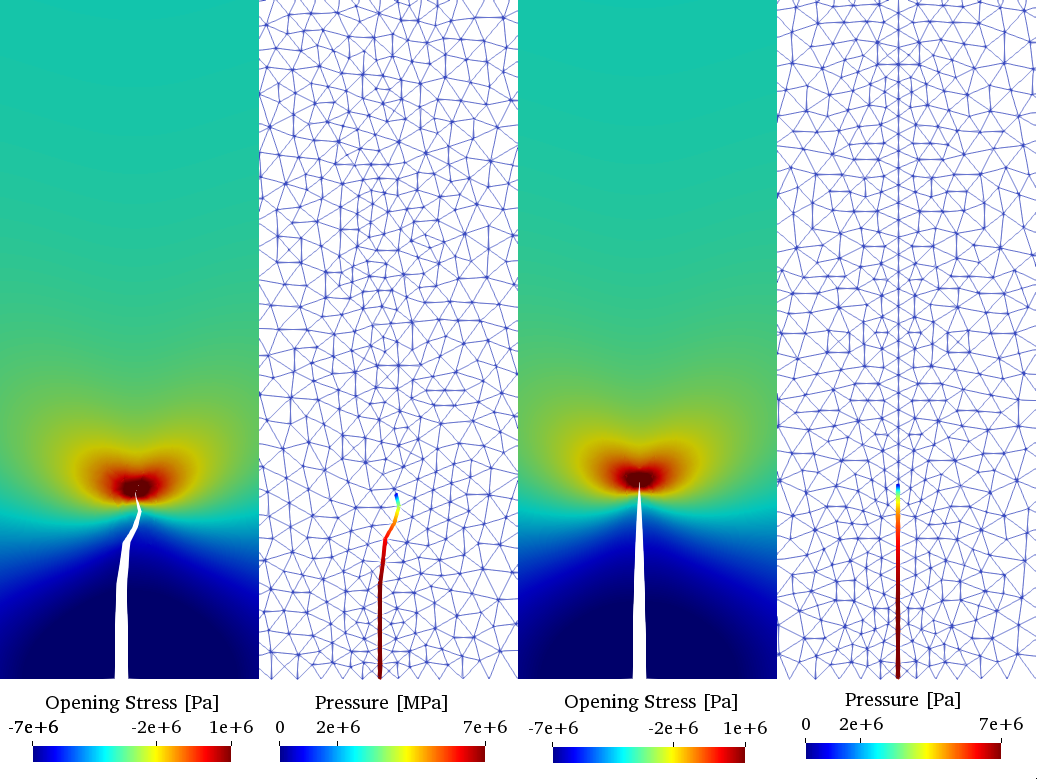}
\end{subfigure}
\begin{subfigure}{\textwidth}
\centering
\adjincludegraphics[width=0.35\textwidth, trim={{.51\width} {.12\height} 0 0},clip]{fluid-solid-time-7}
\hspace*{0.5cm}
\adjincludegraphics[width=0.35\textwidth, trim={0 {.12\height} {.51\width} 0},clip]{fluid-solid-time-7}
\caption{\emph{$t = 1.0s$}}
\end{subfigure}
\begin{subfigure}{\textwidth}
\centering
\adjincludegraphics[width=0.35\textwidth, trim={{.51\width} {.12\height} 0 0},clip]{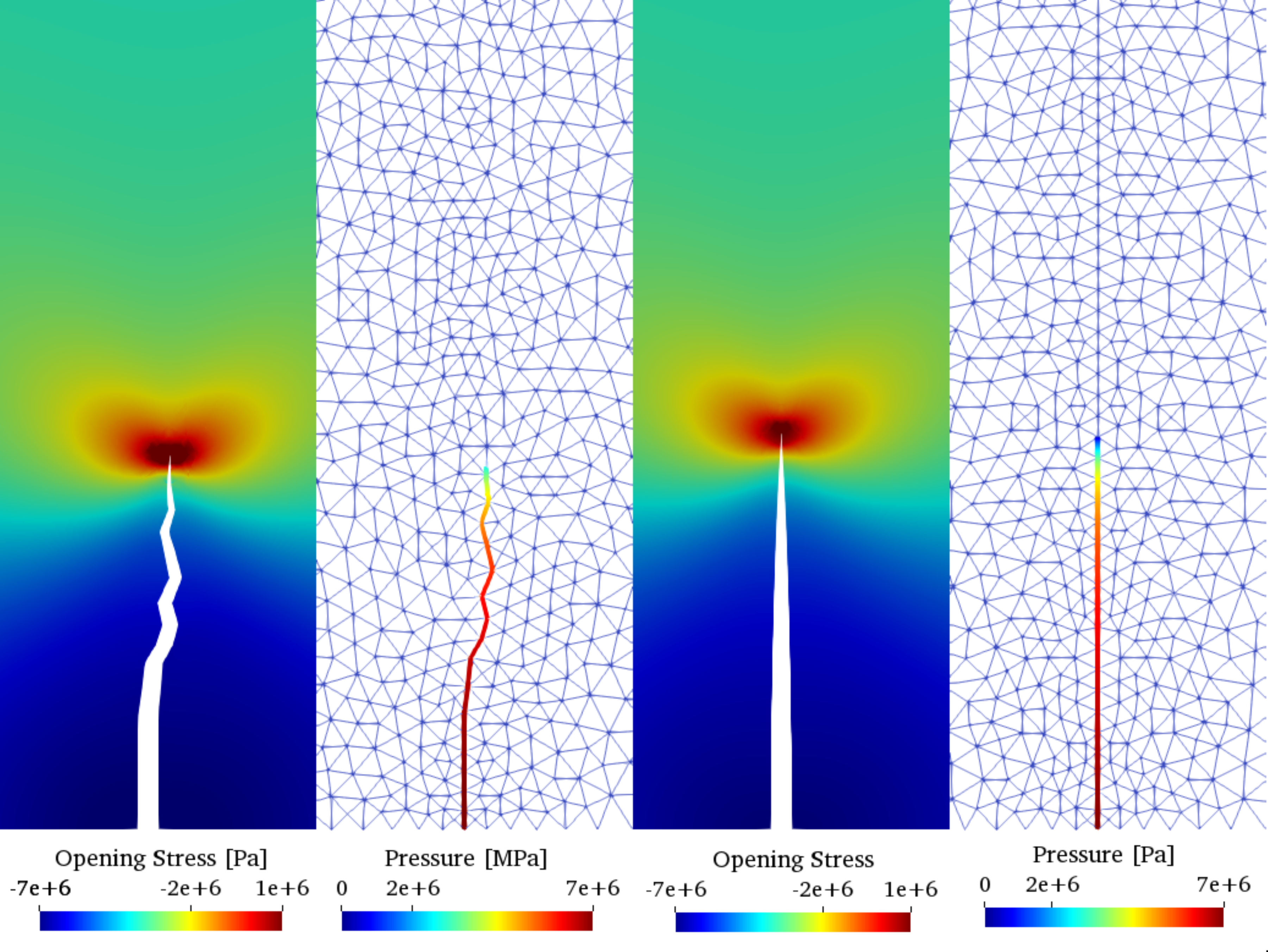}
\hspace*{0.5cm}
\adjincludegraphics[width=0.35\textwidth, trim={0 {.12\height} {.51\width} 0},clip]{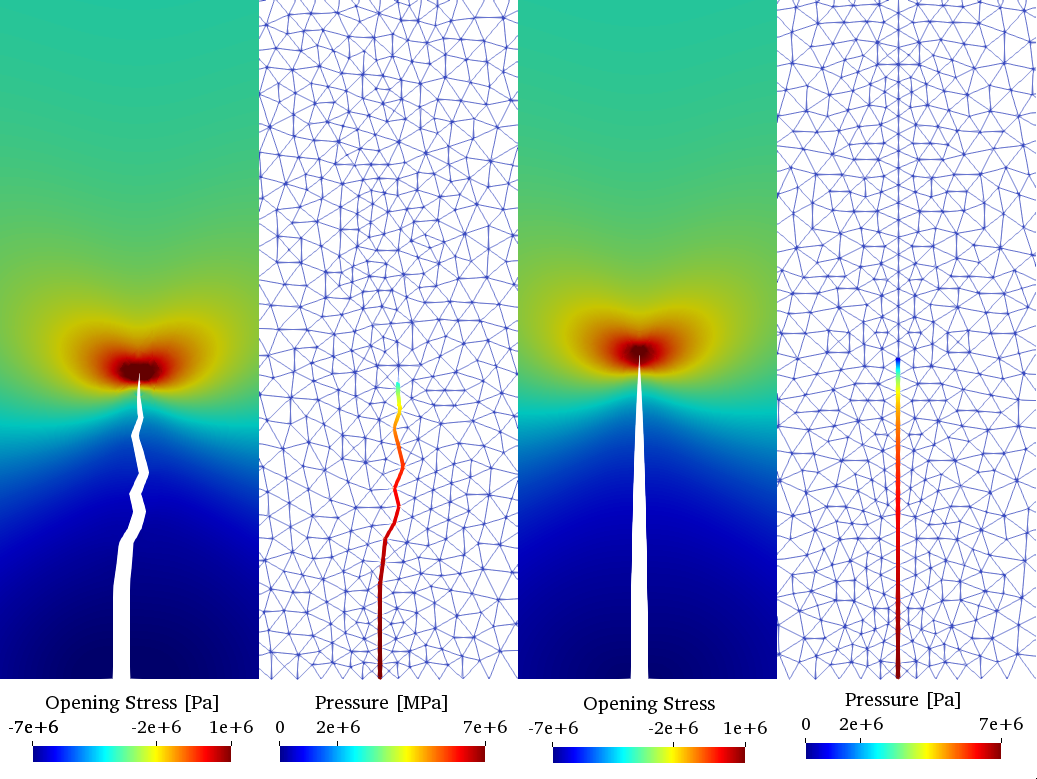}
\caption{\emph{$t = 2.0s$}}
\end{subfigure}
\begin{subfigure}{\textwidth}
\centering
\adjincludegraphics[width=0.35\textwidth, trim={{.51\width} {.12\height} 0 0},clip]{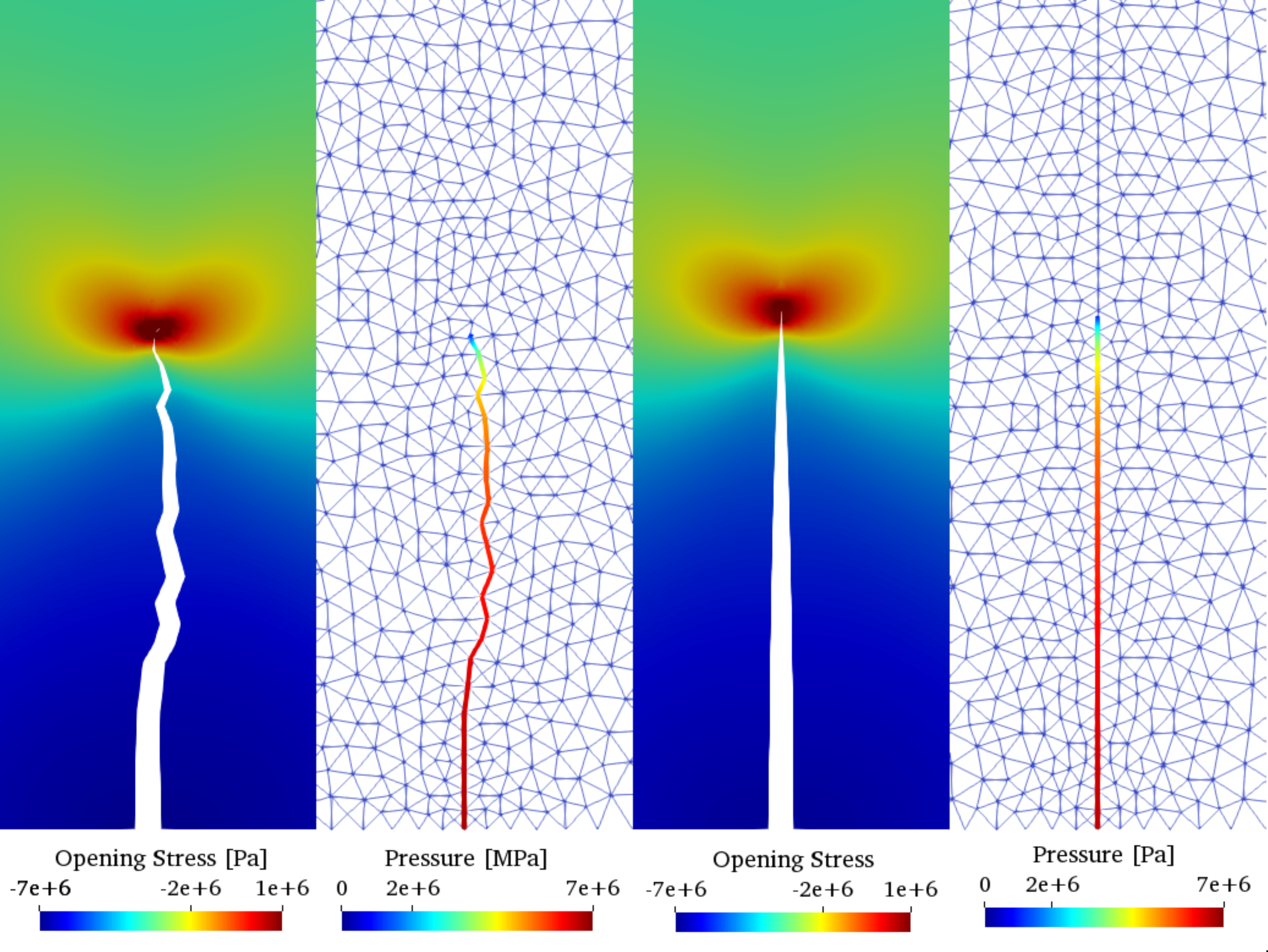}
\hspace*{0.5cm}
\adjincludegraphics[width=0.35\textwidth, trim={0 {.12\height} {.51\width} 0},clip]{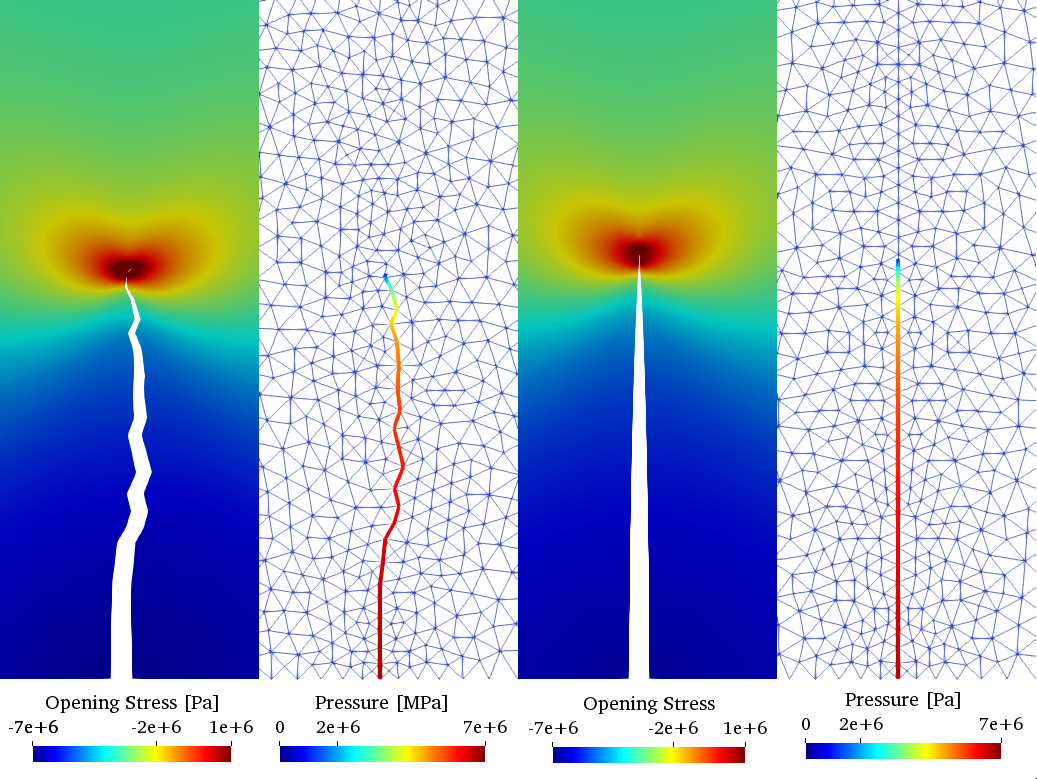}
\caption{\emph{$t = 3.0s$}}
\end{subfigure}
\end{figure} 
\begin{figure}[h]
\centering
\begin{subfigure}{\textwidth}
\centering
\adjincludegraphics[width=0.35\textwidth, trim={{.51\width} {.12\height} 0 0},clip]{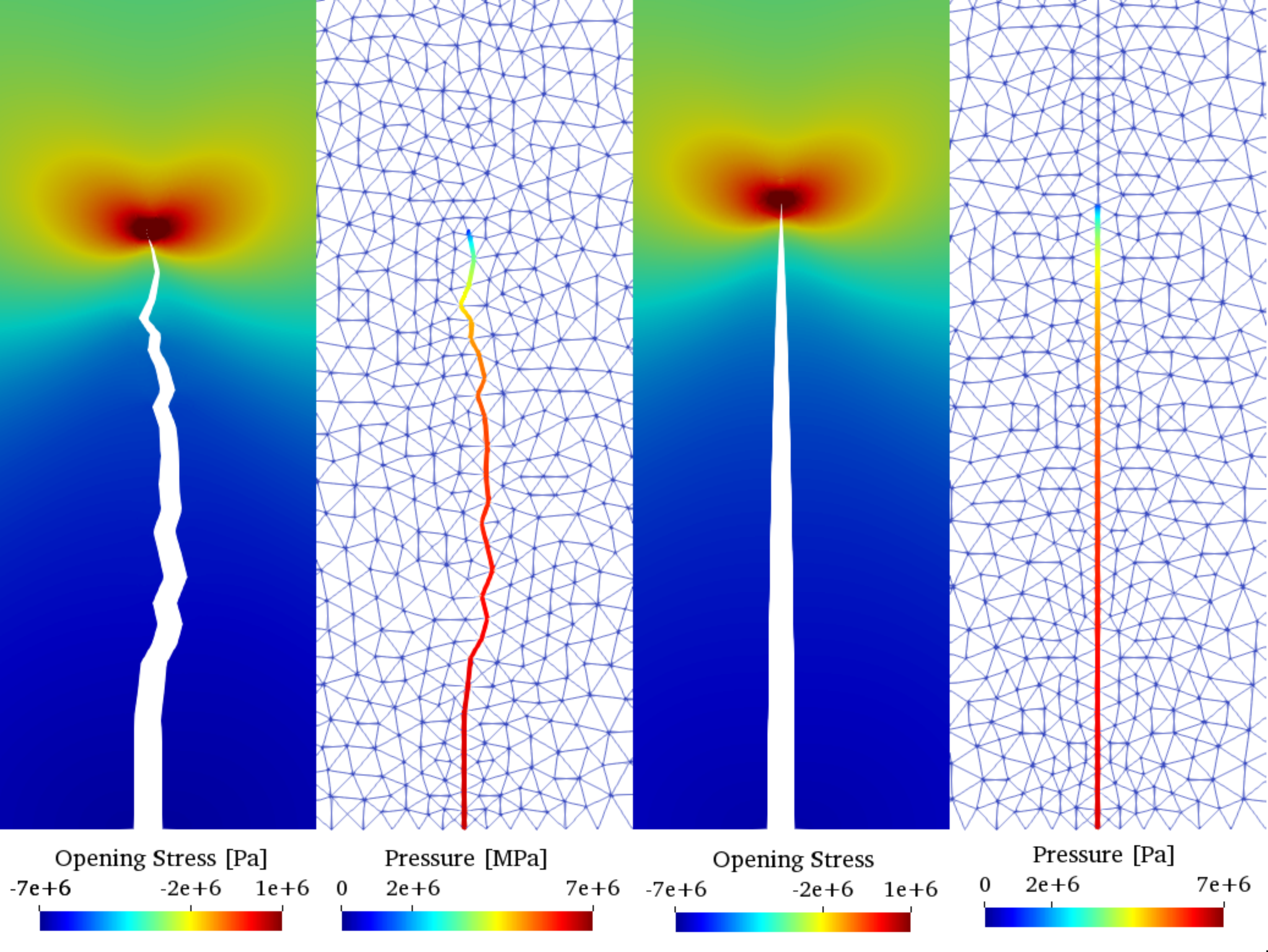}
\hspace*{0.5cm}
\adjincludegraphics[width=0.35\textwidth, trim={0 {.12\height} {.51\width} 0},clip]{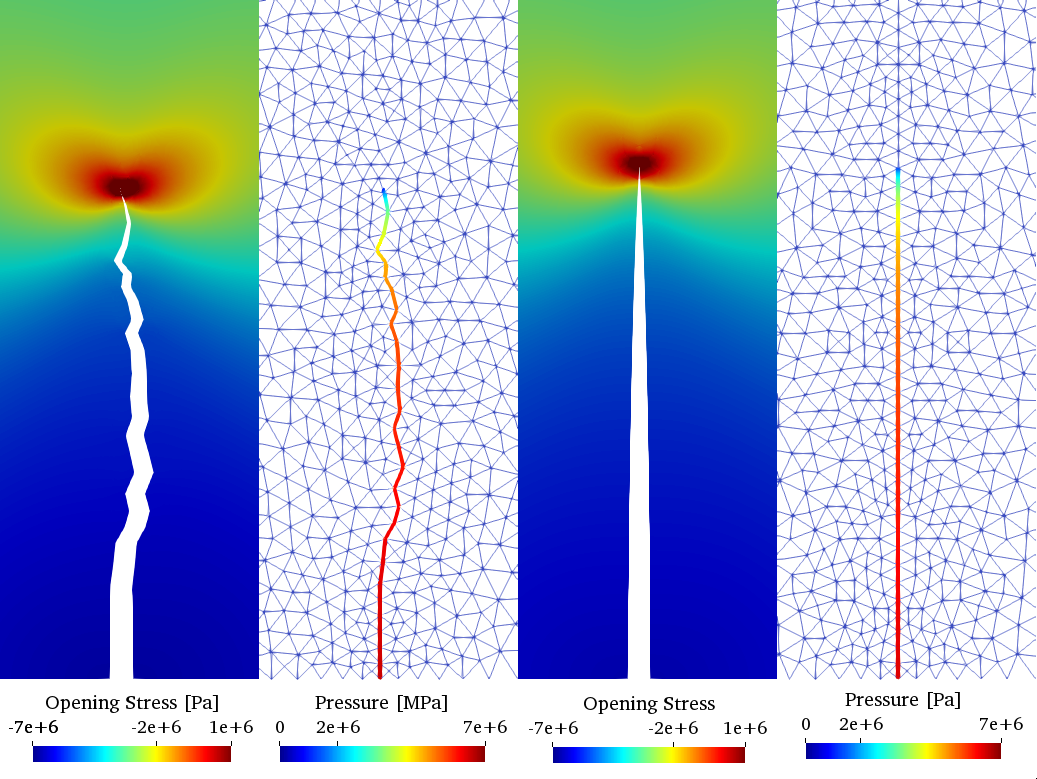}
\caption{\emph{$t = 4.0s$}}
\end{subfigure}
\begin{subfigure}{\textwidth}
\centering
\centering
\adjincludegraphics[width=0.35\textwidth, trim={{.51\width} {.12\height} 0 0},clip]{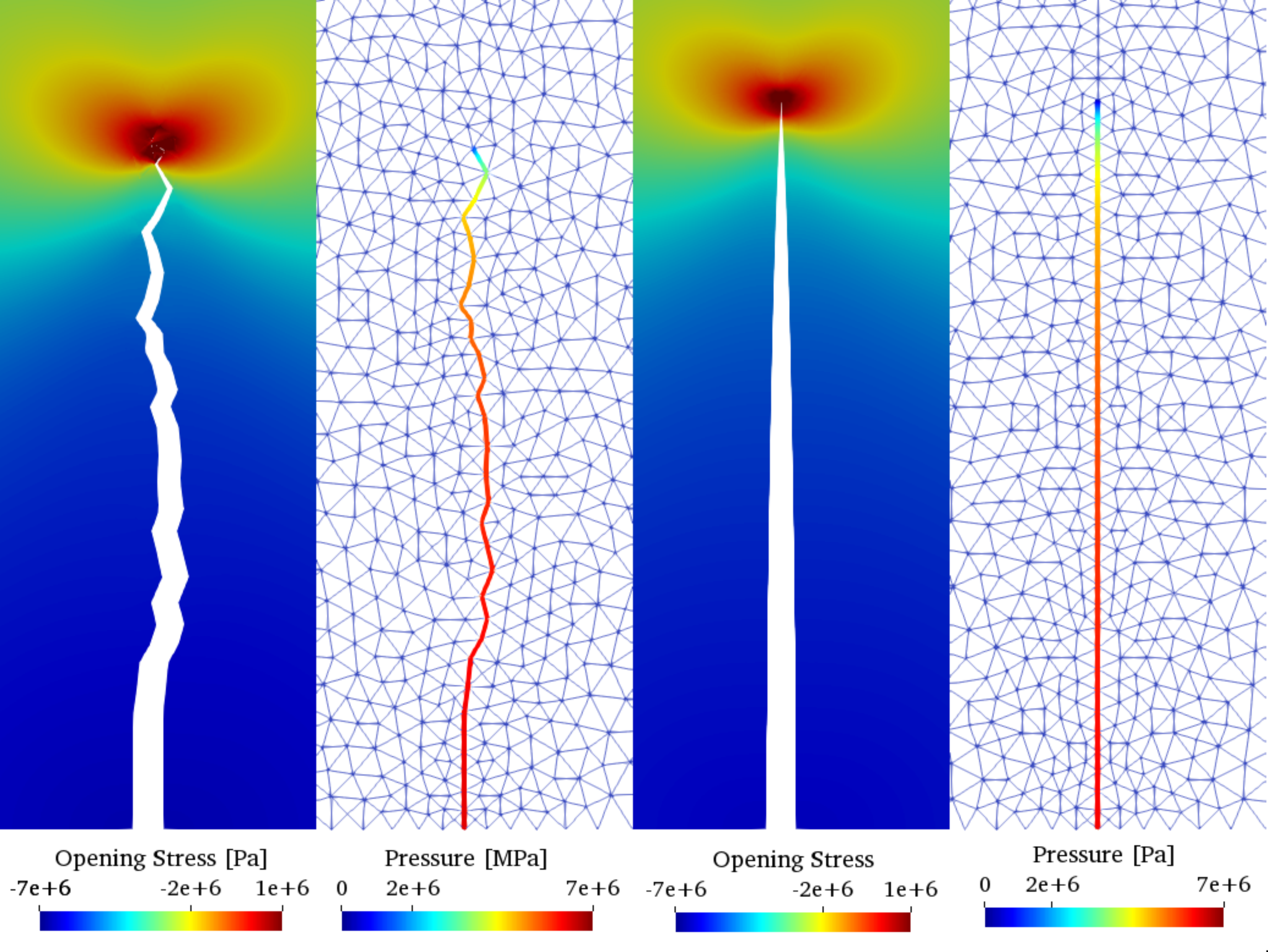}
\hspace*{0.5cm}
\adjincludegraphics[width=0.35\textwidth, trim={0 {.12\height} {.51\width} 0},clip]{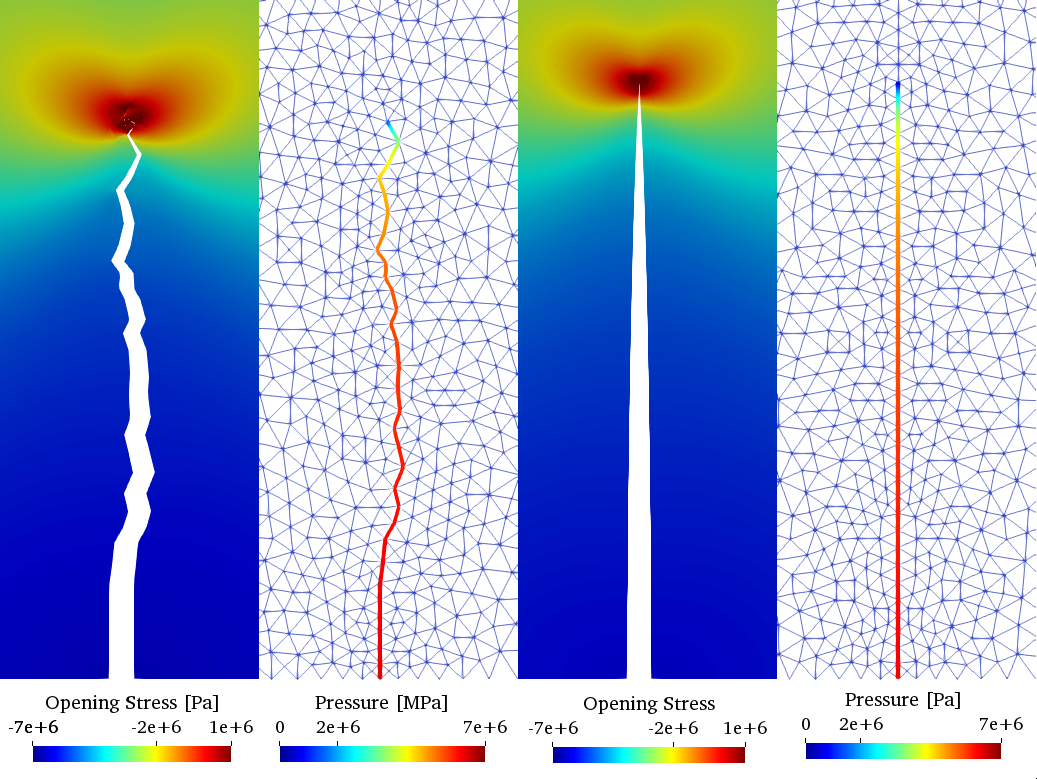}
\caption{\emph{$t = 5.0s$}}
\end{subfigure}
\caption{\emph{
Simulation results corresponding to the viscosity-dominated regime according to the parameters 
given in Table \ref{tab:parametersHM} obtained with two different meshes. 
Each row corresponds to a different simulation time and shows snapshots obtained 
on an arbitrary mesh (left snapshot) and on a biased mesh (right snapshot).
In each snapshot, the contour plot on the left shows the opening stress (the solid deformation is 
magnified by a factor of $100$), while the wireframe on the right shows the fluid domain:
the blue segments constitute inactive fluid elements, whereas the colored segments constitute the active 
fluid domain, where the color represents the fluid pressure.
Note that, if the straight path is included in the set of potential paths.}} 
\label{fig:fluid-solid-time}
\end{figure} 

\begin{figure}[h]
\centering
\includegraphics[width=0.5\textwidth]{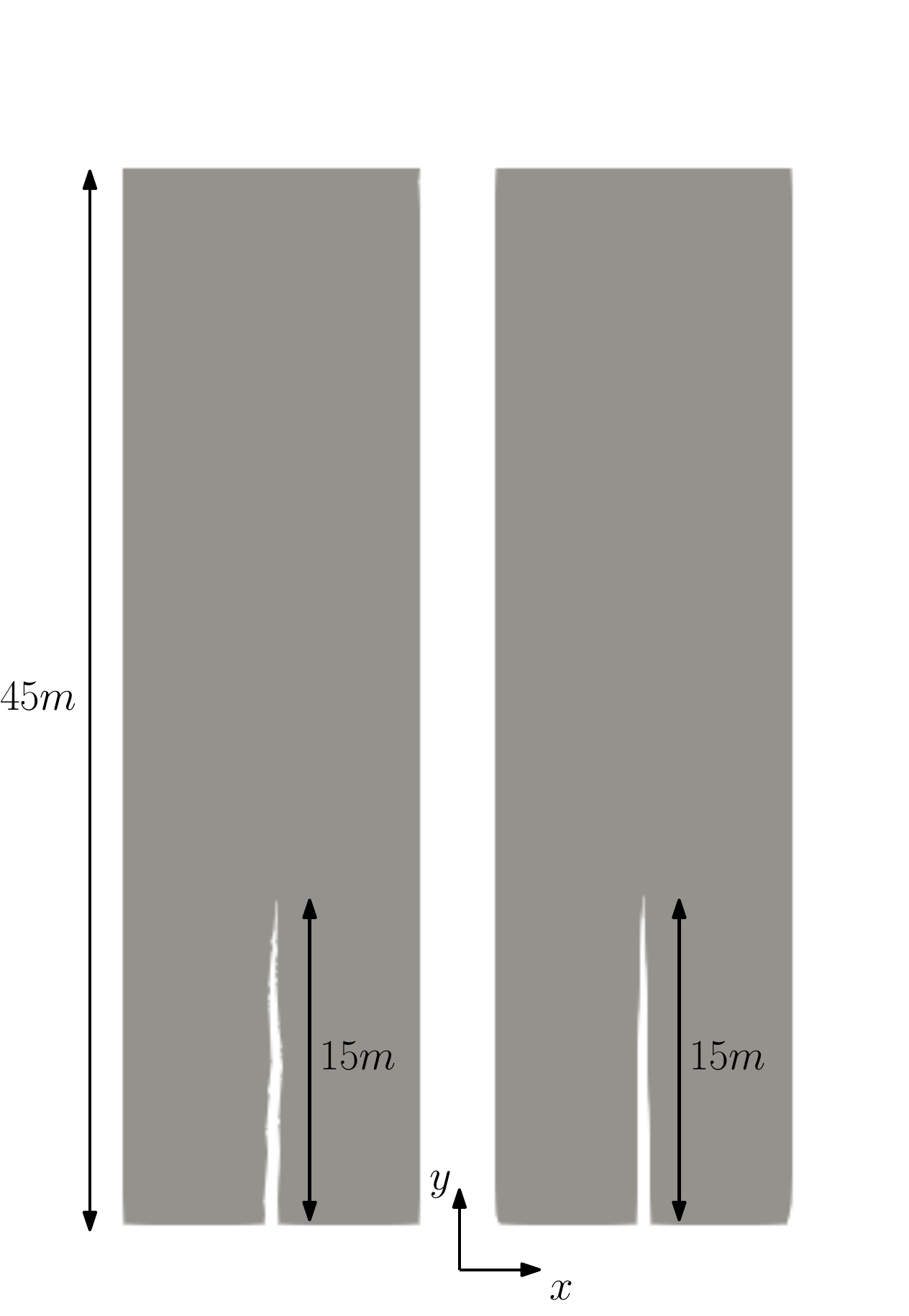} 
\caption{\emph{
Qualitative comparison of the macroscopic crack path obtained with the arbitrary mesh (left picture) 
and on the biased mesh (right picture) at the final time $t=60 \ s$ (final crack length $15 \ m$). 
Note that the crack paths obtained on the two meshes are in very good 
agreement. The crack opening is magnified of a factor of $500$. }}
\label{fig:crack-paths-full-length}
\end{figure}

Figure \ref{fig:fluid-solid-time} shows snapshots comparing side by side the results obtained on the two 
different meshes for selected simulation times in the time range 
$t\in [1 s, 5s]$ in a close-up view of size $1.3 \ m \times 3.5 \ m$ near the injection point. 
The left figures show contours of the opening stress field. In order to expose the crack path, the 
deformations are magnified by a factor of $100$. The right figures show the wireframe of the fluid domain:
the blue segments constitute inactive fluid elements, whereas the colored segments constitute the active 
fluid domain, where the color represents the fluid pressure. 
It is interesting to observe that although the crack at this scale does exhibit a \emph{zig-zag} behavior, 
the macroscopic response qualitatively hardly differs from the straight path, as shown in Figure   
\ref{fig:crack-paths-full-length}. 

\begin{figure}[h]
\includegraphics[width=0.5\textwidth]{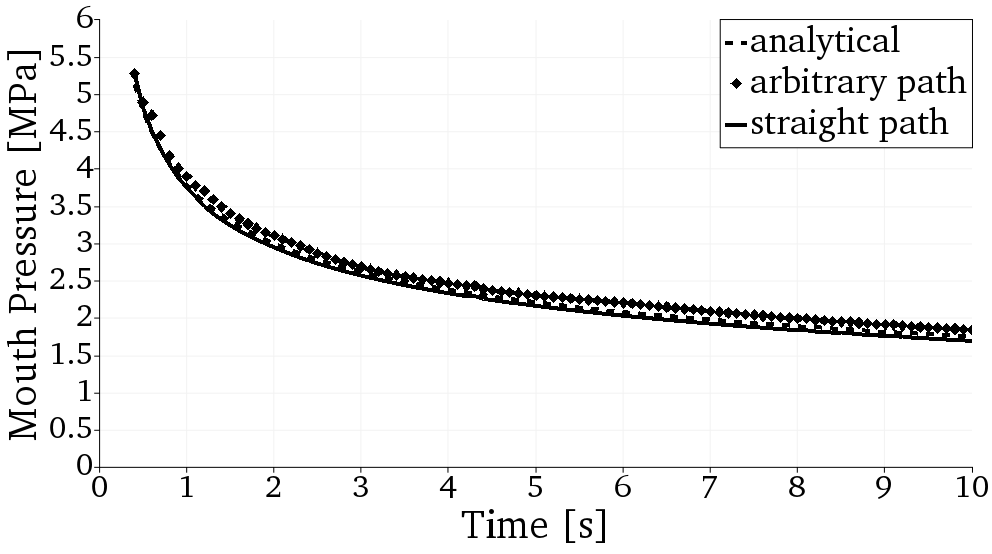} 
\includegraphics[width=0.5\textwidth]{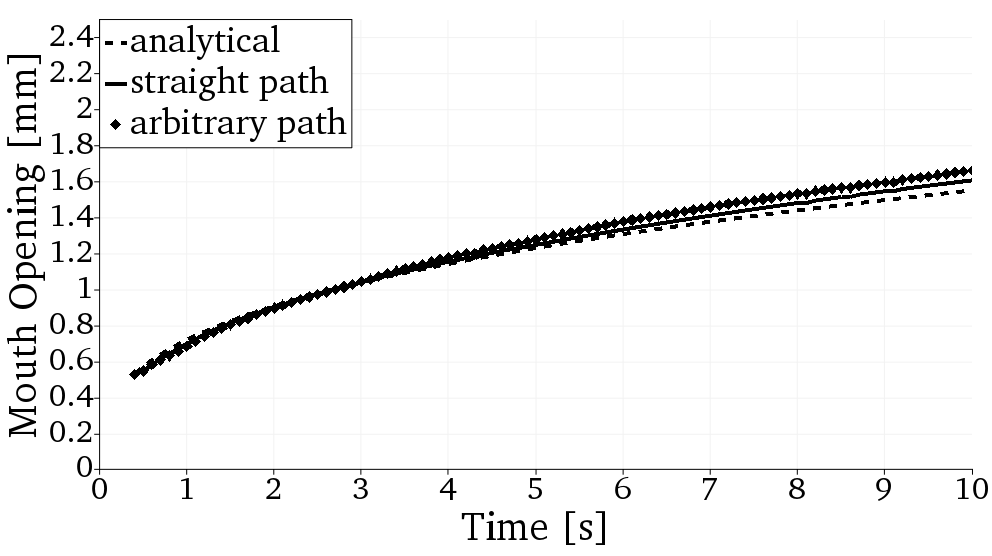} 
\caption{\emph{
Time evolution of the mouth net pressure $p - \sigma_0$ (left) and mouth opening (right)
obtained from the simulations with both the arbitrary and the biased meshes in comparison 
with the analytical solution from \cite{Garagash-asme:2005}.}}
\label{fig:mouth-analytical}
\end{figure}

Figure \ref{fig:mouth-analytical} compares the time evolution of the net pressure, $p - \sigma_0$, and 
opening $w$ at the crack mouth (\emph{i.e.} at the injection point) obtained with the two different meshes 
with the analytical solution from \cite{Garagash-asme:2005}. 
In both cases, the results are in good agreement with the analytical solution.

\clearpage
\subsection{Fully-coupled vs. staggered hydro-mechanics with applied inflow boundary conditions}

As mentioned in the introduction, the concern about a \emph{staggered} coupling approach 
has been brought up previously. Here we investigate the limitations of such approach. 
We revisit the test problem of Section \ref{sec:results:verification-given-length} and we explore 
the staggered approach. In this case, Equations \eqref{eq:discreteElastoStatics} and \eqref{eq:discreteLubrication} 
are solved with respect to their primal variable in a sequential iterative manner in which the other 
unknown required from the equation is taken from the results of the previous iteration:
the solid displacements are obtained from Equation \eqref{eq:discreteElastoStatics} using the 
pressure from the previous iteration; similarly, the pressure is computed from Equation
\eqref{eq:discreteLubrication} using the crack opening from the previous iteration. 
Numerically this corresponds to an \emph{ad hoc} linearization of the problem where
the linear system \eqref{eq:fully-coupled-fully-discrete} above becomes decoupled 
and of the form $A \vect{U} = \vect{E} - B \vect{P}, D \vect{P} = \vect{F} - C \vect{U}$. 

It is important to note that under these conditions the discrete system loses its parabolic character. 
Specifically the decoupled fluid equation becomes elliptic thus requiring a Dirichlet boundary condition, 
\emph{i.e.} the specification of the pressure at least at one point in the domain.
In order to mitigate the problems associated with these
deficiencies in the staggered approach, different numerical artifacts have been proposed,
including: setting the Dirichlet boundary condition at the mouth of the crack, which is seldom
or never representative of hydraulic fracture operations \cite{Boone:1990}, estimating a value
for the mouth pressure from the imposed flow \cite{Khoei:2015}, applying the fictitious
Dirichlet boundary condition near the crack tip using the concept of the fluid lag where the
pressure is assumed to be the vapor pressure (in the case of impermeable rock)
\cite{Vahab:2018}. In \cite{Hirmand:2019}, the problem of system ill-posedness is also avoided
by enforcing the need to apply Dirichlet boundary conditions at the mouth of the crack. 

We consider a plane-strain crack as in Section \ref{sec:results:verification-given-length} 
with a given initial length $\bar{\len}$. 
To be able to impose natural boundary conditions specifying the injection
rate at the crack mouth in the fluid boundary value problem resulting from the staggered solver, we need to 
impose an essential pressure boundary at some point along the crack. Having at disposal the analytical 
solution to this problem \cite{Garagash-asme:2005}, we choose to impose zero pressure at the point behind 
the crack tip where the analytical pressure solution is zero, as the problem degeneracy forbids to impose the 
pressure at the crack tip. 
Similarly to what we did in Section \ref{sec:results:verification-given-length}, we simulate one time step,
advancing from the solution at time $t^n$ to the solution at time $t^{n+1}$. We initialize the staggered solvers 
using as old opening $w^n$ the analytical opening field at time $t^n$. The guesses for opening and 
pressure, $w^{n+1}_0$ and $p^{n+1}_0$, can be freely chosen, depending on how far from the analytical
solution we want to initialize the two iterative procedures.

\begin{figure}
\begin{subfigure}{\textwidth}
\includegraphics[width=0.45\textwidth]{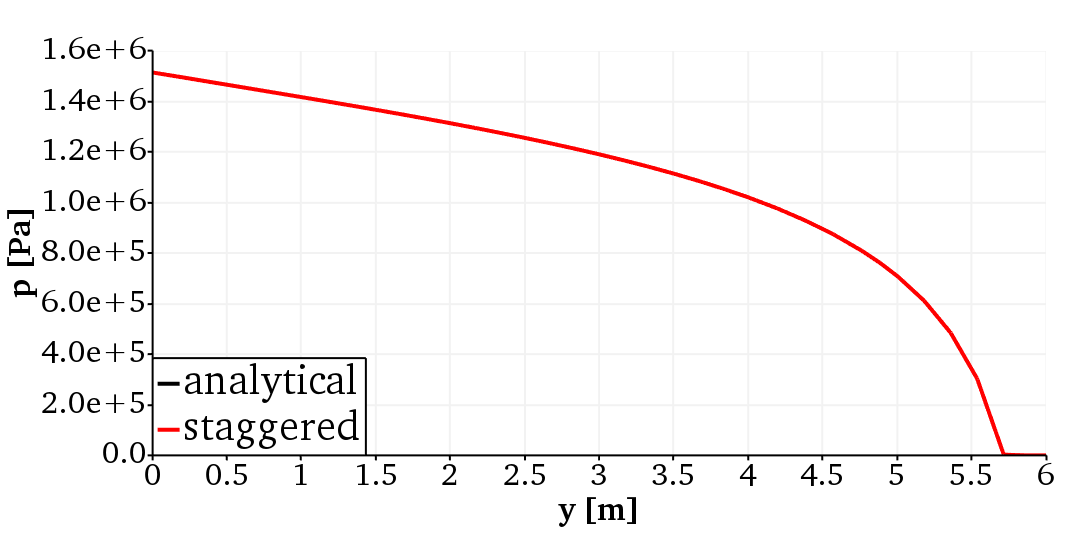}
\hfill
\includegraphics[width=0.45\textwidth]{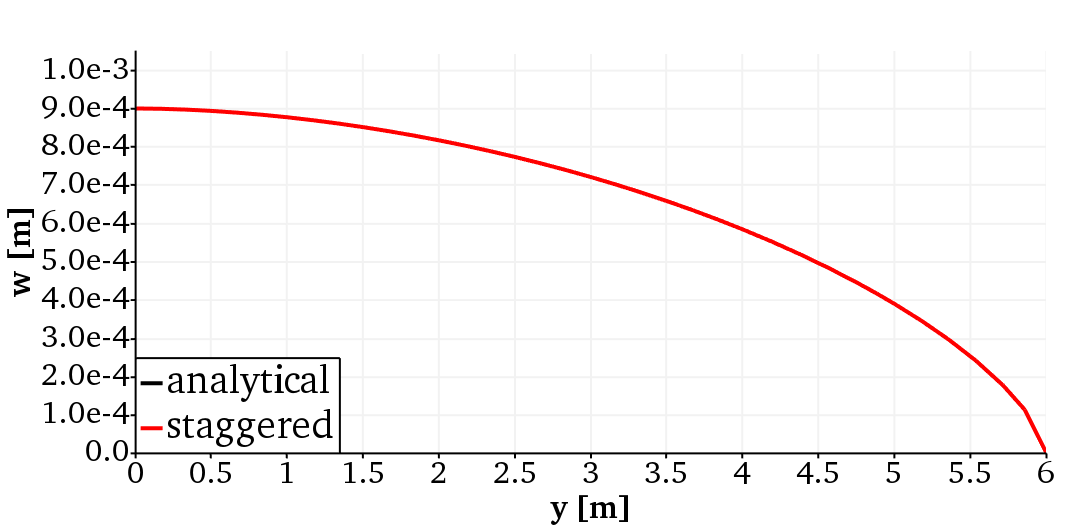} 
\caption{\emph{Initial guess}}
\end{subfigure}
\begin{subfigure}{\textwidth}
\includegraphics[width=0.45\textwidth]{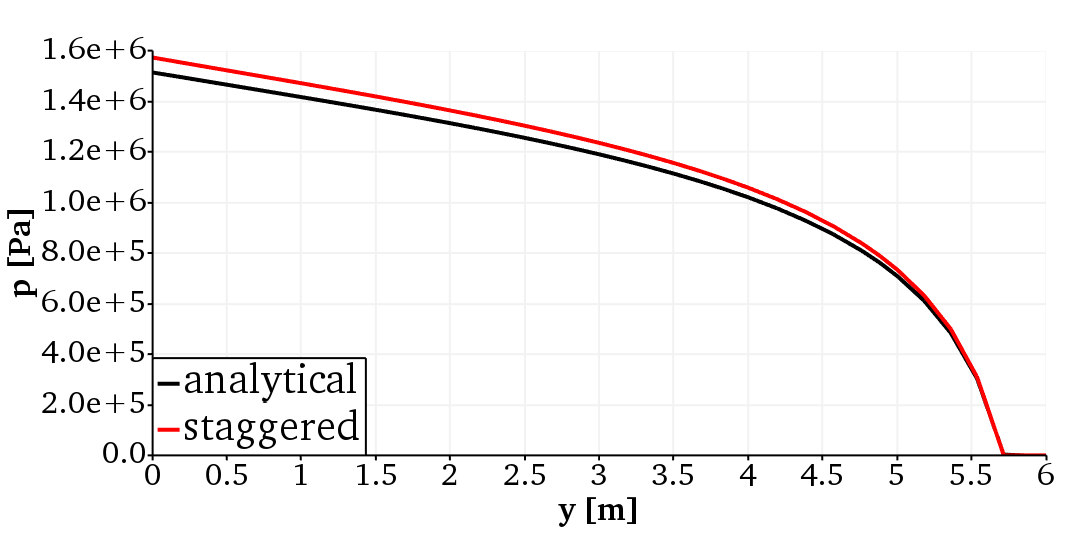}
\hfill
\includegraphics[width=0.45\textwidth]{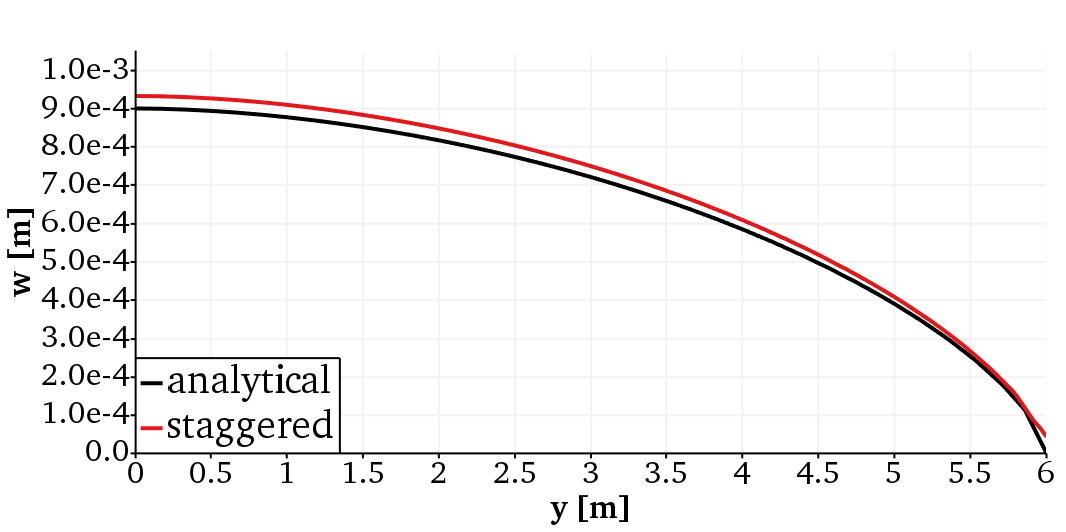}
\caption{\emph{Iteration $1$}}
\end{subfigure}
\begin{subfigure}{\textwidth}
\includegraphics[width=0.45\textwidth]{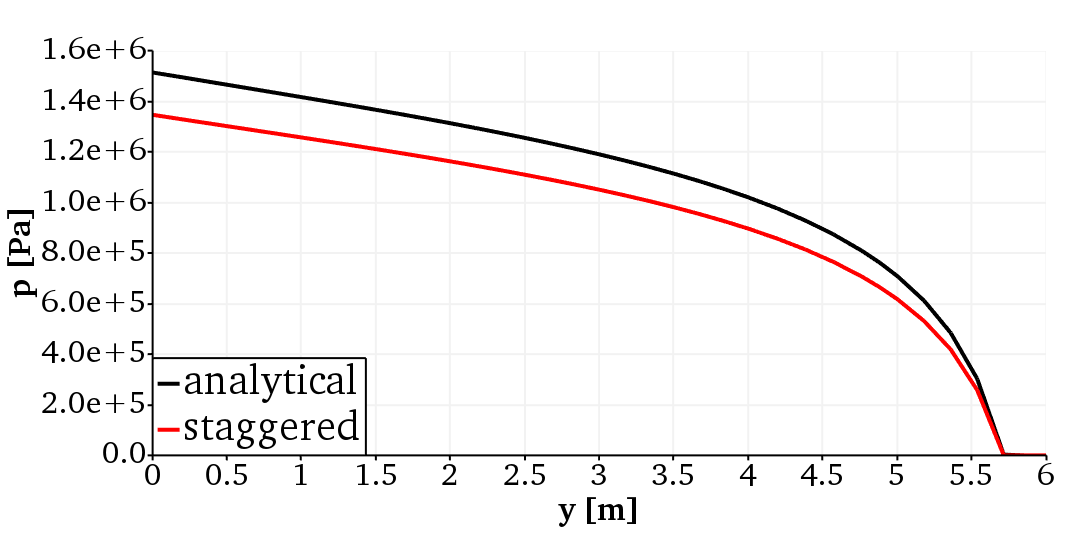}
\hfill
\includegraphics[width=0.45\textwidth]{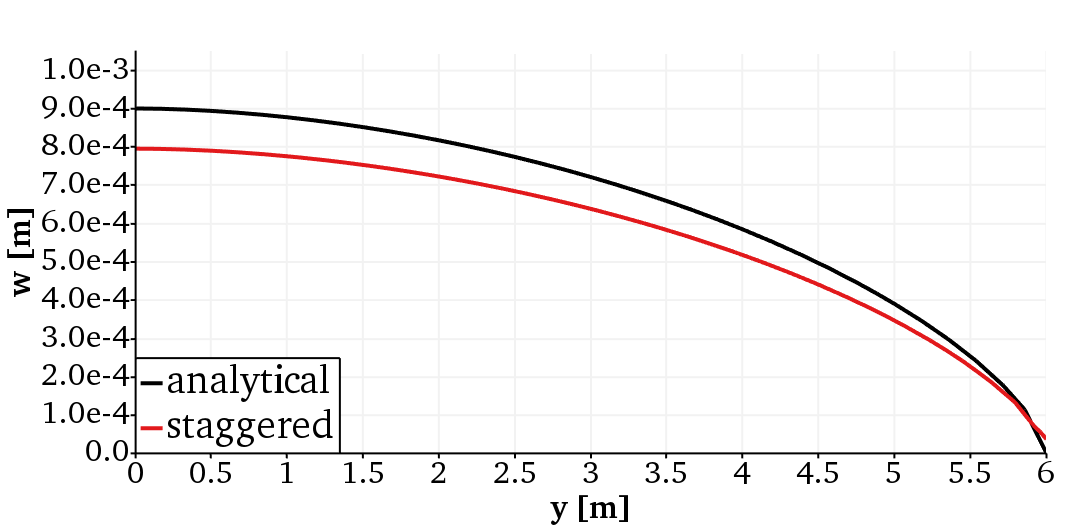} 
\caption{\emph{Iteration $2$}}
\end{subfigure}
\begin{subfigure}{\textwidth}
\includegraphics[width=0.45\textwidth]{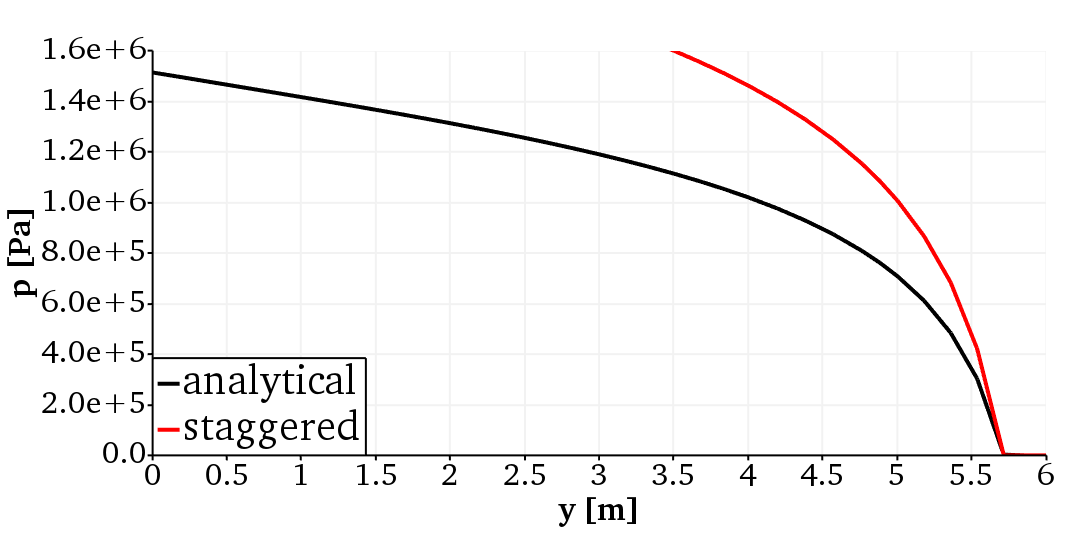}
\hfill
\includegraphics[width=0.45\textwidth]{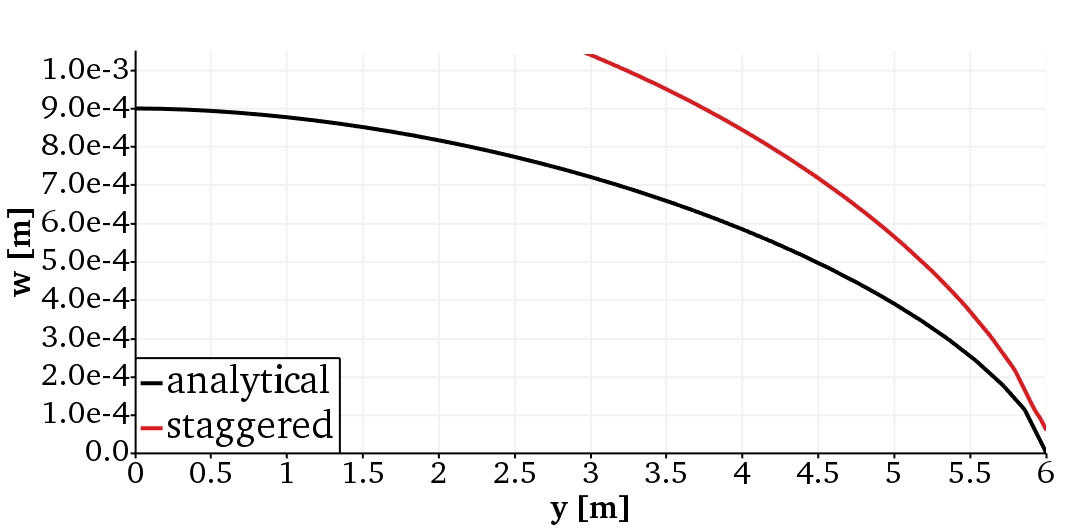}
\caption{\emph{Iteration $3$}}
\end{subfigure}
\begin{subfigure}{\textwidth}
\includegraphics[width=0.45\textwidth]{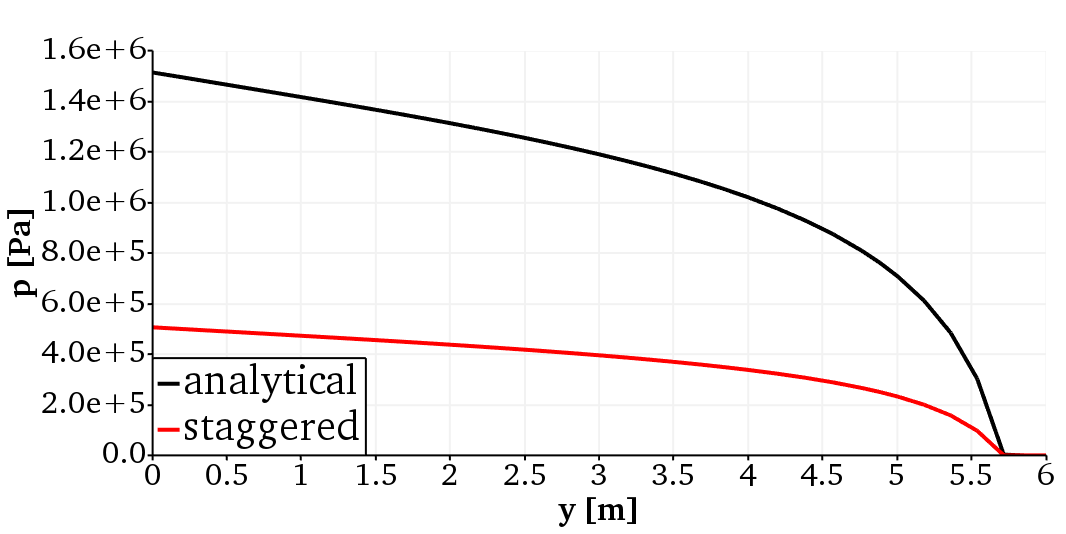}
\hfill
\includegraphics[width=0.45\textwidth]{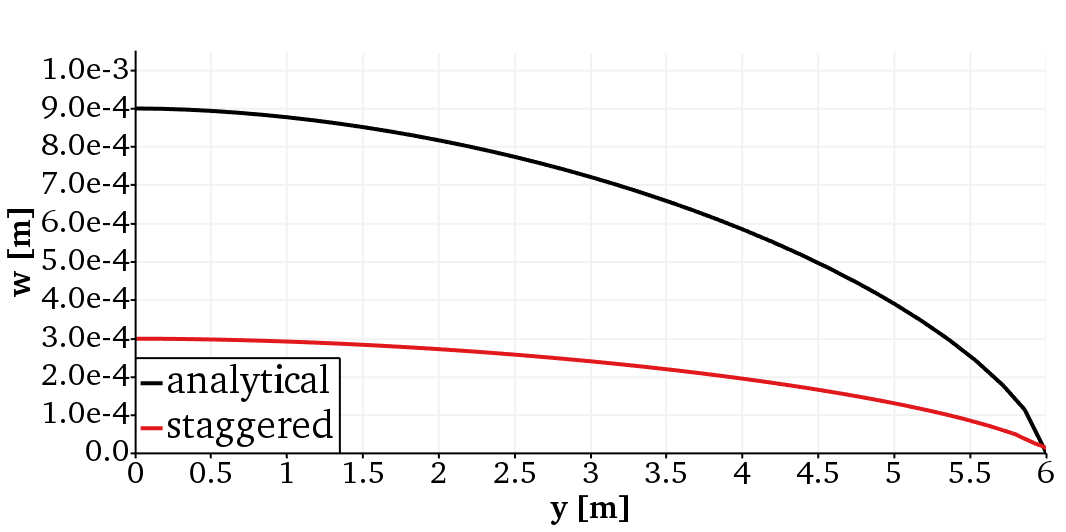}
\caption{\emph{Iteration $4$}}
\end{subfigure}
\caption{\emph{
Snapshots of the pressure (left) and crack opening (right) distributions during the \emph{staggered}
nonlinear iteration procedure. 
The staggered solver diverges even when the initial guess corresponds to the analytical solution.}} 
\label{fig:iterations-staggered}
\end{figure}

\begin{figure}
\begin{subfigure}{\textwidth}
\includegraphics[width=0.45\textwidth]{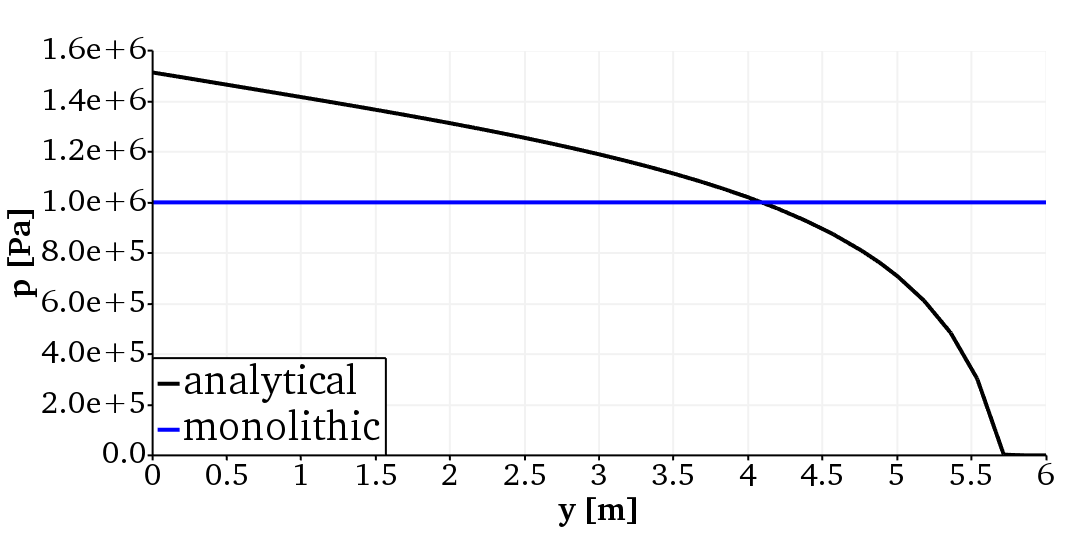}
\hfill
\includegraphics[width=0.45\textwidth]{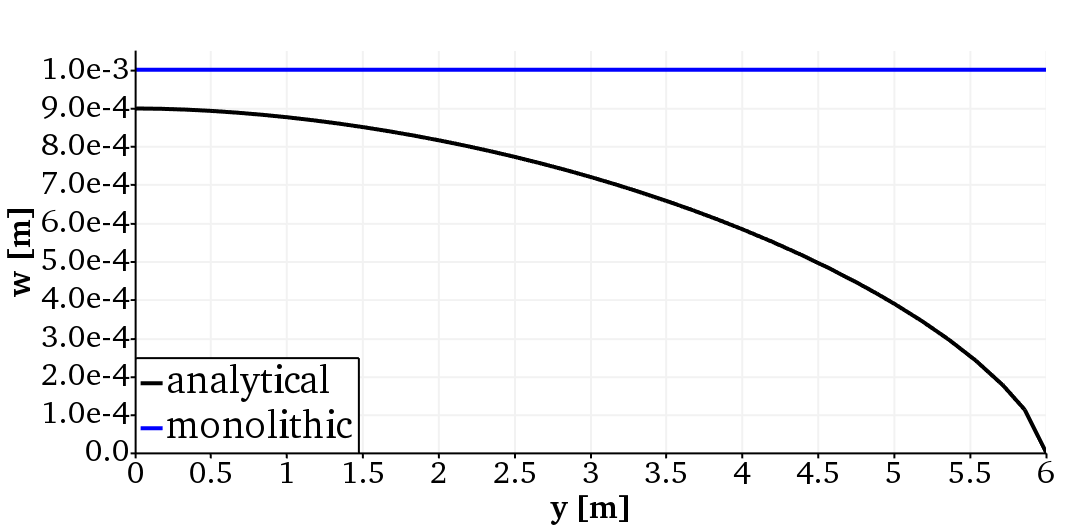} 
\caption{\emph{Initial guess}}
\end{subfigure}
\begin{subfigure}{\textwidth}
\includegraphics[width=0.45\textwidth]{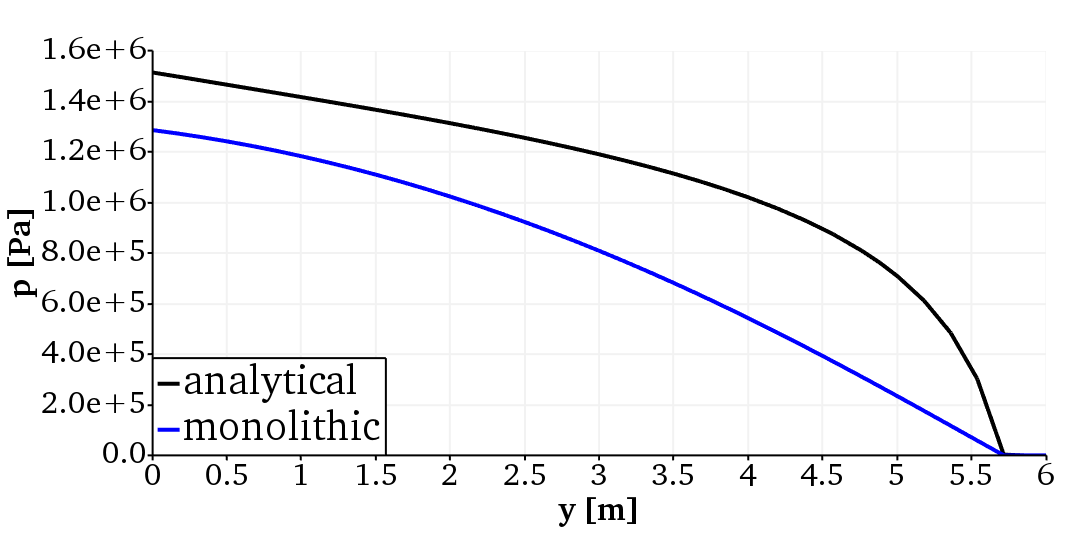}
\hfill
\includegraphics[width=0.45\textwidth]{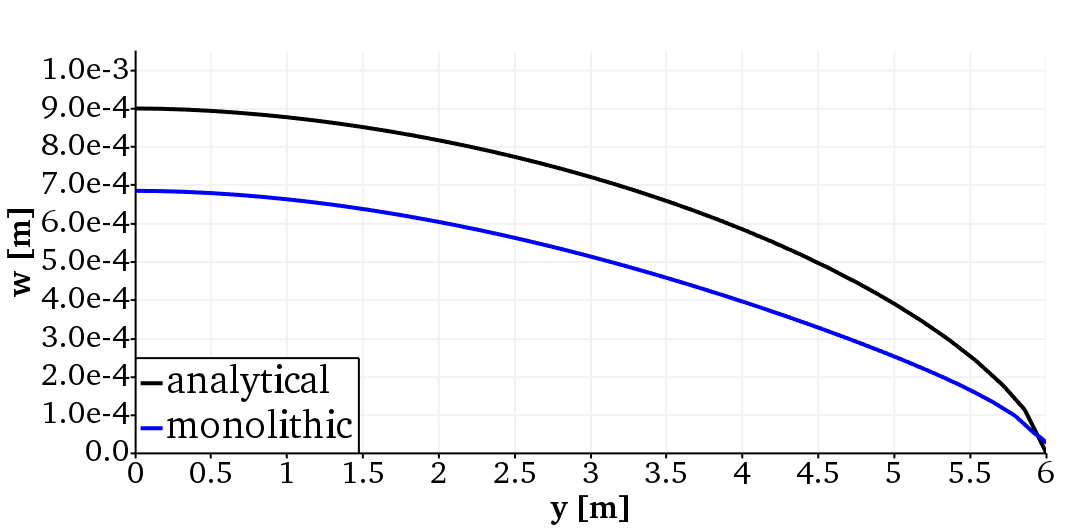} 
\caption{\emph{Iteration $1$}}
\end{subfigure}
\begin{subfigure}{\textwidth}
\includegraphics[width=0.45\textwidth]{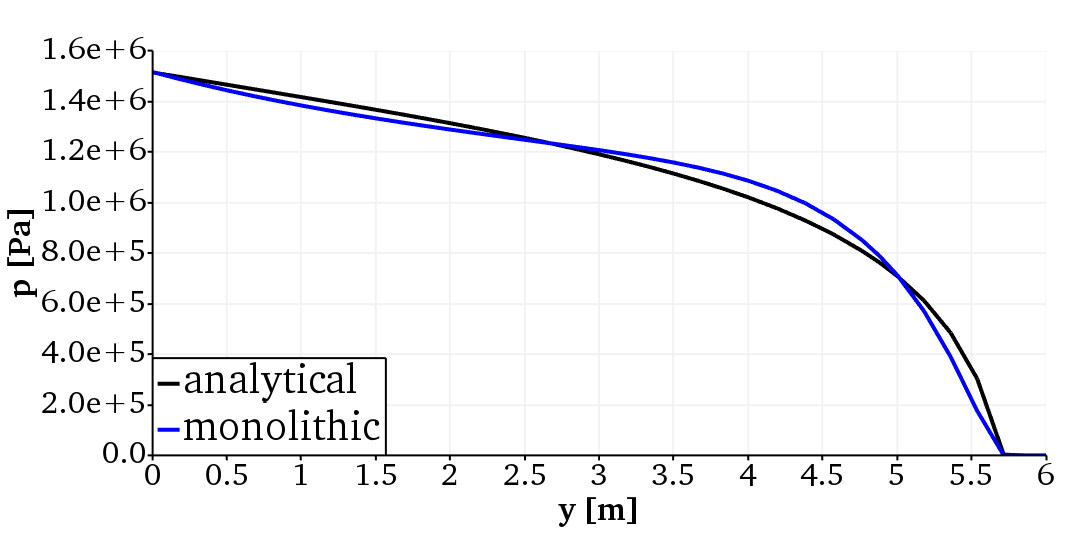}
\hfill
\includegraphics[width=0.45\textwidth]{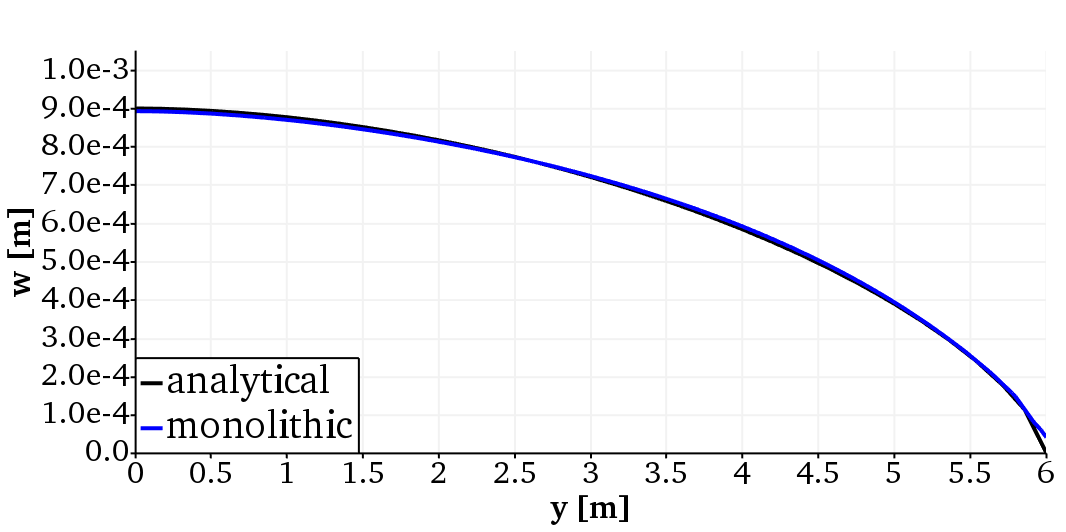} 
\caption{\emph{Iteration $2$}}
\end{subfigure}
\begin{subfigure}{\textwidth}
\includegraphics[width=0.45\textwidth]{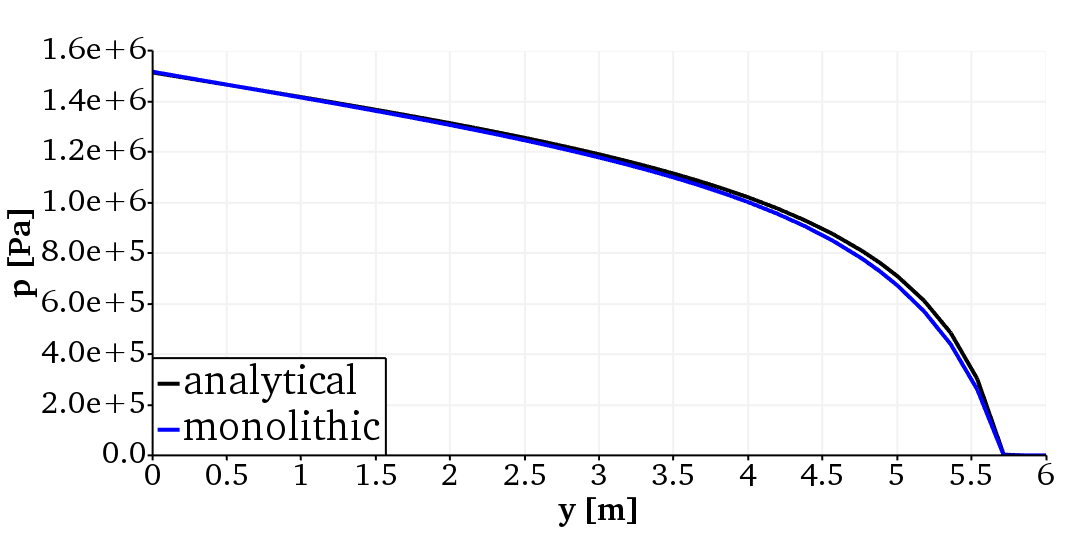}
\hfill
\includegraphics[width=0.45\textwidth]{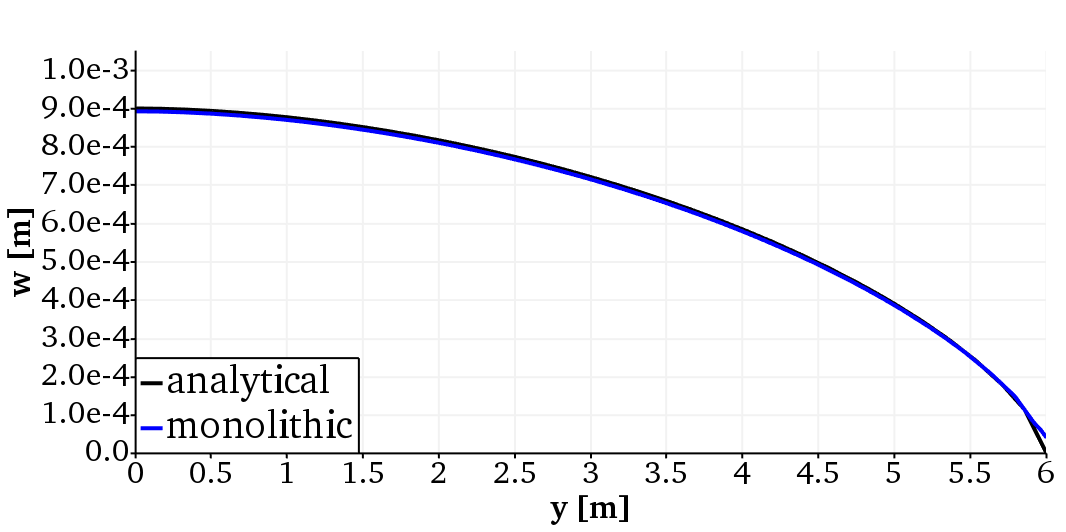} 
\caption{\emph{Iteration $3$}}
\end{subfigure}
\begin{subfigure}{\textwidth}
\includegraphics[width=0.45\textwidth]{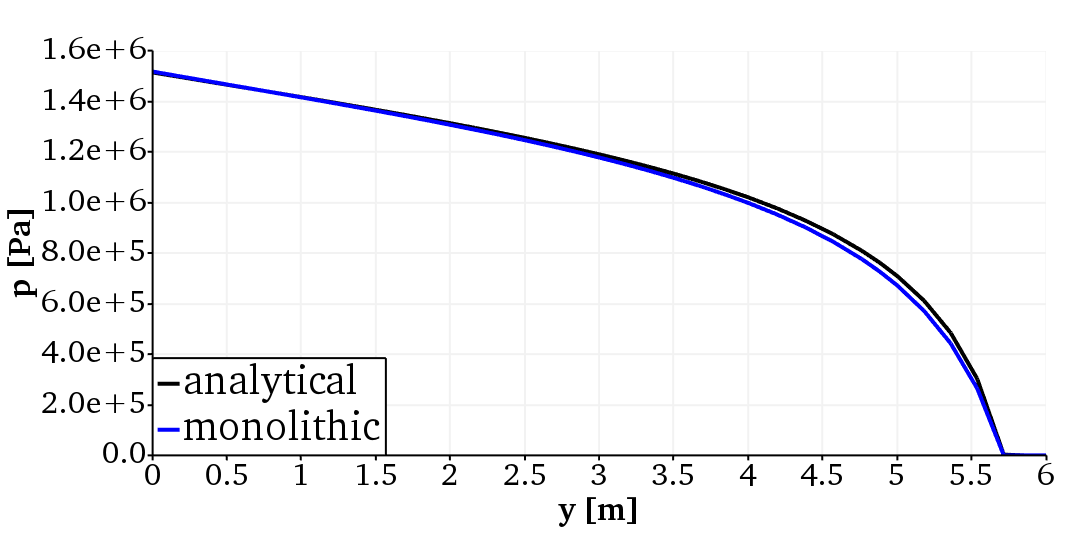}
\hfill
\includegraphics[width=0.45\textwidth]{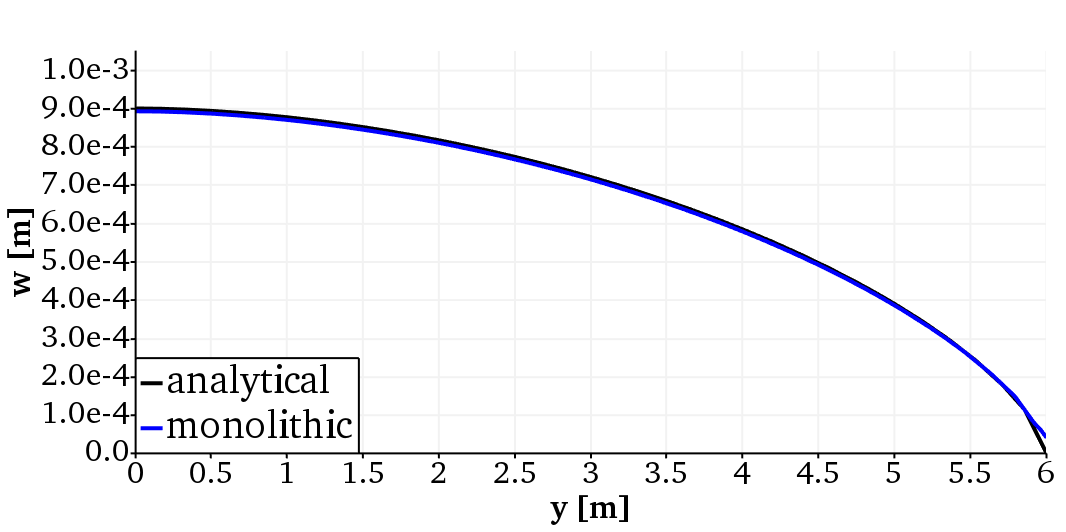} 
\caption{\emph{Iteration $4$}}
\end{subfigure}
\caption{\emph{
Snapshots of the pressure (left) and crack opening (right) distributions during the \emph{fully-coupled} 
nonlinear iteration procedure.
The fully-coupled solver converges robustly with respect to the initial guess.}} 
\label{fig:iterations-fully-coupled}
\end{figure}

Figures \ref{fig:iterations-staggered} and \ref{fig:iterations-fully-coupled} 
show the results obtained during the nonlinear iterations with the two solvers, staggered and fully-coupled, 
starting from an 
initial crack length $\bar{\len} = 6m$ and with the set of parameters in Table \ref{tab:parametersHM} 
corresponding to the viscosity-dominated regime. Figure \ref{fig:iterations-staggered} shows that the 
staggered approach does not converge for inflow boundary conditions, not even using the analytical 
solutions as initial guesses. This instability can be explained by the activation of vicious-circle 
mechanism for which a higher pressure obtained from the lubrication equation will result in a higher mouth 
opening in the elasticity equation that will make the pressure drastically drop at the following fluid 
iteration. In turn, a lower pressure obtained in the lubrication will result in a lower mouth 
opening in the elasticity equation that will make the pressure drastically increase at the following fluid 
iteration.
This mechanism disappears in the fully-coupled approach, which computes the opening and pressure at once 
and with the proper tangents. This is confirmed by our numerical experimentation, which shows 
robust convergence even starting from initial guesses that are far from the analytical solution, see
Figure \ref{fig:iterations-fully-coupled}.

\clearpage
\subsection{Verification of the fully-coupled algorithm for a penny-shaped crack in the viscosity dominated regime} 
  \label{sec:results:verification-given-radius}

In order to exercise the proposed computational framework under three dimensional conditions, 
we conduct a simulation of a fluid-filled penny-shaped crack in the viscosity dominated regime 
and compare against the analytical solutions in \cite{Savitski:2002}.

\comment{ 
Fluid-driven radial fractures in impermeable solids
generally propagate in the viscosity regime, while the toughness regime is relevant only in exceptional circumstances
\cite{Savitski:2002}. }

For the radial crack geometry, the nondimensional toughness 
$\mathcal{K}$ and viscosity $\mathcal{M}$ are expressed as:
$$
\mathcal{K} = 
\left(\frac{t^2}{(12 \mu)^5 \ Q_0^3 \ E'^{13}} \right)^{1/18} 
\frac{8}{\sqrt{2 \pi}}K_{IC},
$$
and
$$
\mathcal{M} = 
12 \mu 
\left( \frac{Q_0^3\ E'^{13}}{t^2} \right)^{1/5}
\left( \frac{8}{\sqrt{2 \pi}}K_{IC} \right) ^{-18/5},
$$
see \cite{Savitski:2002}. Note that, similarly to the plane-strain crack geometry, $\mathcal{K}$ and $\mathcal{M}$ 
represent the importance of fracture and viscous flow as the main dissipation mechanism, the competition of 
these mechanisms being highlighted from the fact that $\mathcal{M} = \mathcal{K}^{-18/5}$.
Furthermore, analogously to the case of a plane-strain crack, analytical solutions based on the assumption that the fluid front 
coincides with the crack tip result in a pressure singularity at the tip.
The zero-toughness solution is known to have a weakly singular
near-tip fracture opening with a $2/3$-power dependence on the distance from the tip, while the 
pressure field is still singular at the tip with the $-1/3$-power dependence. In addition to the $-1/3$-power tip singularity,
the pressure field also presents a logarithmic singularity at the injection point \cite{Savitski:2002}.
Threshold values delimiting the two regimes have been obtained with the result that 
$\mathcal{K} < \mathcal{K}_0 = 0.37$ corresponds to the viscosity-dominated regime \cite{Garagash:2000},
and $\mathcal{K} > \mathcal{K_{\infty}} = 3.5$ corresponds to the toughness-dominated regime \cite{Savitski:2002}.
However, differently than in the case of a plane-strain crack, for a radial crack 
$\mathcal{K}$ and $\mathcal{M}$ depend on time, suggesting that transition between regimes may happen
during fracture propagation, albeit typically very slowly \cite{Savitski:2002}.

\begin{table}[h]
\centering
\begin{tabular}{|c|c|} 
\hline
$E$ & $17 \ G Pa$ \\
\hline
$\nu$ & $0.2$ \\
\hline
$\mu$ & $0.15 \ Pa \ s$ \\
\hline
$Q_0$ & $0.03 \ m^3 \ s^{-1}$  \\
\hline
$G_c$ & $100 \ Pa \ m$ \\
\hline
\hline
$\mathcal{K}$ & $0.35$ \\
\hline
$\mathcal{M}$ & $42.4$ \\
\hline 
\end{tabular}
\caption{\emph{The parameters used in the simulations with the 
corresponding nondimensional toughness $\mathcal{K}$ and viscosity $\mathcal{M}$.
Note that $\mathcal{K}$ and $\mathcal{M}$ are time-dependent, hence depend on
the crack radius $\bar{R}$. The values of $\mathcal{K}$ and $\mathcal{M}$ reported 
in the Table correspond to $\bar{R}$, that is for a time $t = t^0 = 17.6 s$ (see analytic
solution in \cite{Savitski:2002}). This set of parameters corresponds to the viscosity-dominated 
regime.}}
\label{tab:parametersHM-penny}
\end{table}

We provide verification of the proposed computational framework in the viscosity-dominated regime,
see parameters in Table \ref{tab:parametersHM-penny}, for an axisymmetric crack. 
Once the parameters are chosen, the analytical solutions provide the crack radius, pressure
distribution, and crack opening distribution as a function of time, respectively
$R(t), p(r,t), w(r,t)$. We extract the functional forms of these solutions from
\cite{Savitski:2002} in the case of the viscosity-dominated regime.
Similarly to the analysis done in Section \ref{sec:results:verification-given-length}, we verify 
that the computed and theoretical pressure and opening distributions are 
consistent  for a crack of given radius $\bar{R}$ embedded in the computational mesh by 
performing the calculations of one time step, which advance the solutions from time $t^n$
to time $t^{n+1}$. We choose the initial 
guesses of the pressure and opening fields ${p}^{n+1}_0$ and ${w}^{n+1}_0$ adopting the analytical 
solution from the previous time step (\emph{i.e.} ${w}^{n+1}_0 = {w}^n$ and ${p}^{n+1}_0 = {p}^n$).

In our calculations, we utilize the computational domain $[0, 50m] \times [0, 50m] \times [0, 50m]$.  
We exploit the symmetry of the problem, being careful to only apply the injection rate $\frac{Q_0}{4}$.
The radial fracture lies on the plane $z = 0$ and has an initial radius $\bar{R} = 10m$. 
The injection point is located at the origin. Consistently with the choice of initial radius, 
the initial time of the simulation is $ t = t^0 = 17.6 s$, as prescribed from the analytical solution
\cite{Savitski:2002}. Notice that, with the set of parameters summarized in Table 
\ref{tab:parametersHM-penny}, the viscosity-dominated regime corresponds 
to earlier times than $26.8 s$, whereas the toughness-dominated regime corresponds to 
larger times than $16 Gs$. This is consistent with the observation in \cite{Savitski:2002} that the 
toughness-dominated regime is relevant only in exceptional circumstances.
The mesh is refined in the proximity of the crack tip, where the mesh size
is $h = 0.2m$, the polynomial order of the finite elements is $p=2$, and the time step 
is $\Delta t = 1 s$. 

\begin{figure}[h]
\center
\includegraphics[width=\textwidth]{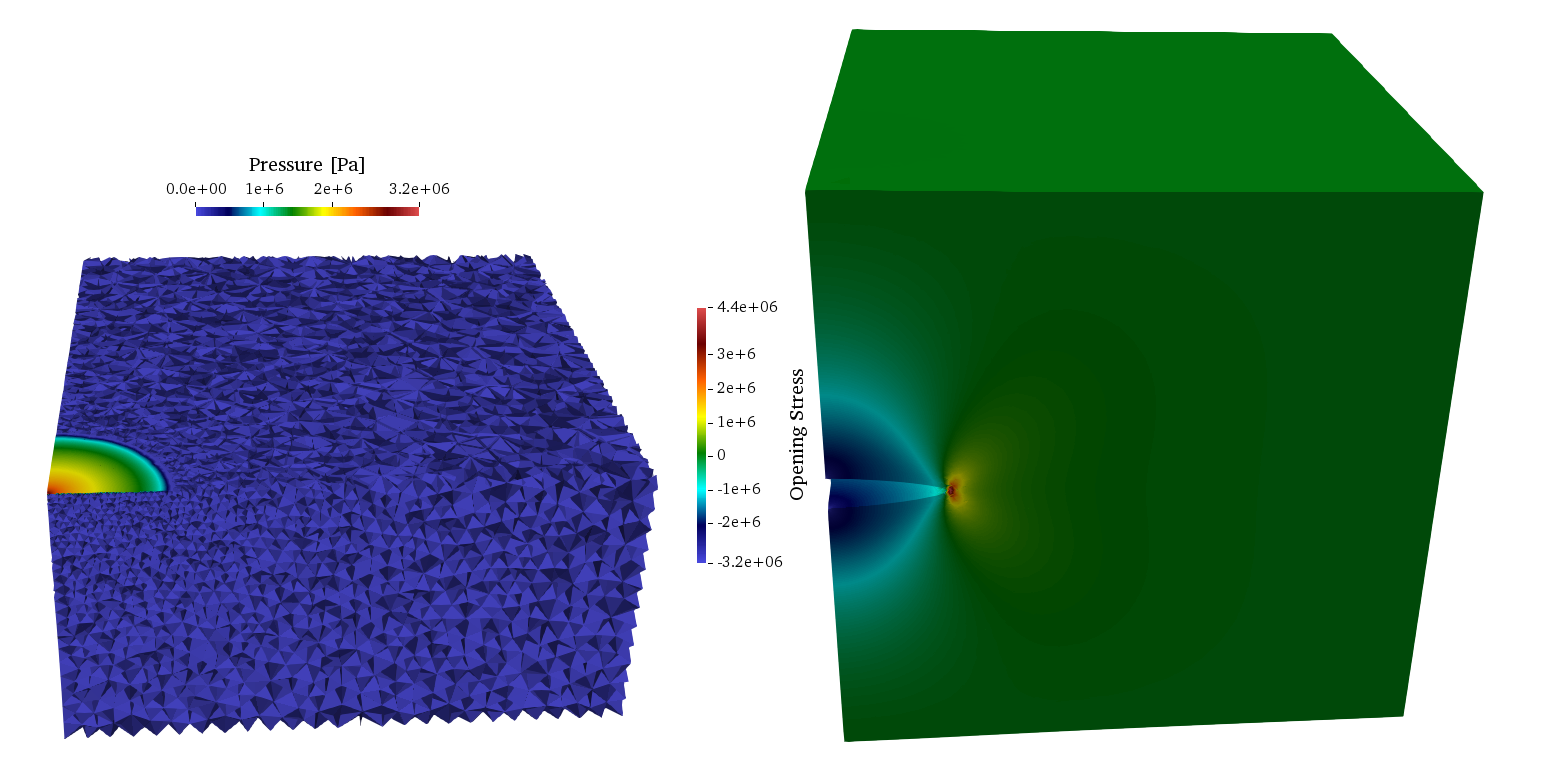} 
\caption{\emph{A slice of the fluid domain (on the left) and the solid domain (on the right) for a 
penny-shaped crack in 3D with contours of the fluid pressure field and the opening stress, 
respectively. The fluid equations are solved on the wireframe of the solid mesh.}} 
\label{fig:penny-shaped-pressure-stress}
\end{figure}

\begin{figure}[h]
\center
\includegraphics[width=0.45\textwidth]{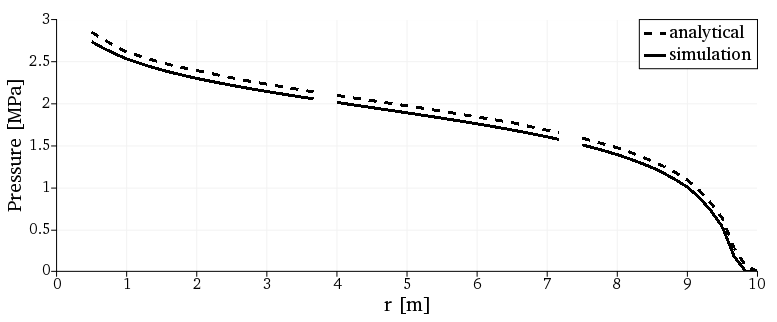} 
\includegraphics[width=0.45\textwidth]{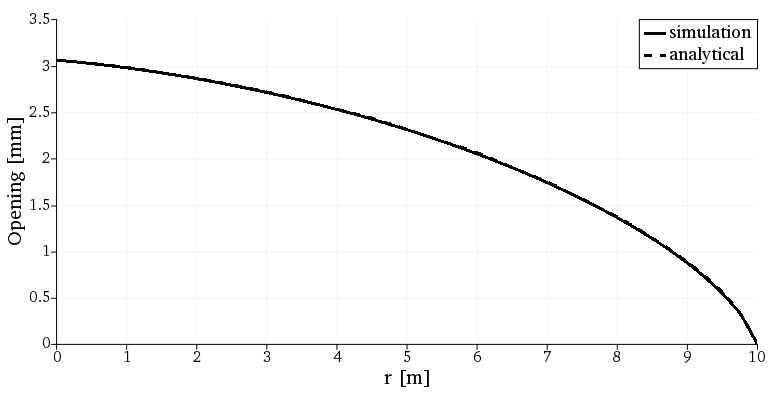} 
\caption{\emph{Verification of the computational predictions of the fluid pressure field 
(left plot) and crack opening field (right plot) for a penny-shaped radial crack 
(analytical solution from \cite{Savitski:2002}).}} \label{fig:penny-shaped-analytical}
\end{figure}

In Figure \ref{fig:penny-shaped-pressure-stress} we show a section of the fluid domain immersed in 
the 3D solid domain. Contours of the fluid pressure and the opening stress show the balance between the 
two at the crack lips, as prescribed by mechanical equilibrium.
Figure \ref{fig:penny-shaped-analytical} shows that the opening and the pressure fields are in excellent 
agreement with the analytical solution from \cite{Savitski:2002}.

%

\newpage
\subsection{Interaction with a pre-existing crack}

The problem of interaction of fluid-driven cracks with pre-existing \emph{natural} 
cracks is of significant interest in oil field operations.
Key questions about this interaction of cracks include whether the natural crack is 
filled by the injected fluid and therefore activated, or if it is simply 
crossed by the propagating crack. Several factors contribute to the problem evolution,
including the angle of interaction, the \emph{in situ} stress, the injection rate and the
viscosity of the fracturing fluid \cite{Chuprakov:2013}. 

\begin{figure}[h]
\centering
\includegraphics[width=0.7\textwidth]{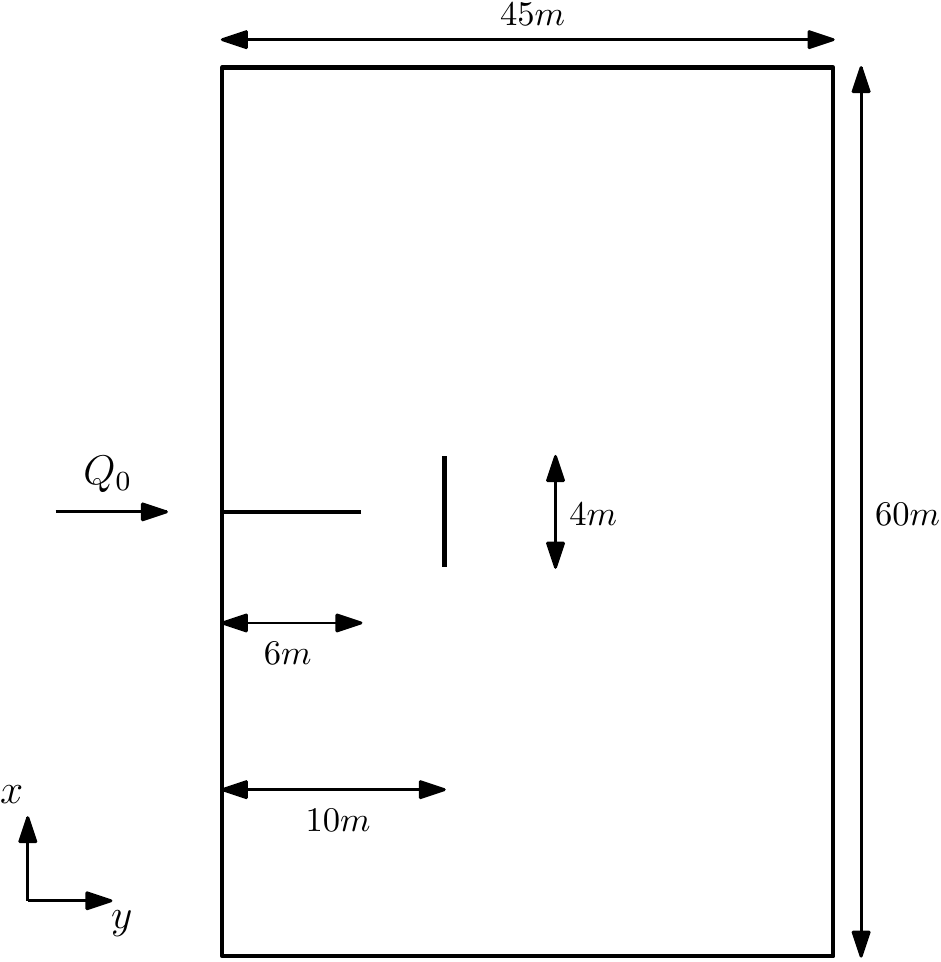} 
\caption{\emph{Schematic of the problem geometry: interaction of a propagating 
fluid-driven crack of $6 \ m$ initial length with a $4 \ m$ pre-existing dry crack
orthogonal to the fluid-driven crack at a $4 \ m$ distance.}}
\label{fig:natural-fracture-geometry}
\end{figure}

In this paper we use the problem of interaction of a fluid-driven crack with a 
natural crack to illustrate the capability of the proposed computational 
approach to effectively describe branching and merging.
We select a case in which a pressurized crack propagates in a domain with a pre-existing 
dry crack, as shown in Figure \ref{fig:natural-fracture-geometry}, with 
material parameters from Table \ref{tab:parametersHM} for the viscosity-dominated
regime.  We initialize the fully-coupled solver as described in Section \ref{sec:results:verification-given-length}
for a crack of initial length $\bar{\len} = 6m$. We assume no far field stress on the remote boundary
and a critical fracture stress $\sigma_c = 6.0MPa$.
The mesh is refined in the neighborhood of the two cracks, where the mesh size is $h = 0.1 m$, 
and the polynomial order of interpolation is $p = 3$, resulting in about $1.6$ million
degrees of freedom. The constant time-step is $\Delta t = 0.025s$.

\begin{figure}[h]
\centering
\begin{subfigure}{\textwidth}
\centering
\includegraphics[trim=0 220 0 220,clip, width=0.7\textwidth]{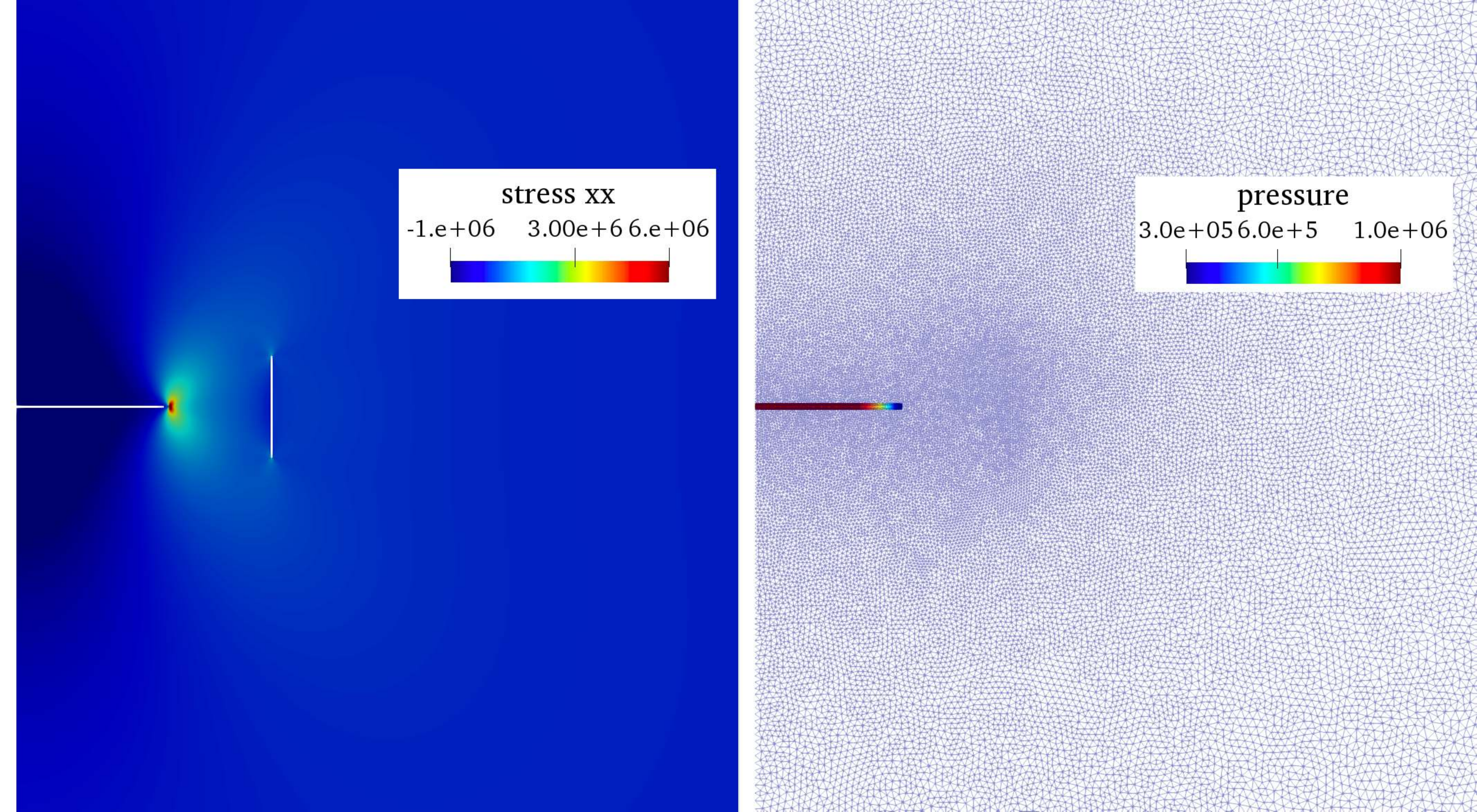} 
\caption{\emph{$t = 0$}}
\end{subfigure}
\begin{subfigure}{\textwidth}
\centering
\includegraphics[trim=0 220 0 220,clip, width=0.7\textwidth]{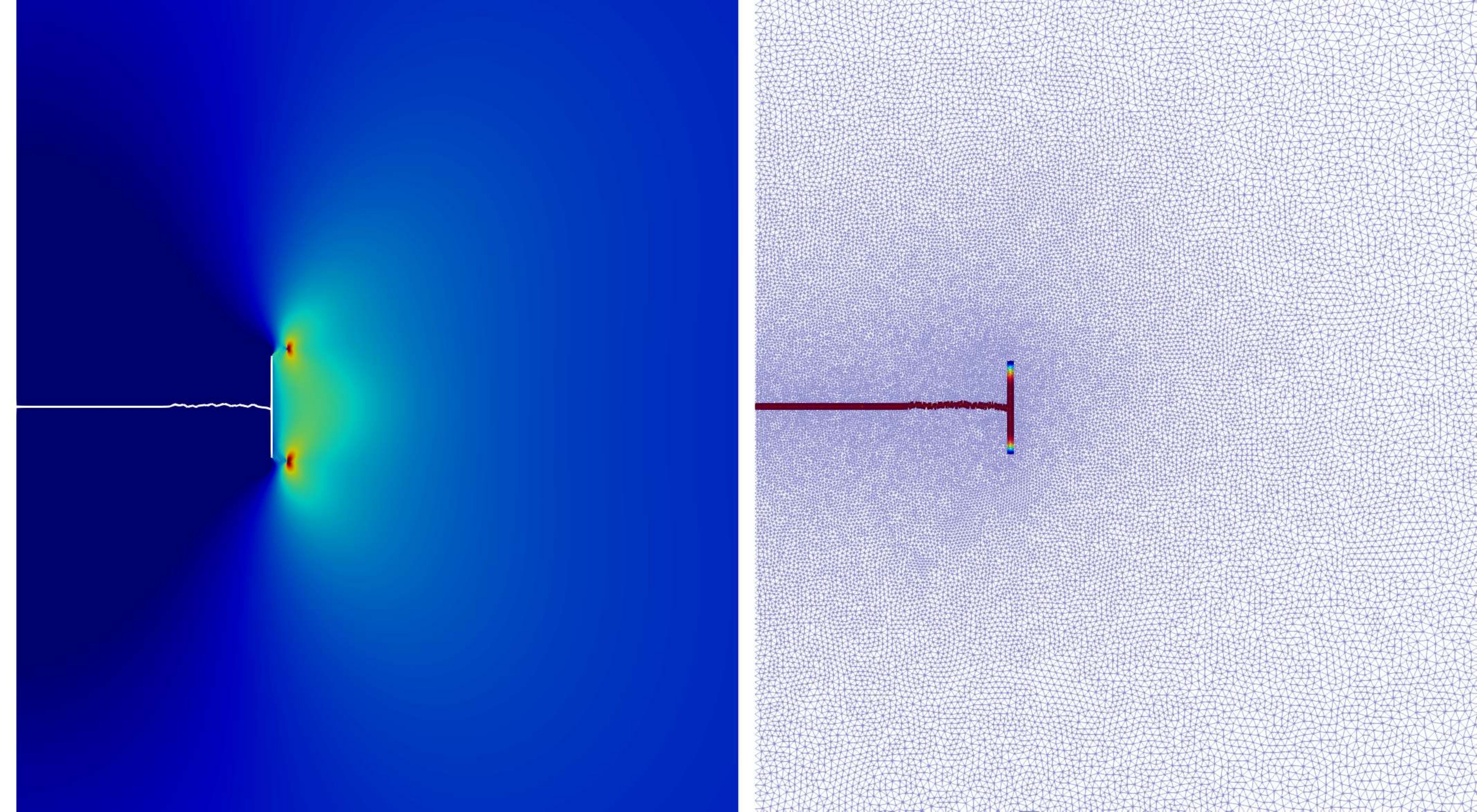} 
\caption{\emph{$t = 29 s$}}
\end{subfigure}
\begin{subfigure}{\textwidth}
\centering
\includegraphics[trim=0 220 0 220,clip, width=0.7\textwidth]{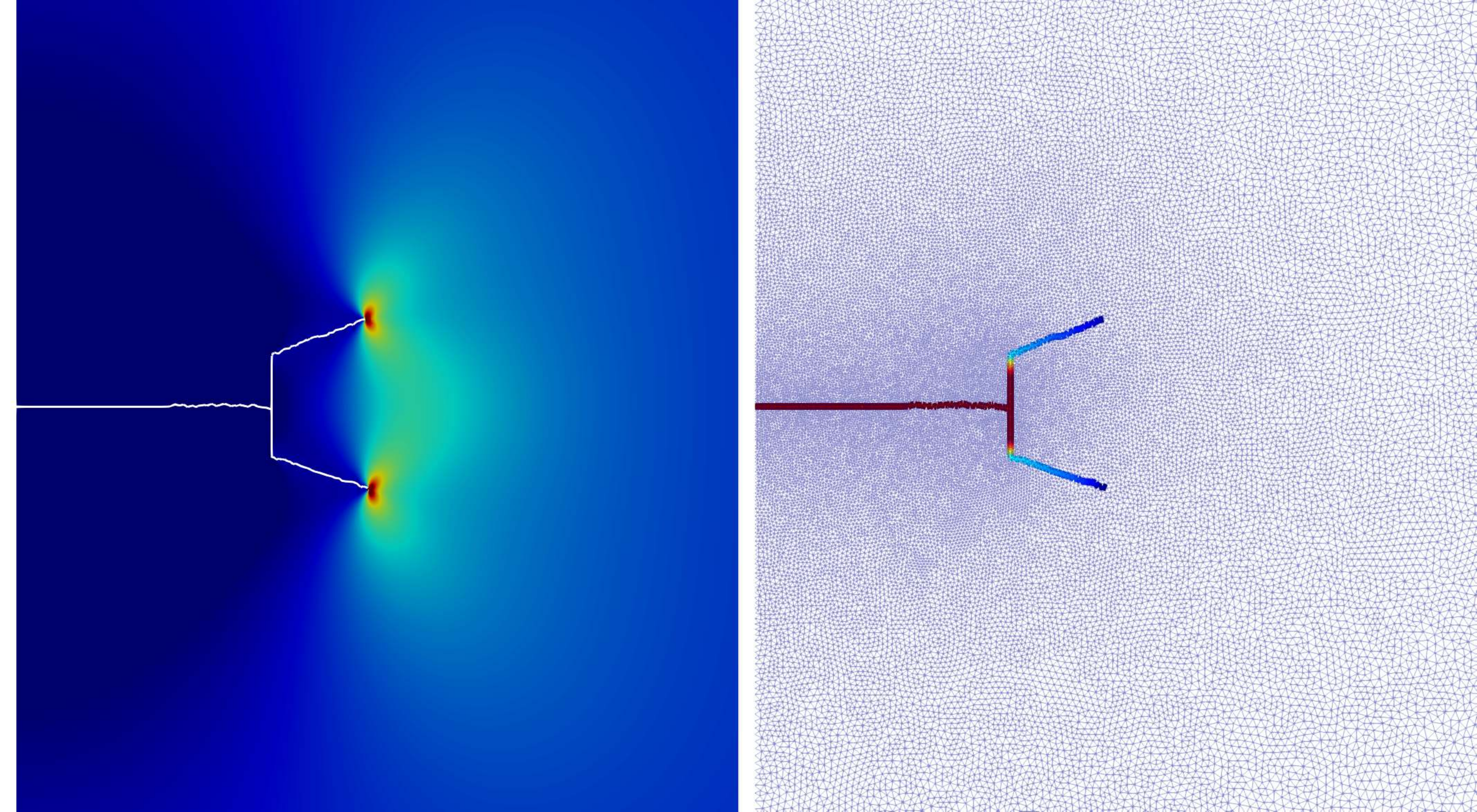} 
\caption{\emph{$t = 58 s$}}
\end{subfigure}
\begin{subfigure}{\textwidth}
\centering
\includegraphics[trim=0 220 0 220,clip, width=0.7\textwidth]{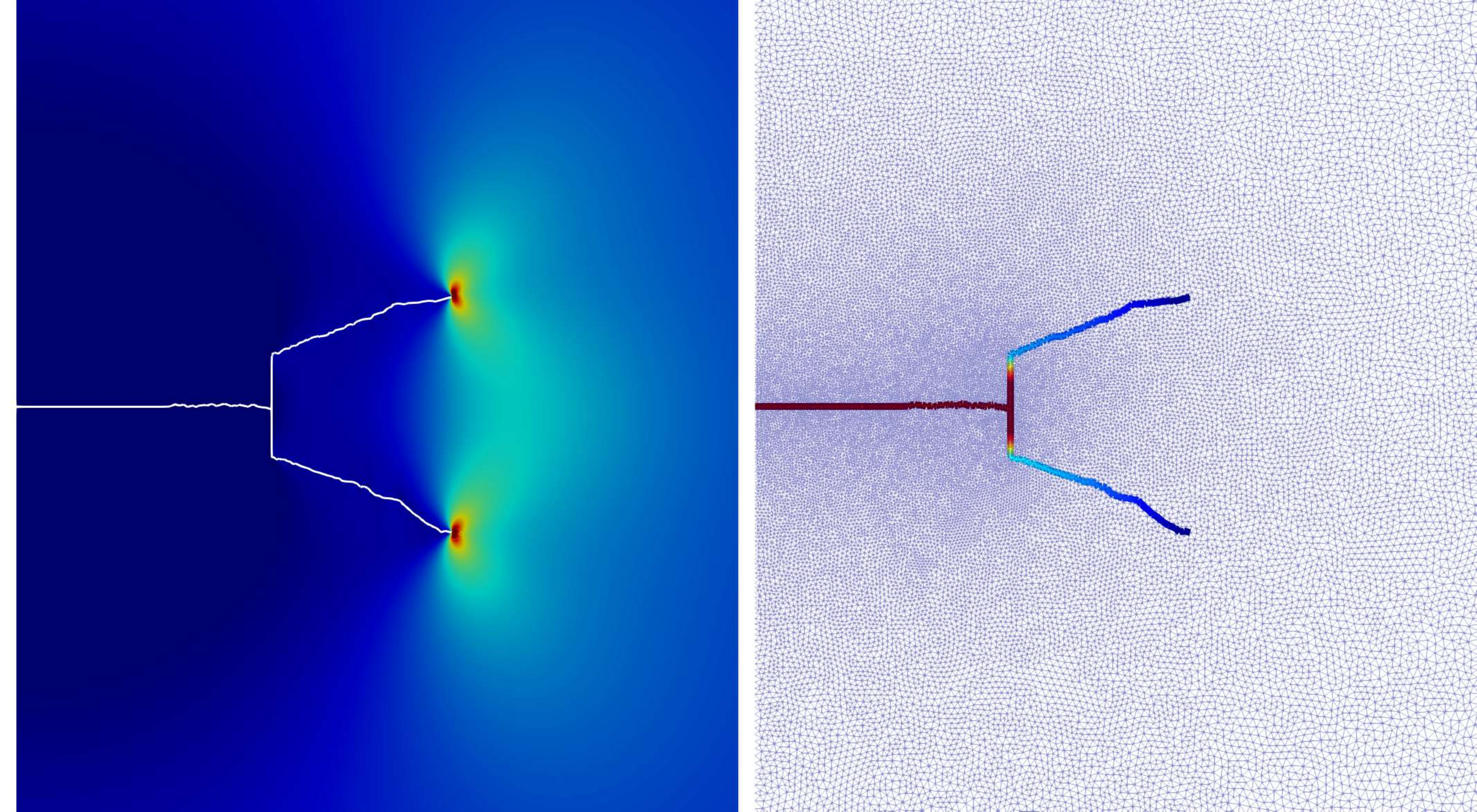} 
\caption{\emph{$t = 87 s$}}
\end{subfigure}
\begin{subfigure}{\textwidth}
\centering
\includegraphics[trim=0 220 0 220,clip, width=0.7\textwidth]{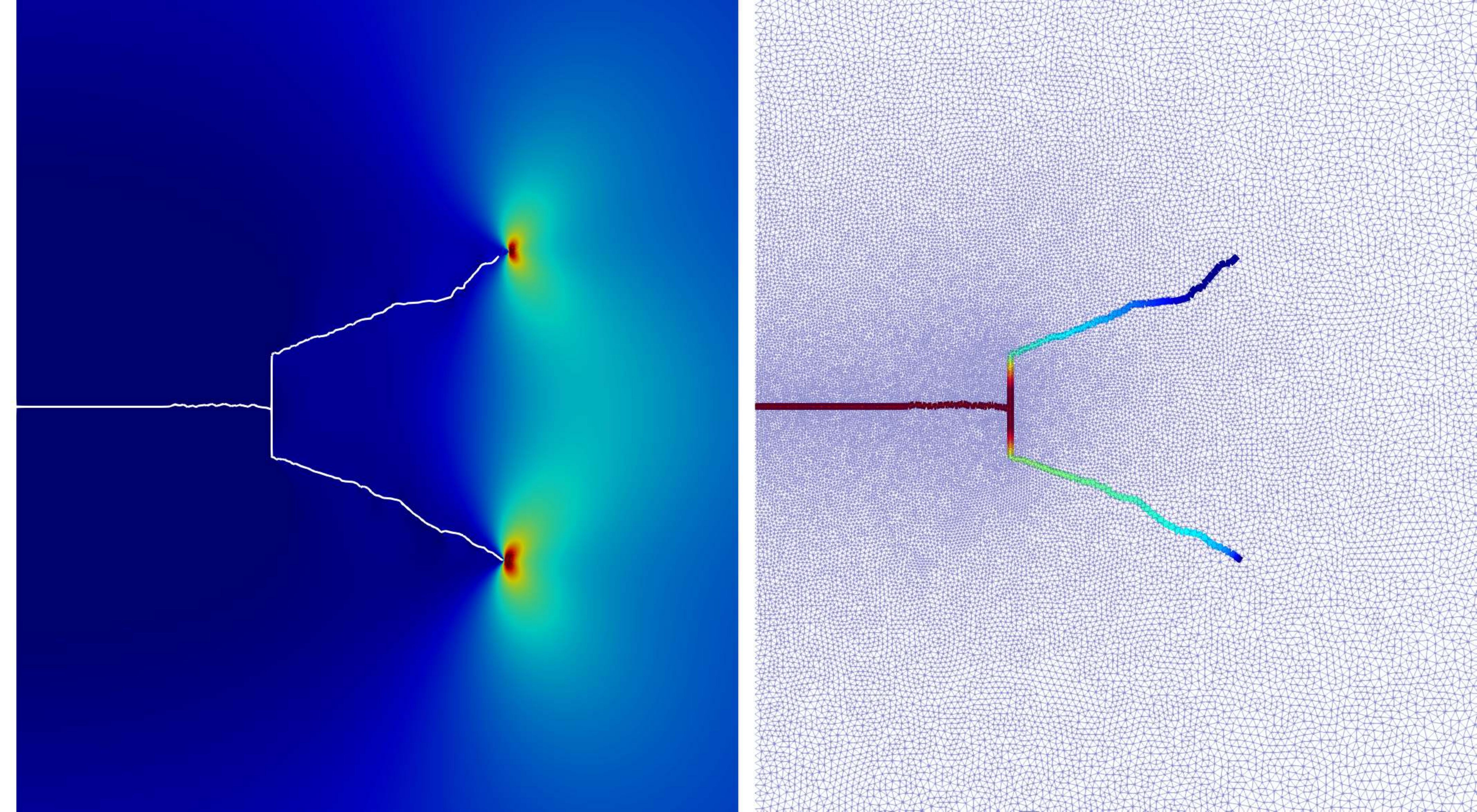} 
\caption{\emph{$t = 116 s$}}
\end{subfigure}
\caption{\emph{
Snapshots of the time evolution of the stress field (left) and fluid pressure field (right).  
Fluid is injected in the $6 \ m$ long initial crack. As opening stress builds up and reaches 
the critical stress, the crack propagates and merges with the pre-existing dry crack. 
The fluid fills up previously empty space in the dry crack, applying increasing pressure at the
crack lips and leading eventually to further propagation.}} 
\label{fig:natural-fracture}
\end{figure}

Figure \ref{fig:natural-fracture} shows the time evolution of the solid stress and 
fluid pressure fields. As fluid is injected in the pressurized crack, opening stress builds
up leading to fracture propagation. The pressurized crack merges with the dry crack,
where fluid fills the previously empty space reaching the crack tips. As pressure builds up in
the newly activated crack, further crack propagation occurs. It is important to point out that, 
although the resulting crack path looks qualitatively symmetric, the unstructured character of 
the mesh leads to actual symmetry breakdown.

\begin{figure}[h]
\centering
\includegraphics[width=0.8\textwidth]{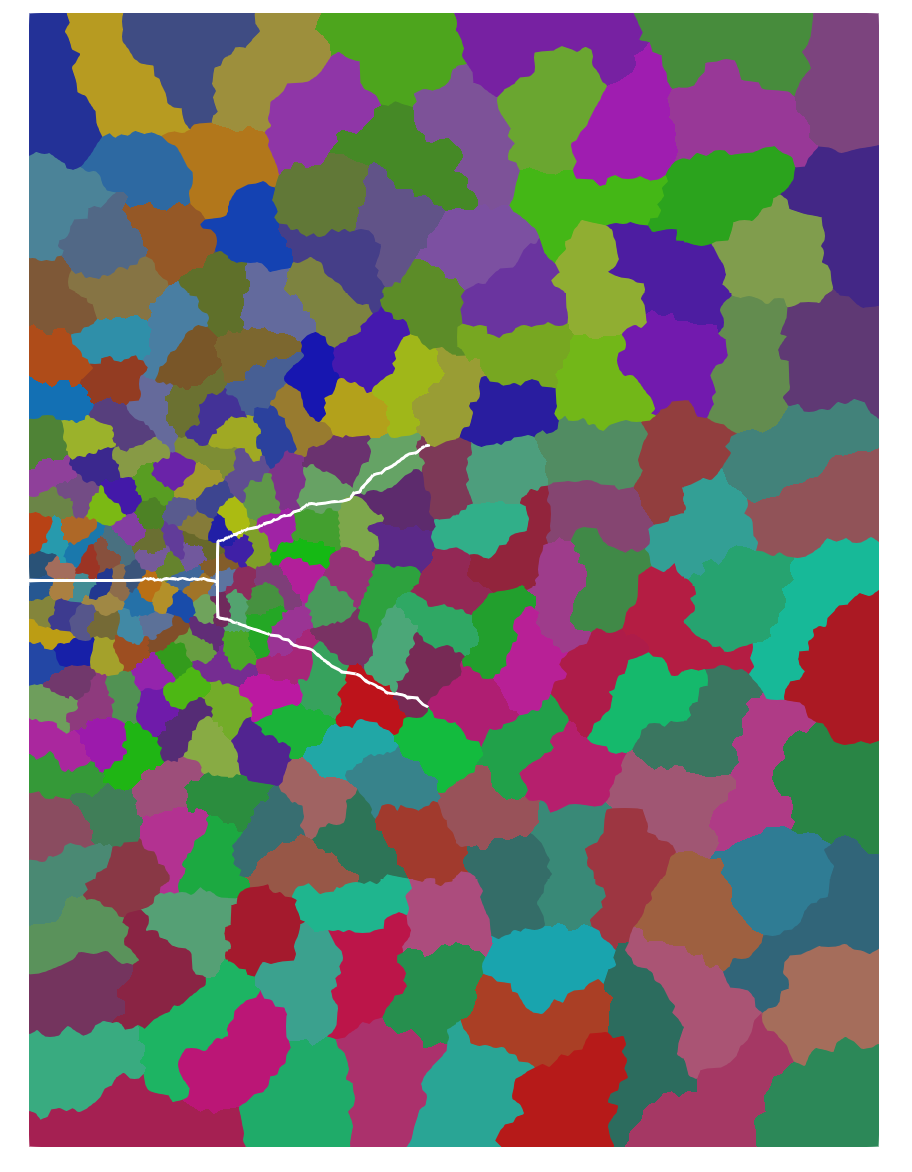} 
\caption{\emph{
Final state of a parallel simulation involving a propagating fluid-filled crack
interacting with a pre-existing crack and further branching. 
Each color represents one of the 256 different mesh partitions in the parallel calculation.}} 
\label{fig:processor-boundary}
\end{figure}

This simulation was run in parallel on 256 processors. Figure \ref{fig:processor-boundary} shows the 
final state of the calculation, highlighting the seamless propagation of cracks across processor boundaries.
\comment{(each partition of the domain is painted with a different color)}
\comment{
Figure \ref{fig:processor-boundary} is a demonstration of the seamless crack propagation across 
processors boundaries.}

\clearpage
\subsection{Assessment of parallel scalability}

\begin{figure}[h]
\centering
\includegraphics[width=0.7\textwidth]{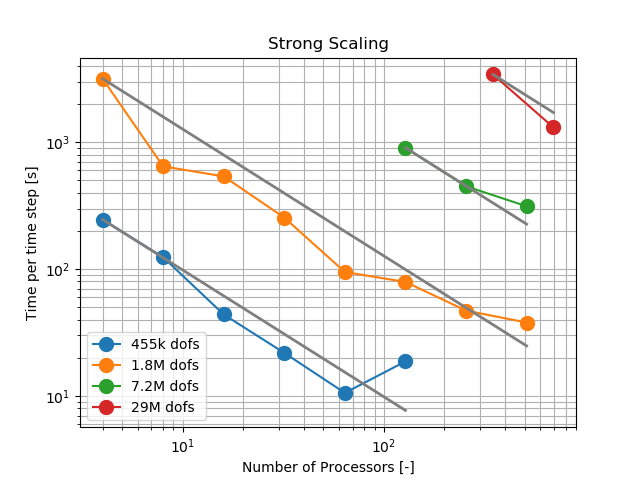} 
\caption{\emph{Strong scaling analysis of the wall time per time step against the number of processors
employed in the parallel calculation.}} \label{fig:strong-scaling}
\end{figure}

In order to assess the parallel scalability of the proposed computational framework 
we performed a strong scaling test on a cluster consisting of 30 nodes, each node having 24 
Intel 2.3 GHz Xeon E5-2670 64-bit processors with 62.8 GB memory.
In Figure \ref{fig:strong-scaling} we report the wall time required to perform one time
step of the benchmark presented in Section \ref{sec:results:verification-propagation} 
(typically consisting of 5-10 iterations of the nonlinear solver), for four different 
problem sizes ($455,000$, $1,800,000$, $7,200,000$, and $29,000,000$ degrees of freedom
in the fully-coupled system of equations) and for a number of processor ranging from 4 to 700.
Figure \ref{fig:strong-scaling} shows that in the proposed framework parallel
computations can lead to a significant reduction in the simulation time, \emph{e.g.} the
computation time per time step can be reduced 
from $53$ minutes to $47$ seconds
by employing 256 processors instead of 4 for a problem size of $1.8$ million degrees of freedom,
and from $57$ to $22$ minutes by employing 700 processors instead of 350 for a problem size of 
$29$ million degrees of freedom. Possible sources of parallel inefficiencies include the slight 
imbalance of workload among processors due to the inactive degrees of freedom in the fluid 
domain of most of the processors.

\clearpage
\section{Conclusions} \label{sec:conclusions}

We presented a computational framework to simulate fluid-driven fracture propagation 
in impermeable solids, addressing the issue of enabling propagation of cracks with 
arbitrarily intricate paths in a suitable framework for parallel large-scale simulations. 

We adopted a hybrid approach in which the solid is described within the DG/CZM framework 
\cite{seagraves:2010}, which facilitates the description of crack propagation,coupled 
with standard lubrication interface finite elements placed on all mesh interfaces, where 
fluid flow is activated in newly created channels.
We solved the discretized fully-coupled nonlinear system with a conventional Newton-Raphson 
procedure, which we found robust, provided that the fluid domain is kept fixed during the 
nonlinear iterations and updated only upon convergence.
We showed that a \emph{fully-coupled} treatment of the solid and fluid problems is 
essential, as a \emph{staggered} solution strategy suffers from both ill-posedness and 
convergence issues, as it hides the nonlinearity, which appears only in the fluid-solid coupling.

We extensively verified the proposed framework in simplified configurations that admit
analytical solutions, namely in the cases of plane strain and penny-shaped cracks. 
We also showed that the method is able to describe intricate crack paths, 
including crack branching and merging, in a robust way.
We demonstrated that the proposed approach can handle the limitation of constraining the 
paths to the interelement boundary with a proper mesh resolution, 
while still retaining the correct description of the fluid equations and the massive scalability,
which we have assessed up to 30 million degrees
of freedom on a 700 processors distributed memory machine.

\bibliography{literature}

\begin{thebibliography}{10}
\expandafter\ifx\csname url\endcsname\relax
  \def\url#1{\texttt{#1}}\fi
\expandafter\ifx\csname urlprefix\endcsname\relax\def\urlprefix{URL }\fi
\expandafter\ifx\csname href\endcsname\relax
  \def\href#1#2{#2} \def\path#1{#1}\fi

\bibitem{Lecampion:2017}
B.~Lecampion, A.~Bunger, X.~Zhang, {Numerical methods for hydraulic fracture
  propagation: a review of recent trends}, Journal of Natural Gas Science {\&}
  Engineering 49 (2017) 66--83.
\newblock \href {http://dx.doi.org/10.1016/j.jngse.2017.10.012}
  {\path{doi:10.1016/j.jngse.2017.10.012}}.

\bibitem{EconomidesBook:2000}
M.~Economides, K.~Nolte, Reservoir Stimulation, 3rd Edition, John Wiley \&
  Sons, Chichester, UK, 2000.

\bibitem{Khristianovich:1955}
S.~Khristianovich, Y.~Zheltov, Formation of vertical fractures by means of
  highly viscous liquid, in: Proceedings Fourth World Petroleum Congress,
  Section II/T.O.P., Paper 3, Rome, 1955, pp. 579--586.

\bibitem{Geertsma:1969}
J.~Geertsma, F.~de~Klerk, A rapid method of predicting width and extent of
  hydraulically induced fractures, Journal of Petroleum Technology 21~(12)
  (1969) 1571--1581.
\newblock \href {http://dx.doi.org/10.2118/2458-PA}
  {\path{doi:10.2118/2458-PA}}.

\bibitem{Adachi:2001}
J.~I. Adachi, Fluid-driven fracture in permeable rock, Ph.D. thesis, University
  of Minnesota (2001).

\bibitem{Adachi:2002}
J.~I. Adachi, E.~Detournay, {Self-similar solution of a plane-strain fracture
  driven by a power-law fluid}, International Journal for Numerical and
  Analytical Methods in Geomechanics 26~(6) (2002) 579--604.
\newblock \href {http://dx.doi.org/10.1002/nag.213}
  {\path{doi:10.1002/nag.213}}.

\bibitem{Detournay:2004}
E.~Detournay, \href{file:///tmp/literature/2004-Detournay.pdf}{Propagation
  regimes of fluid-driven fractures in impermeable rocks}, International
  Journal of Geomechanics 4 (2004) 35--45.
\newblock \href {http://dx.doi.org/10.1061/(ASCE)1532-3641(2004)4:1(35)}
  {\path{doi:10.1061/(ASCE)1532-3641(2004)4:1(35)}}.

\bibitem{Garagash-asme:2005}
D.~Garagash, E.~Detournay, Plane-strain propagation of a fluid-driven fracture:
  Small toughness solution, Journal of Applied Mechanics 72~(6) (2005)
  916--928.
\newblock \href {http://dx.doi.org/10.1115/1.2047596}
  {\path{doi:10.1115/1.2047596}}.

\bibitem{Bunger:2005}
A.~P. Bunger, E.~Detournay, D.~I. Garagash,
  \href{file:///tmp/literature/2005-Bunger.pdf}{Toughness-dominated hydraulic
  fracture with leak-off}, International Journal of Fracture 134 (2005)
  175--190.
\newblock \href {http://dx.doi.org/10.1007/s10704-005-0154-0}
  {\path{doi:10.1007/s10704-005-0154-0}}.

\bibitem{Adachi:2008}
J.~I. Adachi, E.~Detournay, {Plane strain propagation of a hydraulic fracture
  in a permeable rock}, Engineering Fracture Mechanics 75~(16) (2008)
  4666--4694.
\newblock \href {http://dx.doi.org/10.1016/j.engfracmech.2008.04.006}
  {\path{doi:10.1016/j.engfracmech.2008.04.006}}.

\bibitem{Boone:1990}
T.~Boone, A.~Ingraffea, \href{file:///tmp/literature/1990-ijnamg-boone.pdf}{A
  numerical procedure for simulation of hydraulically-driven fracture
  propagation in poroelastic media}, International Journal for Numerical and
  Analytical Methods in Geomechanics 14 (1990) 27--47.
\newblock \href {http://dx.doi.org/10.1002/nag.1610140103}
  {\path{doi:10.1002/nag.1610140103}}.

\bibitem{Carrier:2012}
B.~Carrier, S.~Granet, \href{file:///tmp/literature/2012-Carrier.pdf}{Numerical
  modeling of hydraulic fracture problem in permeable medium using cohesive
  zone model}, Engineering Fracture Mechanics 79 (2012) 312--328.
\newblock \href {http://dx.doi.org/10.1016/j.engfracmech.2011.11.012}
  {\path{doi:10.1016/j.engfracmech.2011.11.012}}.

\bibitem{Settgast:2017}
R.~R. Settgast, P.~Fu, S.~D.~C. Walsh, J.~A. White, C.~Annavarapu, F.~J.
  Ryerson, \href{file:///tmp/literature/2017-Settgast.pdf}{A fully coupled
  method for massively parallel simulation of hydraulically driven fractures in
  3-dimensions}, International Journal for Numerical and Analytical Methods in
  Geomechanics 41 (2017) 627--653.
\newblock \href {http://dx.doi.org/10.1002/nag.2557}
  {\path{doi:10.1002/nag.2557}}.

\bibitem{ChenZ:2009}
Z.~Chen, A.~P. Bunger, X.~Zhang, R.~G. Jeffrey,
  \href{file:///tmp/literature/2009-Chen.pdf}{Cohesive zone finite
  element-based modeling of hydraulic fractures}, Acta Mechanica Solida Sinica
  22 (2009) 443--452.
\newblock \href {http://dx.doi.org/10.1016/S0894-9166(09)60295-0}
  {\path{doi:10.1016/S0894-9166(09)60295-0}}.

\bibitem{Sarris:2011}
E.~Sarris, P.~Papanastasiou,
  \href{file:///tmp/literature/2011-ijf-SarrisPapanastasiou.pdf}{The influence
  of the cohesive process zone in hydraulic fracturing modelling},
  International Journal of Fracture 167 (2011) 33--45.
\newblock \href {http://dx.doi.org/10.1007/s10704-010-9515-4}
  {\path{doi:10.1007/s10704-010-9515-4}}.

\bibitem{Hunsweck:2013}
M.~J. Hunsweck, Y.~Shen, A.~J. Lew, {A finite element approach to the
  simulation of hydraulic fractures with lag}, International Journal for
  Numerical and Analytical Methods in Geomechanics 37~(9) (2013) 993--1015.
\newblock \href {http://dx.doi.org/10.1002/nag} {\path{doi:10.1002/nag}}.

\bibitem{Belytschko:1999}
T.~Belytschko, T.~Black, {Elastic crack growth in finite element with minimal
  remeshing}, International Journal for Numerical Methods in Engineering 45~(5)
  (1999) 601--620.
\newblock \href
  {http://dx.doi.org/10.1002/(SICI)1097-0207(19990620)45:5<601::AID-NME598>3.0.CO;2-S}
  {\path{doi:10.1002/(SICI)1097-0207(19990620)45:5<601::AID-NME598>3.0.CO;2-S}}.

\bibitem{moes:1999}
N.~Mo\"{e}s, J.~Dolbow, T.~Belytschko,
  \href{file:///tmp/literature/1999-ijnme-moes-p131.pdf}{A finite element
  method for crack growth without remeshing}, International Journal for
  Numerical Methods in Engineering 46 (1999) 131--150.

\bibitem{Khoei:2015}
A.~R. Khoei, M.~Hirmand, M.~Vahab, M.~Bazargan,
  \href{file:///tmp/literature/2015-Khoei.pdf}{An enriched fem technique for
  modeling hydraulically driven cohesive fracture propagation in impermeable
  media with frictional natural faults: Numerical and experimental
  investigations}, International Journal for Numerical Methods in Engineering
  104 (2015) 439--468.
\newblock \href {http://dx.doi.org/10.1002/nme.4944}
  {\path{doi:10.1002/nme.4944}}.

\bibitem{Mohammadnejad:2016}
T.~Mohammadnejad, J.~E. Andrade,
  \href{file:///tmp/literature/2016-Mohammadnejad.pdf}{Numerical modeling of
  hydraulic fracture propagation, closure and reopening using xfem with
  application to in-situ stress estimation}, International Journal for
  Numerical and Analytical Methods in Geomechanics 40 (2016) 2033--2060.
\newblock \href {http://dx.doi.org/10.1002/nag.2512}
  {\path{doi:10.1002/nag.2512}}.

\bibitem{Gordeliy:2013}
E.~Gordeliy, A.~Peirce,
  \href{file:///tmp/literature/2013-Gordeliy.pdf}{Coupling schemes for modeling
  hydraulic fracture propagation using the xfem}, Computer Methods in Applied
  Mechanics and Engineering 253 (2013) 305--322.
\newblock \href {http://dx.doi.org/10.1016/j.cma.2012.08.017}
  {\path{doi:10.1016/j.cma.2012.08.017}}.

\bibitem{Gupta:2014}
P.~Gupta, C.~A. Duarte, \href{file:///tmp/literature/2014-Gupta.pdf}{Simulation
  of non-planar three-dimensional hydraulic fracture propagation},
  International Journal for Numerical and Analytical Methods in Geomechanics 38
  (2014) 1397--1430.
\newblock \href {http://dx.doi.org/10.1002/nag} {\path{doi:10.1002/nag}}.

\bibitem{Gupta:2016}
P.~Gupta, C.~A. Duarte, {Coupled formulation and algorithms for the simulation
  of non-planar three-dimensional hydraulic fractures using the generalized
  finite element method}, International Journal for Numerical and Analytical
  Methods in Geomechanics 40 (2016) 1402--1437.
\newblock \href {http://dx.doi.org/10.1002/nag.2485}
  {\path{doi:10.1002/nag.2485}}.

\bibitem{Gupta:2018}
P.~Gupta, C.~A. Duarte, {Coupled hydromechanical-fracture simulations of
  nonplanar three-dimensional hydraulic fracture propagation}, International
  Journal for Numerical and Analytical Methods in Geomechanics 42~(1) (2018)
  143--180.
\newblock \href {http://dx.doi.org/10.1002/nag.2719}
  {\path{doi:10.1002/nag.2719}}.

\bibitem{Karihaloo:2003}
B.~L. Karihaloo, Q.~Z. Xiao, {Modelling of stationary and growing cracks in FE
  framework without remeshing: A state-of-the-art review}, Computers and
  Structures 81~(3) (2003) 119--129.
\newblock \href {http://dx.doi.org/10.1016/S0045-7949(02)00431-5}
  {\path{doi:10.1016/S0045-7949(02)00431-5}}.

\bibitem{seagraves:2010}
R.~Radovitzky, A.~Seagraves, M.~Tupek, L.~Noels, A scalable {3D} fracture and
  fragmentation algorithm based on a hybrid, discontinuous {G}alerkin,
  {C}ohesive {E}lement {M}ethod, Computer Methods in Applied Mechanics and
  Engineering 200 (2011) 326--344.
\newblock \href {http://dx.doi.org/10.1016/j.cma.2010.08.014}
  {\path{doi:10.1016/j.cma.2010.08.014}}.

\bibitem{francfort:1998}
G.~A. Francfort, J.-J. Marigo, Revisiting brittle fracture as an energy
  minimization problem, Journal of the Mechanics and Physics of Solids 46
  (1998) 1319--1342.
\newblock \href {http://dx.doi.org/10.1016/S0022-5096(98)00034-9}
  {\path{doi:10.1016/S0022-5096(98)00034-9}}.

\bibitem{bourdin:2000}
B.~Bourdin, G.~A. Francfort, J.-J. Marigo, Numerical experiments in revisited
  brittle fracture, Journal of the Mechanics and Physics of Solids 48~(4)
  (2000) 797--826.
\newblock \href {http://dx.doi.org/10.1016/S0022-5096(99)00028-9}
  {\path{doi:10.1016/S0022-5096(99)00028-9}}.

\bibitem{Bourdin:2007}
B.~Bourdin, {The variational formulation of brittle fracture: numerical
  implementation and extensions}, IUTAM Symposium on Discretization Methods for
  Evolving Discontinuities (2007) 381--393.

\bibitem{Wheeler:2014}
M.~F. Wheeler, T.~Wick, W.~Wollner,
  \href{file:///tmp/literature/2014-Wheeler.pdf}{An augmented-lagrangian method
  for the phase-field approach for pressurized fractures}, Computer Methods in
  Applied Mechanics and Engineering 271 (2014) 69--85.
\newblock \href {http://dx.doi.org/10.1016/j.cma.2013.12.005}
  {\path{doi:10.1016/j.cma.2013.12.005}}.

\bibitem{Verhoosel:2013}
C.~V. Verhoosel, R.~de~Borst, {A phase-field model for cohesive fracture},
  International Journal for Numerical Methods in Engineering 96 (2013) 43--62.
\newblock \href {http://dx.doi.org/10.1002/nme.4553}
  {\path{doi:10.1002/nme.4553}}.

\bibitem{Landis:2014}
Z.~Wilson, C.~Landis, {Phase-field modeling of hydraulic fracture}, Journal of
  the Mechanics and Physics of Solids 96 (2016) 264--290.
\newblock \href {http://dx.doi.org/10.1016/j.jmps.2016.07.019}
  {\path{doi:10.1016/j.jmps.2016.07.019}}.

\bibitem{Miehe:2015}
C.~Miehe, S.~Mauthe, {Phase field modeling of fracture in multi-physics
  problems. Part III. Crack driving forces in hydro-poro-elasticity and
  hydraulic fracturing of fluid-saturated porous media}, Computer Methods in
  Applied Mechanics and Engineering 304 (2015) 619--655.
\newblock \href {http://dx.doi.org/10.1016/j.cma.2015.09.021}
  {\path{doi:10.1016/j.cma.2015.09.021}}.

\bibitem{Lee:2016}
S.~Lee, A.~Mikeli{\'{c}}, M.~F. Wheeler, T.~Wick, Phase-field modeling of
  proppant-filled fractures in a poroelastic medium, Computer Methods in
  Applied Mechanics and Engineering 312 (2016) 509--541.
\newblock \href {http://dx.doi.org/10.1016/j.cma.2016.02.008}
  {\path{doi:10.1016/j.cma.2016.02.008}}.

\bibitem{Giovanardi:2017}
B.~Giovanardi, A.~Scotti, L.~Formaggia,
  \href{file:///tmp/literature/2017-Giovanardi.pdf}{A hybrid xfem -phase field
  (xfield) method for crack propagation in brittle elastic materials}, Computer
  Methods in Applied Mechanics and Engineering 320 (2017) 396--420.
\newblock \href {http://dx.doi.org/10.1016/j.cma.2017.03.039}
  {\path{doi:10.1016/j.cma.2017.03.039}}.

\bibitem{Seagraves:2015a}
A.~Seagraves, R.~Radovitzky, {Large-scale 3D modeling of projectile impact
  damage in brittle plates}, Journal of the Mechanics and Physics of Solids 83
  (2015) 48--71.
\newblock \href {http://dx.doi.org/10.1016/j.jmps.2015.06.001}
  {\path{doi:10.1016/j.jmps.2015.06.001}}.

\bibitem{Serebrinsky:2016}
S.~A. Serebrinsky, M.~S{\'a}nchez, D.~Smilovich, R.~Toscano, A.~Rosolen, M.~B.
  Goldschmit, E.~N. Dvorkin, R.~Radovitzky, {Desarrollo y Validaci{\'{o}}n de
  un Simulador de Fracturamiento Hidr{\'{a}}ulico Orientado al Petr{\'{o}}leo y
  Gas}, in: Congreso sobre M{\'{e}}todos Num{\'{e}}ricos y sus Aplicaciones,
  Vol. XXXIV, 2016, pp. 8--11.

\bibitem{Hirmand:2019}
M.~Hirmand, M.~Vahab, K.~Papoulia, N.~Khalili, {Robust simulation of dynamic
  fluid-driven fracture in naturally fractured impermeable media}, Computer
  Methods in Applied Mechanics and Engineering 357 (2019) 112574.
\newblock \href {http://dx.doi.org/10.1016/j.cma.2019.112574}
  {\path{doi:10.1016/j.cma.2019.112574}}.

\bibitem{Vahab:2018}
M.~Vahab, N.~Khalili, {Computational algorithm for the anticipation of the
  fluid-lag zone in hydraulic fracturing treatments}, International Journal of
  Geomechanics 18~(10) (2018) 1--15.
\newblock \href {http://dx.doi.org/10.1061/(ASCE)GM.1943-5622.0001273}
  {\path{doi:10.1061/(ASCE)GM.1943-5622.0001273}}.

\bibitem{Howard:1957}
G.~Howard, C.~R. Fast, \href{file:///tmp/literature/1957-Howard.pdf}{Optimum
  fluid characteristics for fracture extension}, Proceedings of the American
  Petroleum Institute (1957) 261--270.

\bibitem{Giovanardi:2018}
B.~Giovanardi, L.~Formaggia, A.~Scotti, P.~Zunino,
  \href{file:///tmp/literature/2018-Giovanardi.pdf}{Unfitted fem for modelling
  the interaction of multiple fractures in a poroelastic medium}, Lecture Notes
  in Computational Science and Engineering 121.
\newblock \href {http://dx.doi.org/10.13140/RG.2.2.33012.35207}
  {\path{doi:10.13140/RG.2.2.33012.35207}}.

\bibitem{barenblatt:1962}
G.~Barenblatt, \href{file:///tmp/literature/1962-aam-barenblatt-p55.pdf}{The
  mathematical theory of equilibrium cracks in brittle fracture}, Advances in
  Applied Mechanics 7 (1962) 55--129.
\newblock \href {http://dx.doi.org/10.1016/S0065-2156(08)70121-2}
  {\path{doi:10.1016/S0065-2156(08)70121-2}}.

\bibitem{barenblatt:1959}
G.~I. Barenblatt, \href{file:///tmp/literature/1959-amm-barenblatt.pdf}{The
  formation of equilibrium cracks during brittle fracture. general ideas and
  hypotheses. axially-symmetric cracks.}, Journal of Applied Mathematics and
  Mechanics 23~(3) (1959) 622--636.

\bibitem{camacho:1996}
G.~Camacho, M.~Ortiz,
  \href{file:///tmp/literature/1996-ijss-camacho.pdf}{Computational modeling of
  impact damage in brittle materials}, International Journal of Solids and
  Structures 33~(20--22) (1996) 2899--2983.
\newblock \href {http://dx.doi.org/10.1016/0020-7683(95)00255-3}
  {\path{doi:10.1016/0020-7683(95)00255-3}}.

\bibitem{noels:2006}
L.~Noels, R.~Radovitzky, A general discontinuous {G}alerkin method for finite
  hyperelasticity. {F}ormulation and numerical applications, International
  Journal for Numerical Methods in Engineering 68~(1) (2006) 64--97.
\newblock \href {http://dx.doi.org/10.1002/nme.1699}
  {\path{doi:10.1002/nme.1699}}.

\bibitem{noels:2007b}
L.~Noels, R.~Radovitzky, An explicit discontinuous {G}alerkin method for
  non-linear solid dynamics. {F}ormulation, parallel implementation and
  scalability properties., International Journal for Numerical Methods in
  Engineering 74~(9) (2007) 1393--1420.
\newblock \href {http://dx.doi.org/10.1002/nme.2213}
  {\path{doi:10.1002/nme.2213}}.

\bibitem{METIS}
G.~Karypis, V.~Kumar, {MeTis: Unstructured Graph Partitioning and Sparse Matrix
  Ordering System, Version 4.0}, \url{http://www.cs.umn.edu/~metis} (2009).

\bibitem{PETSc:3.7}
S.~Balay, J.~Brown, P.~Brune, K.~Buschelman, L.~Dalcin, V.~Eijkhout, W.~Gropp,
  D.~Karpeyev, D.~Kaushik, M.~Knepley, L.~C. McInnes, B.~Smith, S.~Zampini,
  H.~Zhang, Petsc users manual, Tech. Rep. ANL-95/11 - Revision 3.7, Argonne
  National Laboratory (2016).

\bibitem{Lecampion:2007}
B.~Lecampion, E.~Detournay, {An implicit algorithm for the propagation of a
  hydraulic fracture with a fluid lag}, Computer Methods in Applied Mechanics
  and Engineering 196 (2007) 4863--4880.
\newblock \href {http://dx.doi.org/10.1016/j.cma.2007.06.011}
  {\path{doi:10.1016/j.cma.2007.06.011}}.

\bibitem{Garagash:1999}
D.~I. Garagash, E.~Detournay, {The Tip Region of a Fluid-Driven Fracture in an
  Elastic Medium}, Journal of Applied Mechanics 67~(1) (1999) 183--192.
\newblock \href {http://dx.doi.org/10.1115/1.321162}
  {\path{doi:10.1115/1.321162}}.

\bibitem{Garagash:2006}
D.~I. Garagash, {Propagation of a plane-strain hydraulic fracture with a fluid
  lag: Early-time solution}, International Journal of Solids and Structures
  43~(18-19) (2006) 5811--5835.
\newblock \href {http://dx.doi.org/10.1016/j.ijsolstr.2005.10.009}
  {\path{doi:10.1016/j.ijsolstr.2005.10.009}}.

\bibitem{Garagash:2019}
D.~I. Garagash, {Cohesive-Zone Effects in Hydraulic Fracture Propagation},
  Journal of the Mechanics and Physics of Solids (2019) 103727\href
  {http://dx.doi.org/10.1016/j.jmps.2019.103727}
  {\path{doi:10.1016/j.jmps.2019.103727}}.

\bibitem{Detournay-ijg:2004}
E.~Detournay, Propagation regimes of fluid-driven fractures in impermeable
  rocks, International Journal of Geomechanics 4~(1) (2004) 35--45.
\newblock \href {http://dx.doi.org/10.1061/(ASCE)1532-3641(2004)4:1(35)}
  {\path{doi:10.1061/(ASCE)1532-3641(2004)4:1(35)}}.

\bibitem{irwin:1957}
G.~Irwin, Analysis of stresses and strains near the end of a crack traversing a
  plate, Journal of Applied Mechanics 24 (1957) 361--364.

\bibitem{Garagash:2000}
D.~Garagash, {Hydraulic fracture propagation in elastic rock with large
  toughness}, in: Proc. 4th North American Rock Mechanics Symp., 2000, pp.
  221--228.

\bibitem{Savitski:2002}
A.~Savitski, E.~Detournay, {Propagation of a penny-shape hydraulic fracture in
  an impermeable rock}, International Journal of Solids and Structures 39
  (2002) 6311--6337.
\newblock \href
  {http://dx.doi.org/http://dx.doi.org/10.1016/S0020-7683(02)00492-4}
  {\path{doi:http://dx.doi.org/10.1016/S0020-7683(02)00492-4}}.

\bibitem{Chuprakov:2013}
D.~Chuprakov, O.~Melchaeva, R.~Prioul, {Hydraulic Fracture Propagation Across a
  Weak Discontinuity Controlled by Fluid Injection}, in: Effective and
  Sustainable Hydraulic Fracturing, IntechOpen, 2013, Ch.~8.
\newblock \href {http://dx.doi.org/http://dx.doi.org/10.5772/57353}
  {\path{doi:http://dx.doi.org/10.5772/57353}}.

\end{thebibliography}


\end{document}